\documentclass[10pt,english,floatfix,twocolumn,superscriptaddress,aps,prd]{revtex4-1}
\usepackage{amsthm}
\usepackage{dsfont}
\usepackage[T1]{fontenc}
\usepackage{geometry}
\usepackage{amsmath}
\usepackage{amssymb}
\usepackage{graphicx}
\usepackage{color}

\makeatletter

\usepackage{babel}

\makeatother

\begin{document}

\begin{flushright}
ADP-15-13/T915
\end{flushright}

\title{
 Statistics of Stationary Points of Random Finite Polynomial Potentials}

\author{Dhagash Mehta}
\email{dmehta@nd.edu}
\affiliation{Department of Applied and Computational Mathematics and Statistics, University of Notre Dame, Notre Dame, IN 46556, USA}
\affiliation{Centre for the Subatomic Structure of Matter, Department of Physics, School of Physical Sciences,
University of Adelaide, Adelaide, South Australia 5005, Australia.}

\author{Matthew Niemerg}
\email{research@matthewniemerg.com}
\affiliation{The Institute for Interdisciplinary Information Sciences,
Tsinghua University, Beijing, China 100084}

\author{Chuang Sun}
\email{chuang.sun@physics.ox.ac.uk}
\affiliation{Rudolf Peierls Centre for Theoretical Physics, University of Oxford, Oxford, OX1 3NP, UK}

\begin{abstract} \noindent The stationary points (SPs) of the potential energy landscapes (PELs) 
of multivariate random potentials (RPs) have found many applications in many areas of Physics, 
Chemistry and Mathematical Biology. 
However, there are few reliable methods available which can find all the 
SPs accurately. Hence, one has to rely on indirect methods such as Random Matrix theory. 
With a combination of the numerical polynomial homotopy continuation 
method and a certification method, we obtain all the certified SPs of
the most general polynomial RP for each sample chosen from the Gaussian distribution with mean 0 and variance 1.
While obtaining many novel results for the finite size case of the RP, we also 
discuss the implications of our results on mathematics of random systems and string theory landscapes.
\end{abstract}

\maketitle

\section{Introduction} 
The surface drawn by a potential,
$V(x)$, with $x=(x_{1},\dots,x_{N})$,
is called the potential energy landscape (PEL) of the corresponding
physical or chemical system \cite{Wales:04,RevModPhys.80.167}. The
stationary points (SPs) of a PEL, defined by the solutions of the equations
$\partial V(x)/\partial x_{i}=0$ for $i=1,\dots,N$, provide important
information about the physics and chemistry of the corresponding system. If the parameters
or the coefficients of $V(x)$ are chosen from some random distribution, then $V(x)$ is called a random potential (RP).
In recent years more attempts to understand the statistical patterns 
behind SPs for random PELs have been made because of its applications in such 
diverse areas as string theory \cite{Denef:2004ze,Douglas:2006es}, cosmology 
(see e.g.~, \cite{Aazami:2005jf,Tye:2008ef,Marsh:2011aa,Aravind:2014aza,Battefeld:2012qx,Bach:2014}),
statistical mechanics \cite{PhysRevLett.92.240601,PhysRevLett.93.149901,PhysRevLett.98.150201,PhysRevLett.109.167203,auf:2013a,auf:2013b,2013JSP...tmp..205F}, neural networks \cite{wainrib2013topological}. 
Similar problems appear in statistics \cite{Cheng:2015} and in a
pure mathematics context, e.g. in topology  \cite{Nicolaescu:2014,FLL:2014,PhysRevLett.108.170601}. In fact,
the mathematical question \textit{how many real solutions, on an average, does a random system of polynomial equations have} is a classic problem.
For random univariate polynomials of degree $D$ with coefficients taking i.~i.~d.~values from the Gaussian distribution with mean 0 and variance 1,
called the Kac formulation, the mean number of real solutions is $\frac{2}{\pi}\ln D$ for $D\rightarrow \infty$ \cite{kac1948average,farahmand1986average}. If
the $i$-th coefficient of a random polynomial is allowed to take i.~i.~d.~values
from the Gaussian distribution with mean 0 and variance ${D \choose i}$, called the Kostlan formulation,
then the mean number of real solutions is $\sqrt{D}$
as $D\rightarrow \infty$
\cite{bogomolny1992distribution,edelman1995many}. There have been various attempts to extend these results to the multivariate case
\cite{kostlan2002expected,azais2005roots,armentano2009random,malajovich2004high,rojas1996average}.

However, progress in studying the statistics of the SPs of PELs has been slow due to conceptual as well as technical
problems. First, from a computational viewpoint, the stationary equations are usually nonlinear. 
Hence, solving these equations is usually extremely difficult. One may
use the Newton-Raphson method, its sophisticated variants, or many other numerical methods
to solve these equations. These methods can find \textit{many} SPs
at best, as opposed to \textit{all} of them. Theoretically, the most recent progress 
was based on successful applications of random matrix theory (RMT) for
exploring the PELs of certain types of RPs. In this approach, the Hessian 
matrix $H_{i,j}=\partial^{2}V(x)/\partial x_{i}\partial x_{j}$,
is treated as a random matrix. 
Then, standard
RMT results can be employed to extract valuable information about the eigenvalue distribution of $H$.
In recent years, there are many results available on the probability that random matrices of various types has indefinite spectrum, see e.g.
\cite{dean2006large,tao2012random,bhargava2015probability}.
However, constraining this analysis for the SPs rather than
on arbitrary points of the $N$-dimensional space amounts to taking the stationary equations under consideration into the matrix 
approach. This can be successfully done only under assumptions of gaussianity, and either conditions of statistical isotropy and 
translational invariance 
\cite{PhysRevLett.92.240601,PhysRevLett.93.149901,PhysRevLett.109.167203} (though this condition can be made milder in certain situations, 
see e.g.~, Sec.~4 in \cite{2013arXiv1307.2379F}) or isotropy and restriction to a spherical surface 
\cite{auf:2013a,auf:2013b,2013arXiv1307.2379F}.  The latter context is natural 
for studying various aspects of glassy transitions such as the number of minima and their complexity, mainly 
in the thermodynamic limit of high dimension as $N\to \infty$ where RMT produces the most explicit and universal results.

There are, however, a few issues with this approach as well.  Many cases exist where $N$
is actually a fundamental parameter in the physical description and may be finite.
Indeed, many physical, chemical or biological systems have a finite number of fields,
particles, neurons, etc, which is represented as the number of variables $N$. 
Moreover, with the RMT approach, it is not yet possible to obtain further information about an individual SP. 
Computing the variance of the number of SPs using RMT is 
also quite difficult, and, as of yet, is still largely an unsolved problem\footnote{See \cite{bleher2015two} for
recent progress on this problem}.
We note that recent progress has been made in
which the average number of SPs of each index (defined as the number of negative eigenvalues of the Hessian matrix evaluated
at the given SP), for any finite $N$, is computed analytically \cite{auf:2013a,auf:2013b,PhysRevLett.109.167203,2013JSP...tmp..205F,2013arXiv1307.2379F}.
The shortcoming of these approaches is that the theory only works for the gaussian 
isotropic RPs with either spherical constraints or statistical translational invariance.

In this article, we provide a numerical scheme to overcome all the difficulties to find the SPs of
a completely generic RP. We use the numerical polynomial homotopy continuation method which finds all the isolated solutions
of a multivariate system of polynomial equations with probability one \cite{Li:2003,SW:05,BHSW13}.
To strengthen our results, we also use a certification approach which
\textit{certifies} if a given numerical approximate corresponds to an exact distinct root of the system. The certification approach
is known as Smale's $\alpha$-theory and certifies that the numerical approximation is in the quadratic convergence region of a
unique nonsingular solution of the system \cite{BCSS,smale1986newton,hauenstein2012algorithm,Mehta:2013zia,Mehta:2014gia}.

In the remainder of the paper, after specifying our RP, we describe the computational methods used in our work in
Section \ref{sec:numerical_setup}. In Section \ref{sec:Results}, we present our results and discuss their implications. 
Finally, in Section \ref{sec:conclusions}, we conclude.

\section{The Most General Polynomial Random Potential and Numerical Set up} \label{sec:numerical_setup}
The most general random multivariate potential with polynomial nonlinearity is
\begin{eqnarray}
V(x)= \sum_{ | \alpha | \leq D} ^{}  a_\alpha x_1^{\alpha_1}...x_N^{\alpha_N}
\label{eq:V}
\end{eqnarray}
where $N$ is the number of scalar fields, $D$ is the highest degree of the monomials,
$\alpha = (\alpha_1,...,\alpha_N) \in  \mathbb{N}^N $ is a multi-integer, and $|\alpha| = \alpha_1 + ... + \alpha_N$. $a_{\alpha}$
are random coefficients which take i.~i.~d.~values from the Gaussian distribution with mean $0$ and variance $1$, known as 
the Kac formulation for the multivariate case.

Finding all the SPs of $V(x)$ boils down to simultaneously solving the system of equations $\frac{\partial V(x)}{\partial x_i} = 0$, 
$i=1,\dots, N$,
each of which is a polynomial equation of degree $D-1$.
The classical Be\'zout theorem states that for generically chosen coefficients, 
the number of complex (which include real) SPs, counting multiplicities,  
is the product of the degrees of each of the stationary equations, $(D-1)^N$. 
Using this fact, we employ the numerical polynomial homotopy continuation (NPHC) method which guarantees 
that we will find all the complex numerical approximates 
of a system of multivariate polynomial (or, having polynomial-like nonlinearity) equations 
\cite{Li:2003,SW:05,BHSW13}. The NPHC method has been used to 
explore potential energy landscapes of various problems arising in physics and chemistry 
\cite{Mehta:2009,Mehta:2009zv,Mehta:2011xs,Mehta:2011wj,Kastner:2011zz,Maniatis:2012ex,Mehta:2012wk,
Hughes:2012hg,Mehta:2012qr,He:2013yk}. 
Here, one first creates an easier to solve system which has the same variables as the original system as well as the same number of solutions as the classical Be\'zout 
count; then, each solution of the new system is tracked to the original system with a single parameter to 
obtain all the numerical approximates of the original system. It is rigorously proven \cite{morgan1987homotopy,morgan1987computing} that 
by tracking the paths over complex space one can be guaranteed to find all the isolated complex solutions of the system using the NPHC method.
The interested reader is referred to the early references
\cite{Li:2003,SW:05,BHSW13,Mehta:2009,Mehta:2009zv,Mehta:2011xs,Mehta:2011wj,Kastner:2011zz,Maniatis:2012ex,Mehta:2012wk,
Hughes:2012hg,Mehta:2012qr,He:2013yk} for a detailed description of the method. For the NPHC method, 
we use the \emph{Bertini} software \cite{BHSW06,BHSW13}.

Though the NPHC method guarantees one will find all numerical solutions, skeptics may 
wonder if the numerical solutions are good enough.
Smale and others defined a ``good enough'' numerical solution of a system of polynomial equations
as a point which is in the quadratic convergence region
of a nearby exact solution of the system \cite{BCSS,smale1986newton}. Once the numerical solution is good enough in the above sense, 
it can then be 
approximated to arbitrary accuracy. Such a numerical solution
is called a \textit{certified} solution. Smale also showed how to certify a numerical solution using
only data such as the Jacobian and higher derivatives at the numerical solution via $\alpha$-theory.  A nontrivial result of Smale is that 
if $\alpha$, which is computed using the Jacobian and higher derivatives of the system of equations, at
a given numerical solution is less than $(13-3\sqrt{7})/4$, then the numerical solution is within the quadratic convergence region 
of the nearby exact solution of the system.  Hence, this approximate is a certified numerical solution.
We use an implementation of $\alpha$-theory provided in 
{\tt alphaCertified}~\cite{hauenstein2012algorithm} in the present work (see Refs.~\cite{Mehta:2013zia,Mehta:2014gia} 
for certification of the PELs). We call a solution a real finite solution if the imaginary part of each variable at the solution is $10^{-10}$,  
the maximum eigenvalue of the Hessian matrix at that solution is $\leq 10^{13}$, and it is certified using {\tt alphaCertified}.

In practice, we must restrict ourselves to $N = 2, \dots, 11$ and $D = 3, \dots, 5$ and the sample size for each pair of $(N,D)$ is $1000$. 
The limitation comes due to the combined computation of solving equations, certifying solutions and computing Hessian eigenvalues.

\section{Results and Discussion} \label{sec:Results}
%
%

In this Section, we present our results from our numerical experiments. 
We first find all the isolated complex SPs of $V(x)$ for each of the 1000 samples, and then compute
various quantities from the SPs. We present the results for
$N=2\dots11$ and $D=3\dots5$. We also discuss the results in a physical and mathematical context. 

\subsection{Average number of real SPs} 
\begin{figure}[htp]
\includegraphics[width=8.0cm]{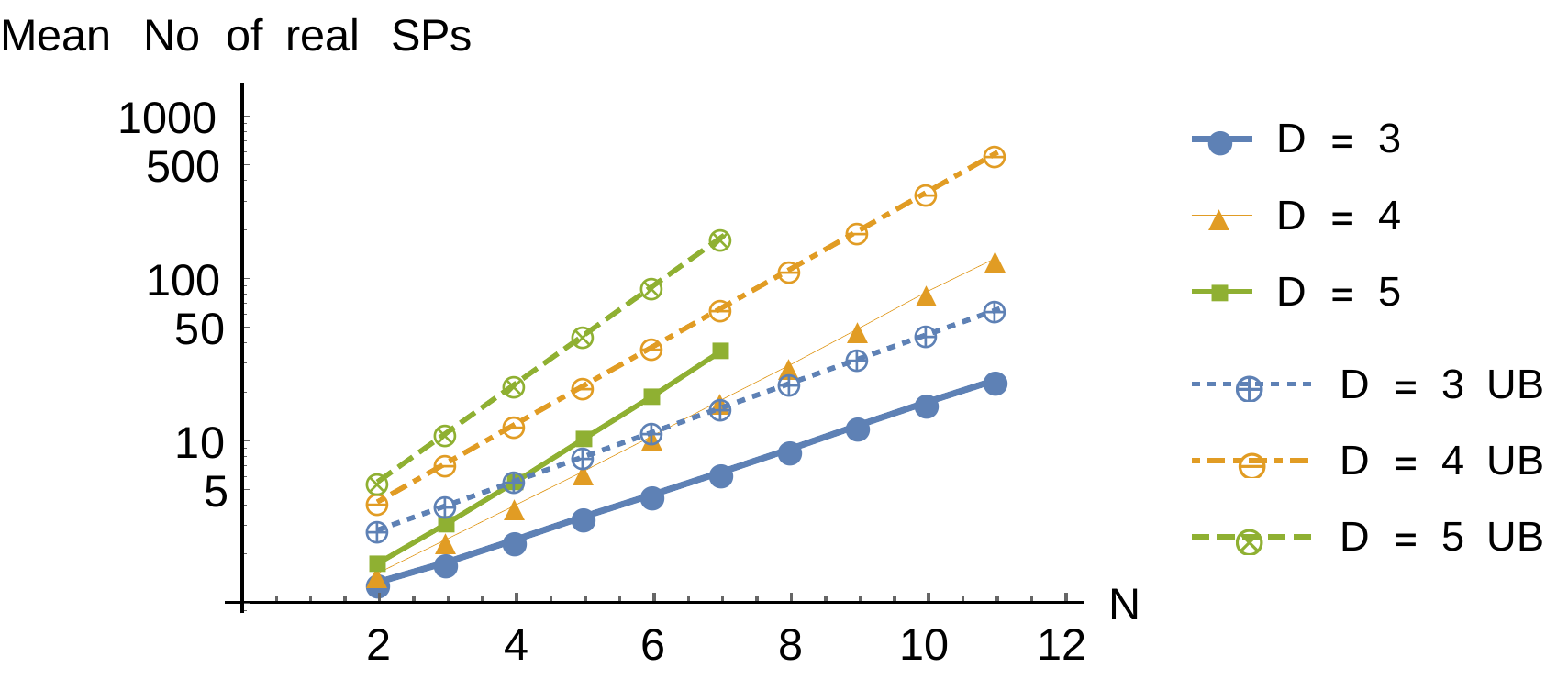}
\caption{Mean number of SPs as a function of $N$ for various values of $D$. Here, 'UB' means 
the upper bound $~\sqrt{2}(D-1)^{(N+1)/2}$ \cite{dedieu2008number}.} \label{MeanRealSP}
\end{figure}

The number of complex solutions of the stationary equations for 
generic coefficients is always $(D-1)^N$ for our RP.
A more interesting quantity is the mean number of real SPs which we plot as a function of
$N$ for different values of $D$ in Figure \ref{MeanRealSP}. For limited values of $D$,
a similar plot was drawn in \cite{Greene:2013ida} for a different RP which was a Taylor expansion and 
coefficients drawn from a uniform distribution. Remarkably, we see similar behaviour of the mean number of SPs, namely, the number
grows rapidly as $N$ increases. In fact, in Figure \ref{MeanRealSP},
we compare our numerical results for the mean number of real SPs as a function of $N$ 
with an analytically computed upper bound, $~\sqrt{2}(D-1)^{(N+1)/2}$, 
computed in \cite{dedieu2008number} where 
the coefficients were i.~i.~d.~picked from the Gaussian distribution
with mean 0 and variance being a multinomial, called Kostlan-Shub-Smale formulation 
\cite{kostlan2002expected, edelman1995many}. 
While the numerical results are in fact far below from the upper bound, this should be expected since it is 
known that mean number of real SPs for the Kostlan formulation is significantly higher than the Kac formulation \cite{edelman1995many}.

\begin{figure}[htp]
\includegraphics[width=8.0cm]{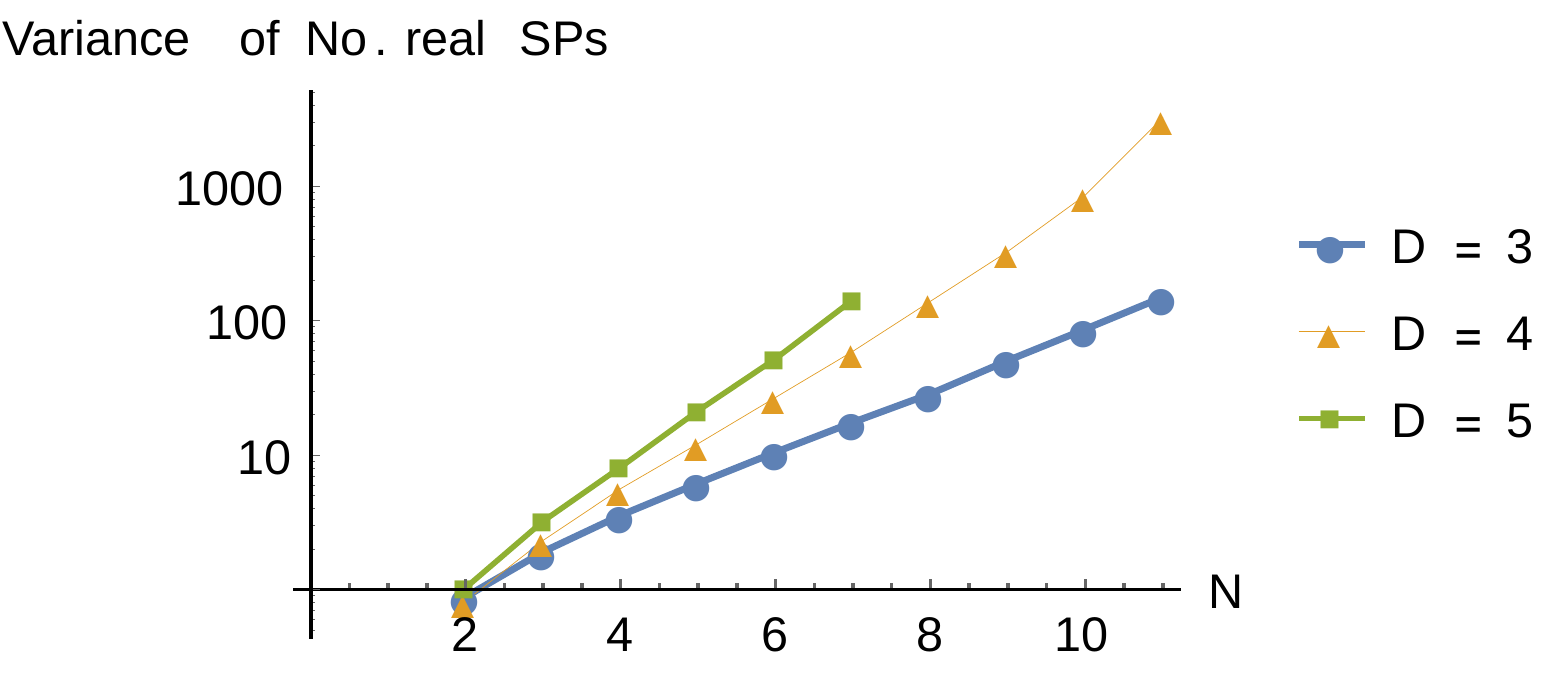}
\caption{Variance of the number of SPs as a function of $N$ for various values of $D$} \label{VarRealSP}
\end{figure}

\begin{figure}[htp]
\includegraphics[width=4cm]{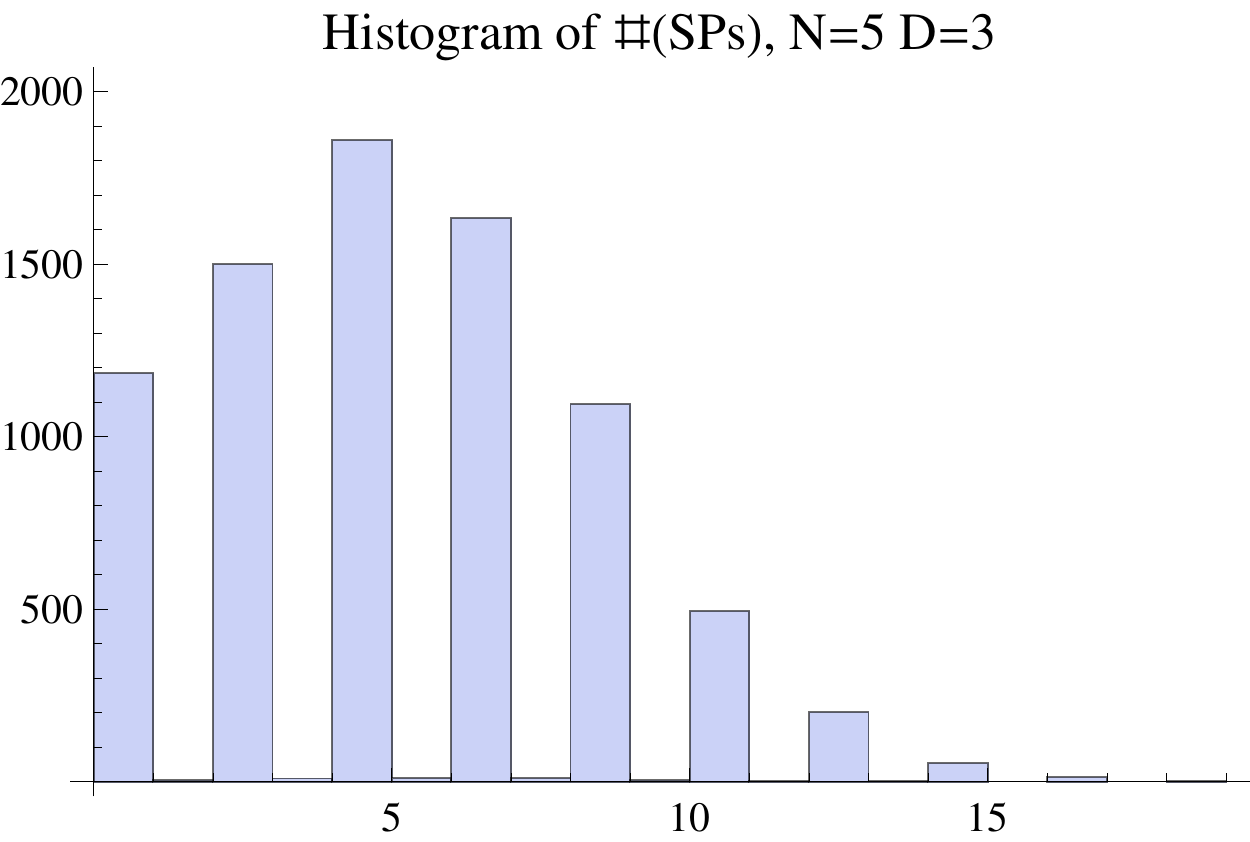}
\includegraphics[width=4cm]{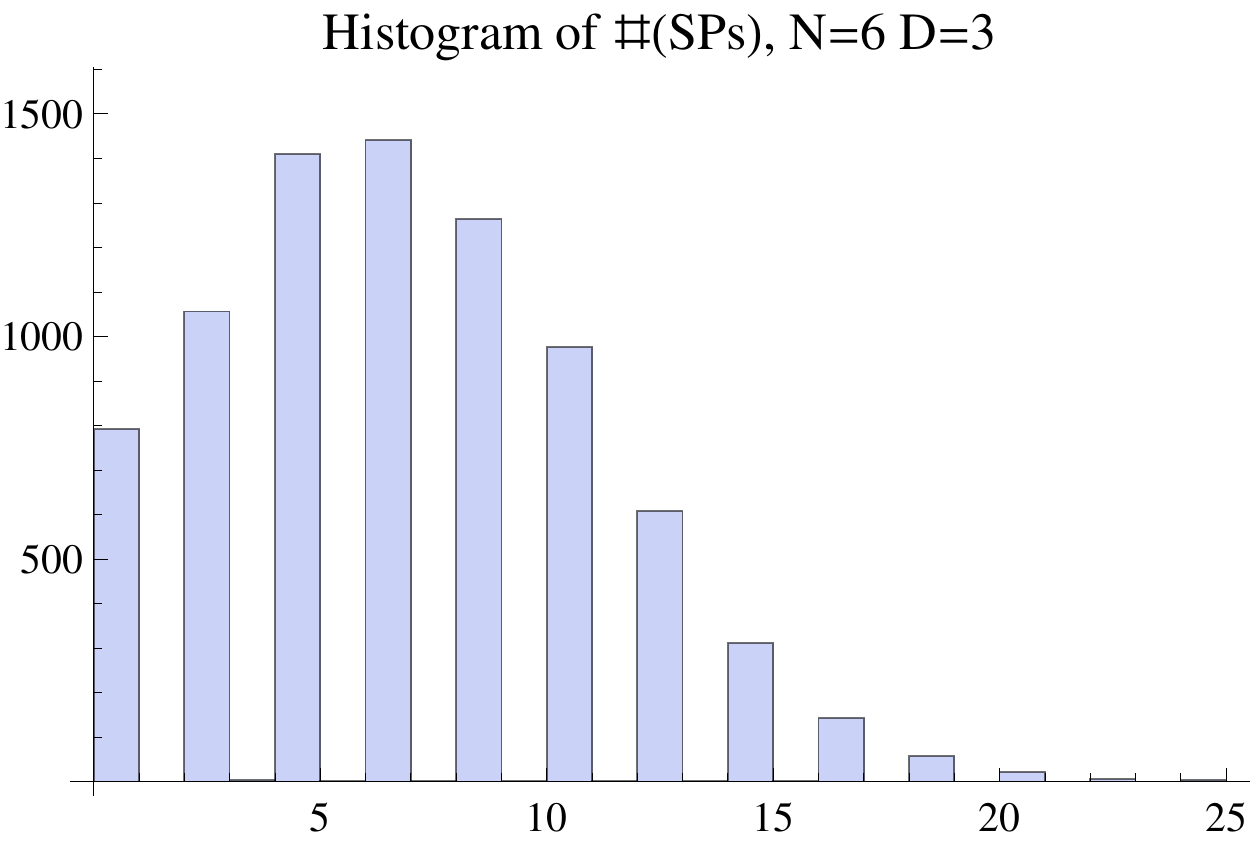}\\
\includegraphics[width=4cm]{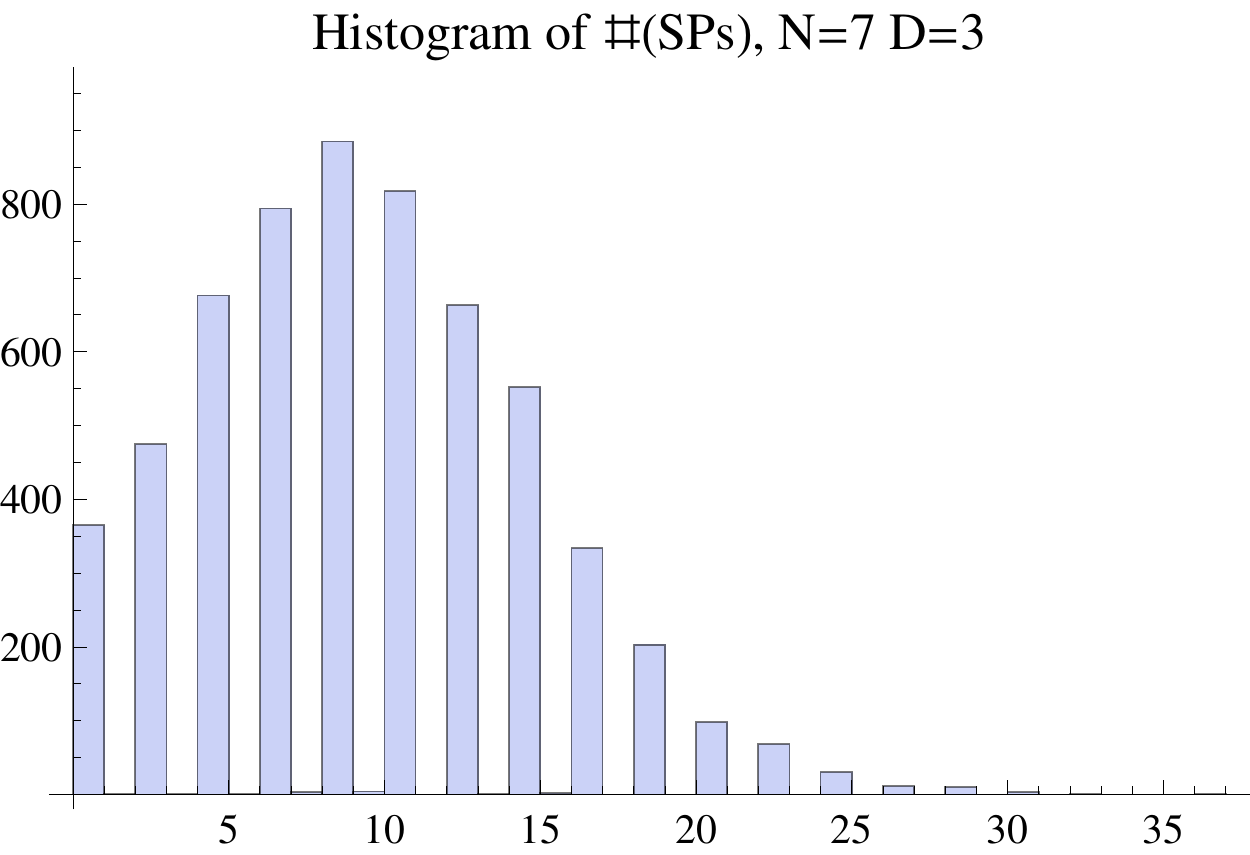}
\includegraphics[width=4cm]{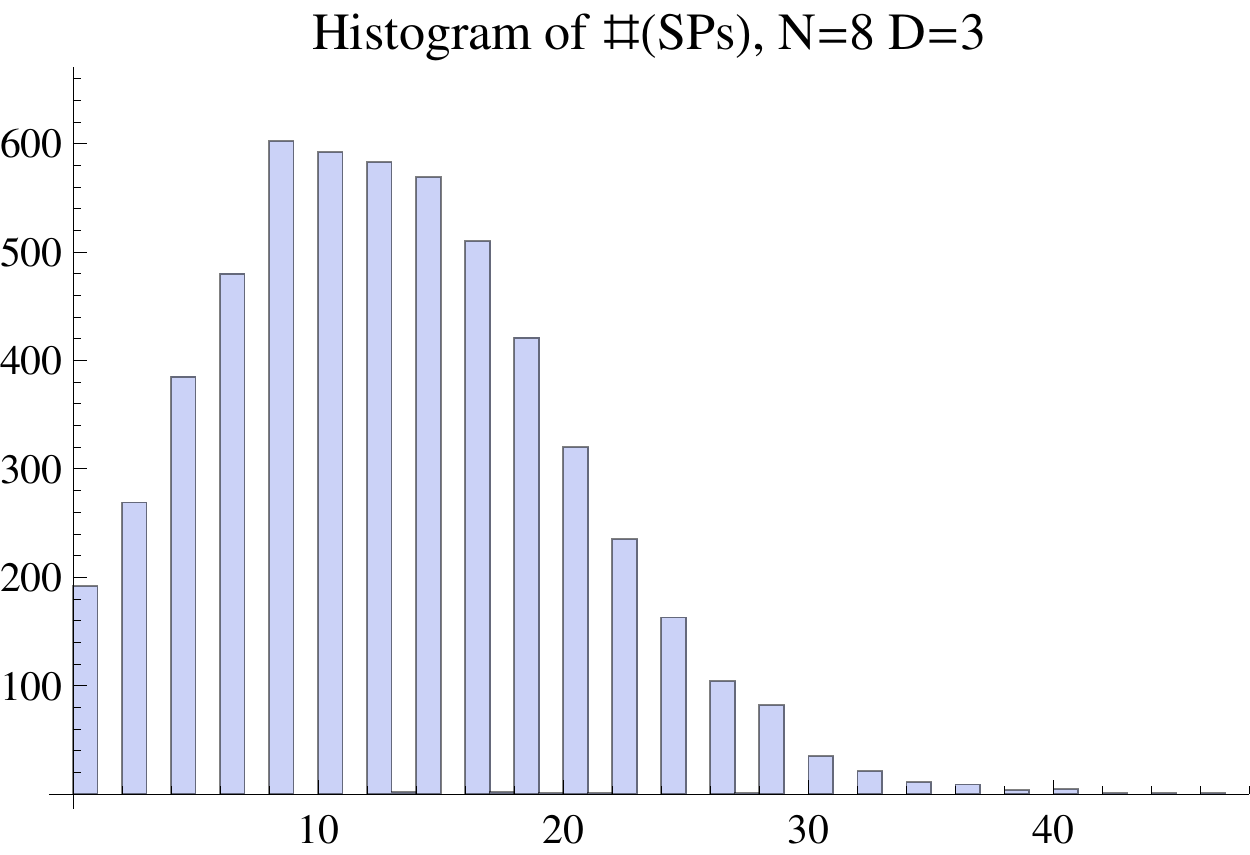}\\
\includegraphics[width=4cm]{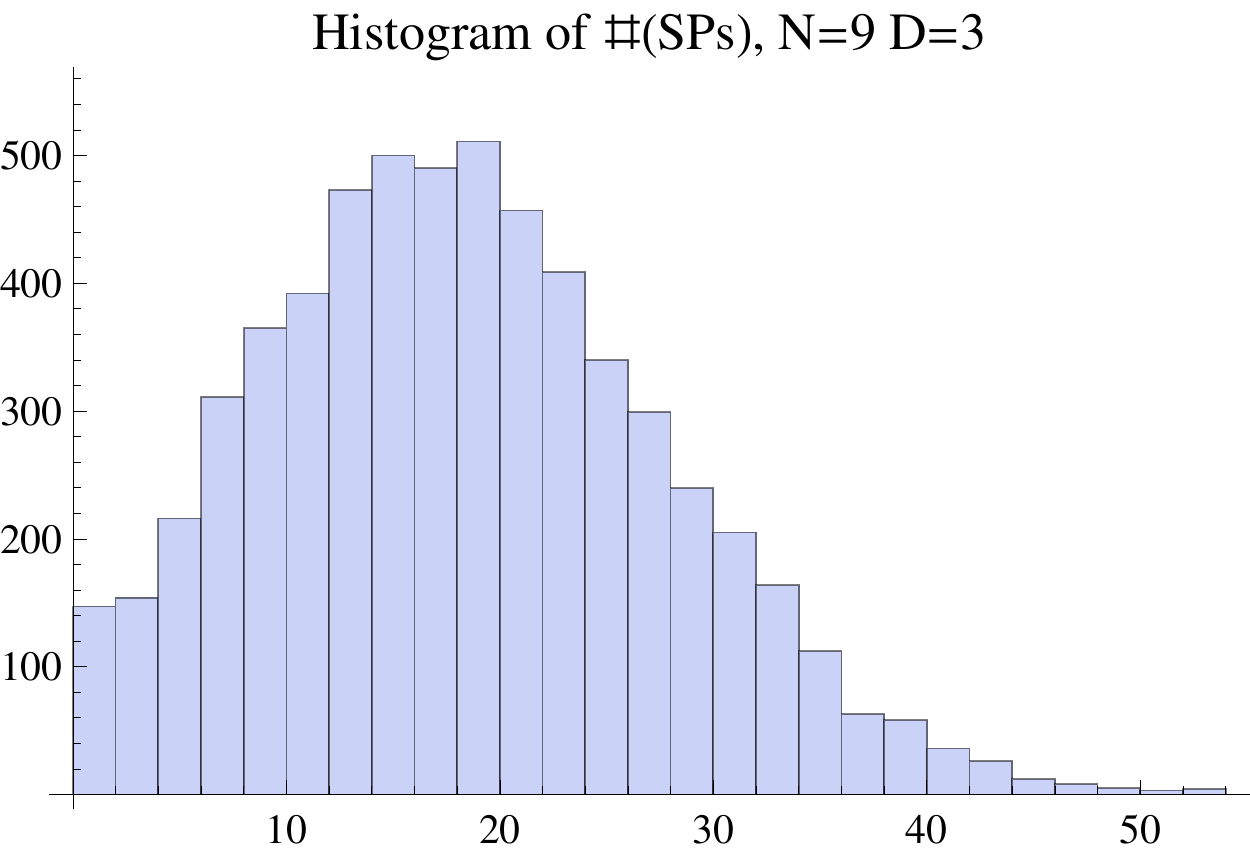}
\includegraphics[width=4cm]{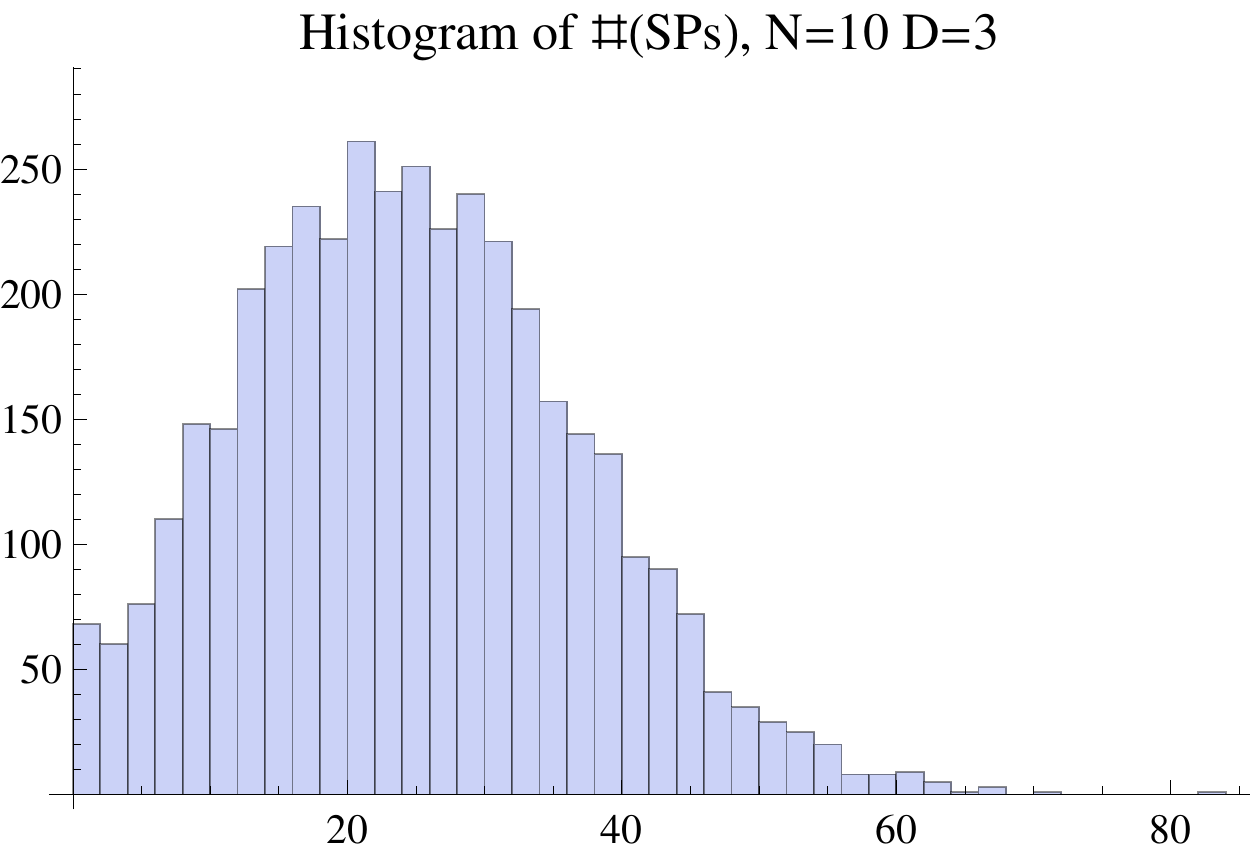}\\
\caption{Histogram of the numbers of SPs for $D=3$. $N=5, 6$ in the top row, $N=7,8$ in the middle row, and 
$N=9,10$ in the bottom row.} \label{HistSPD3}
\end{figure}

\begin{figure}[htp]
\includegraphics[width=4cm]{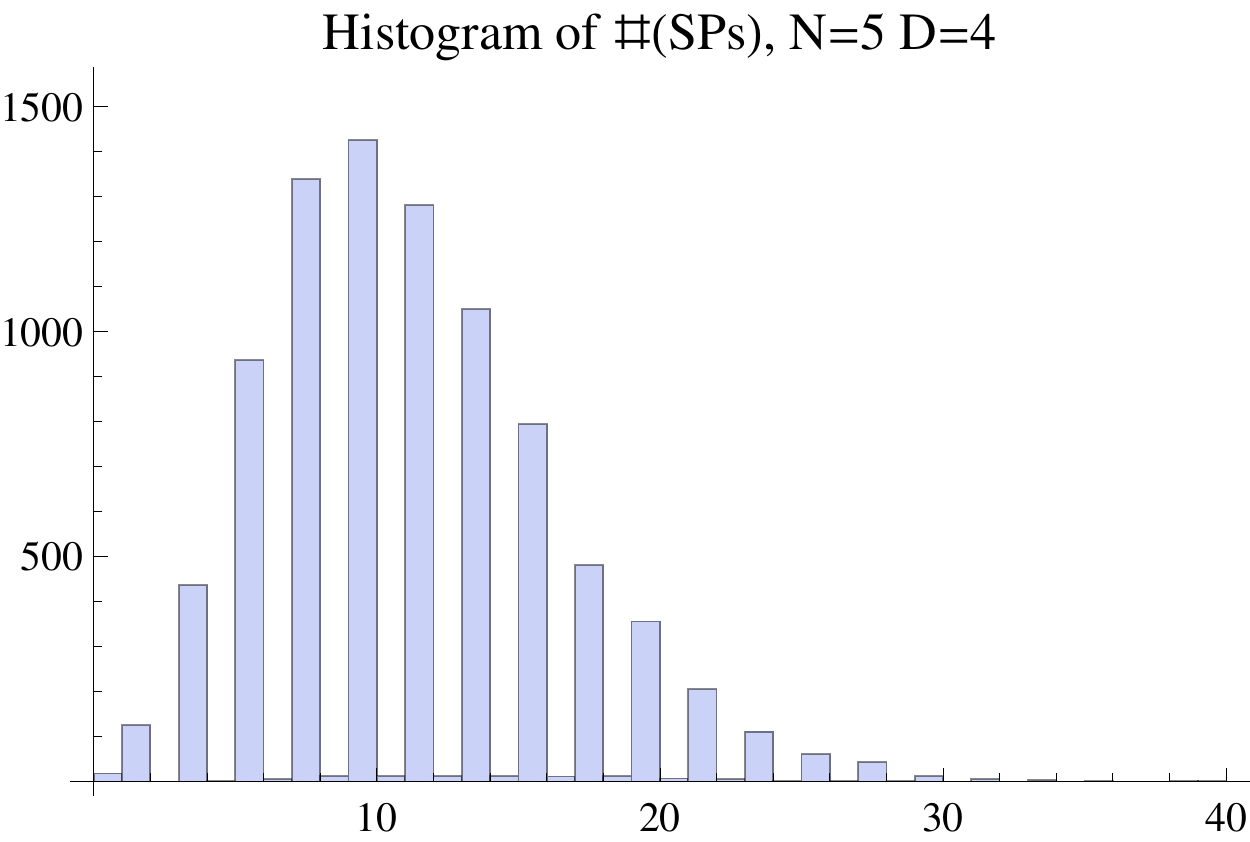}
\includegraphics[width=4cm]{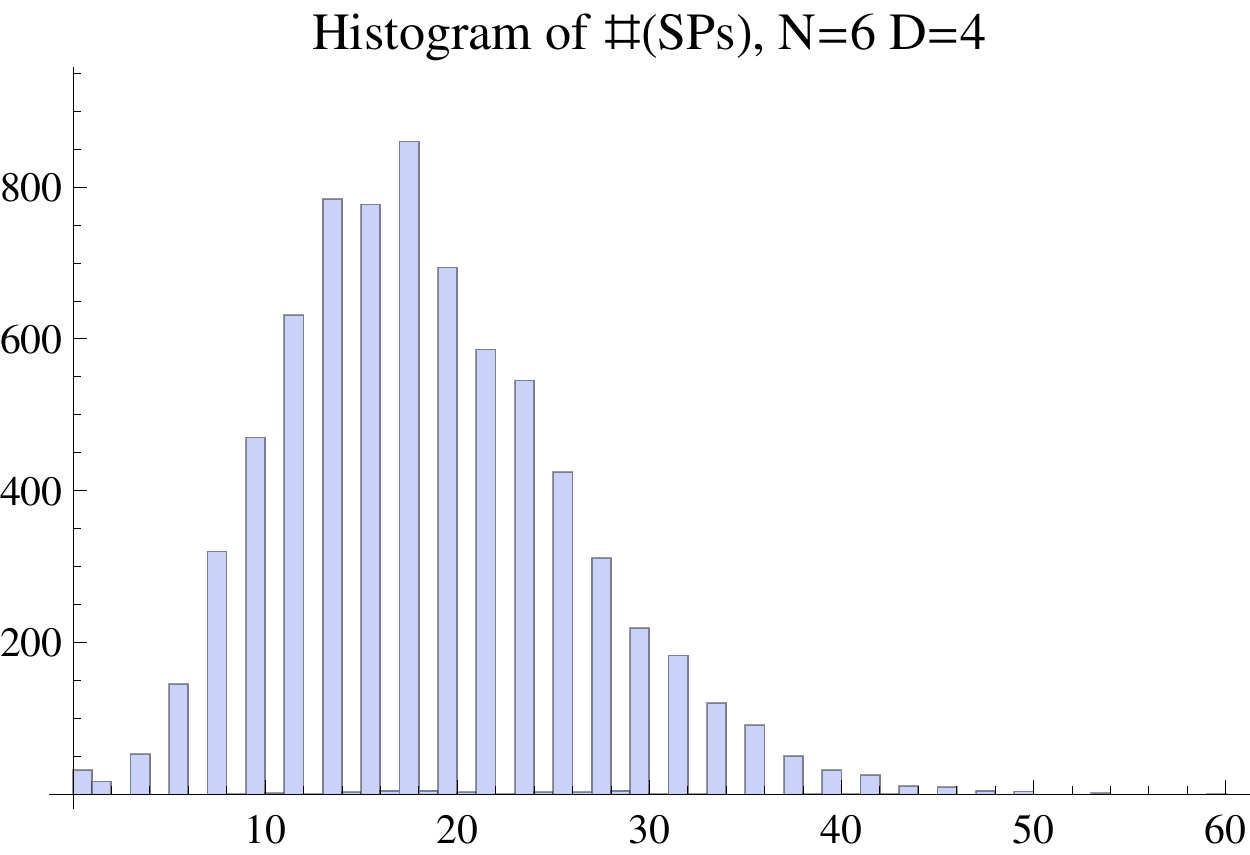}\\
\includegraphics[width=4cm]{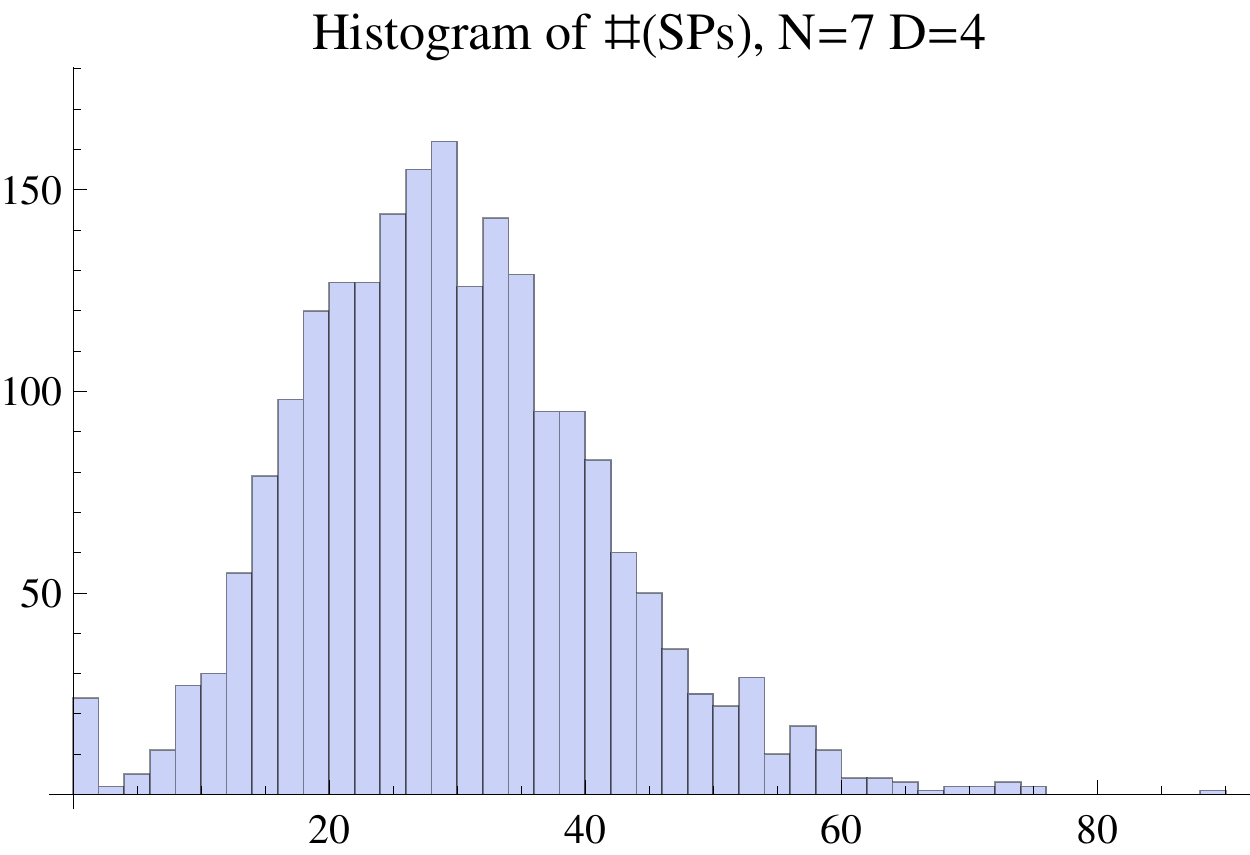}
\includegraphics[width=4cm]{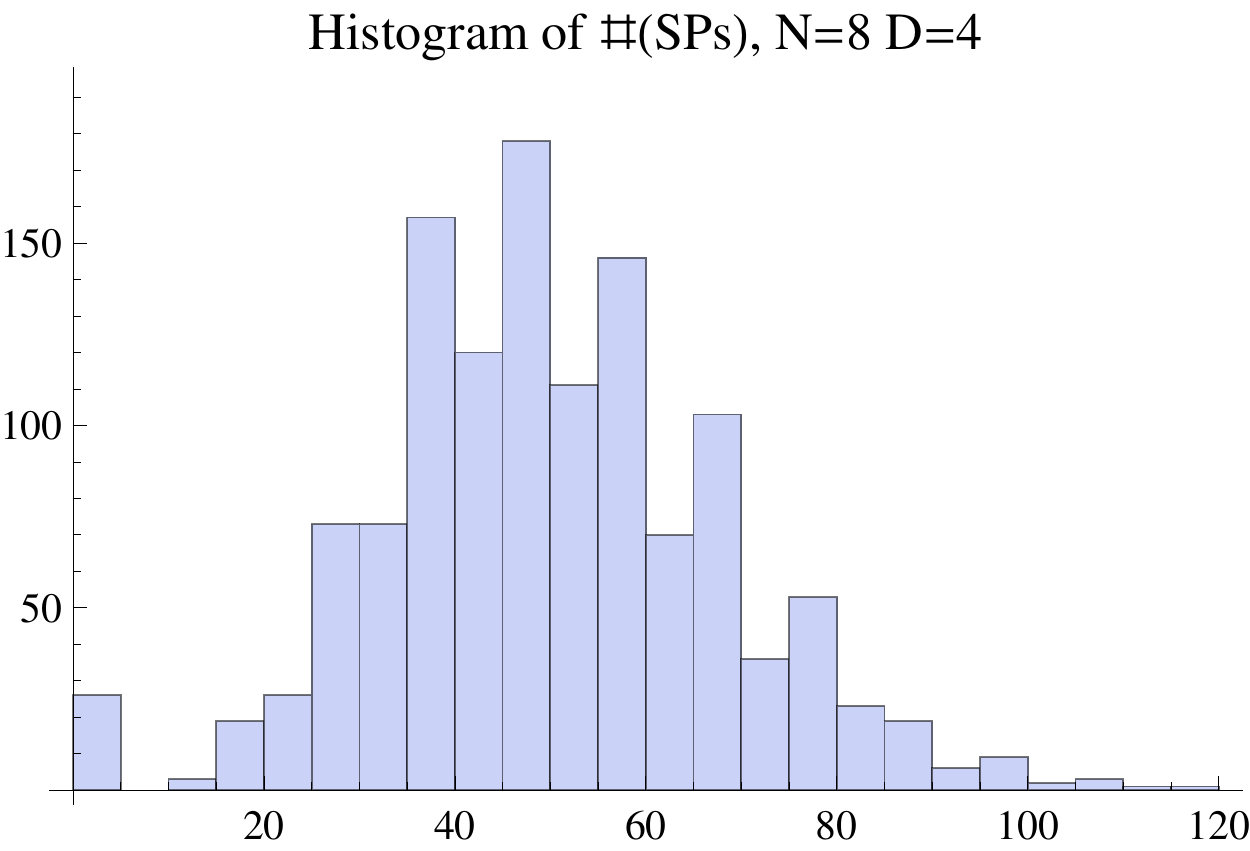}\\
\includegraphics[width=4cm]{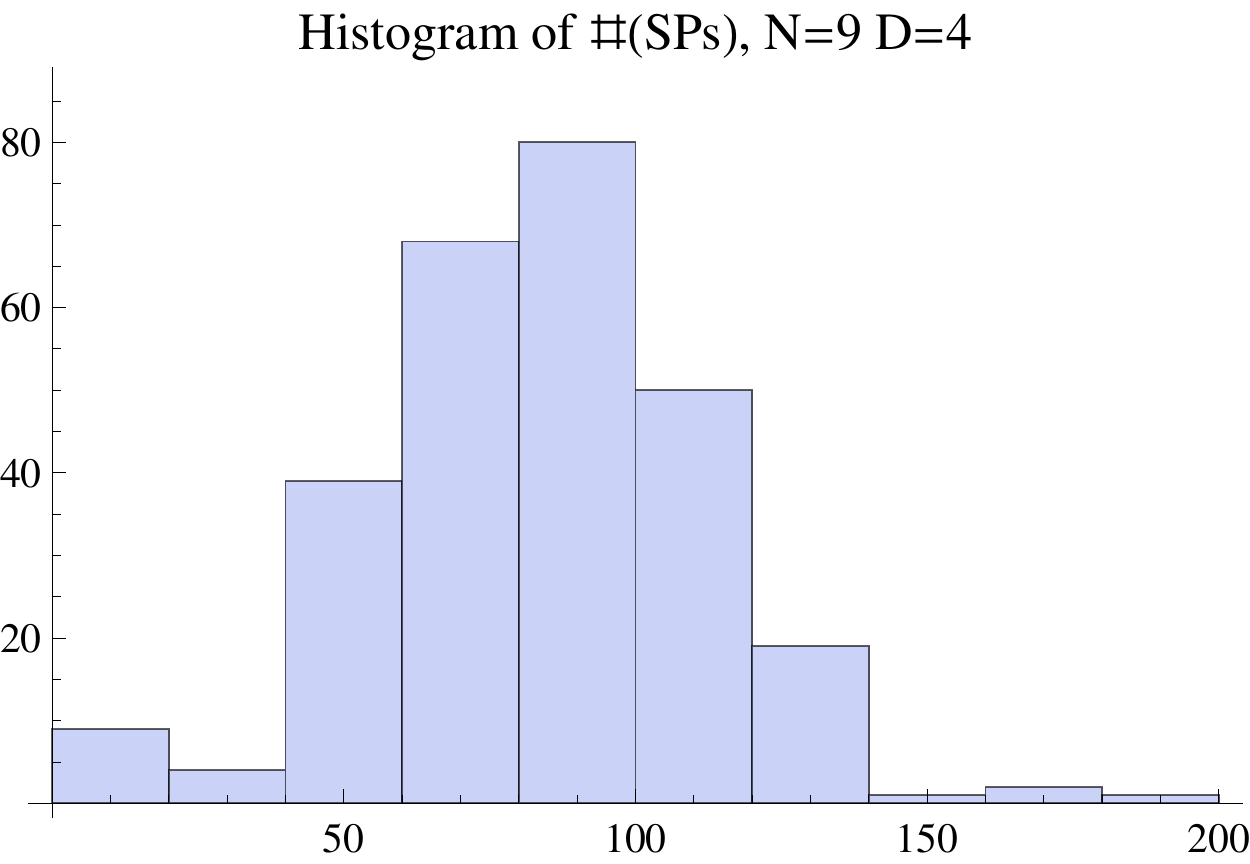}
\includegraphics[width=4cm]{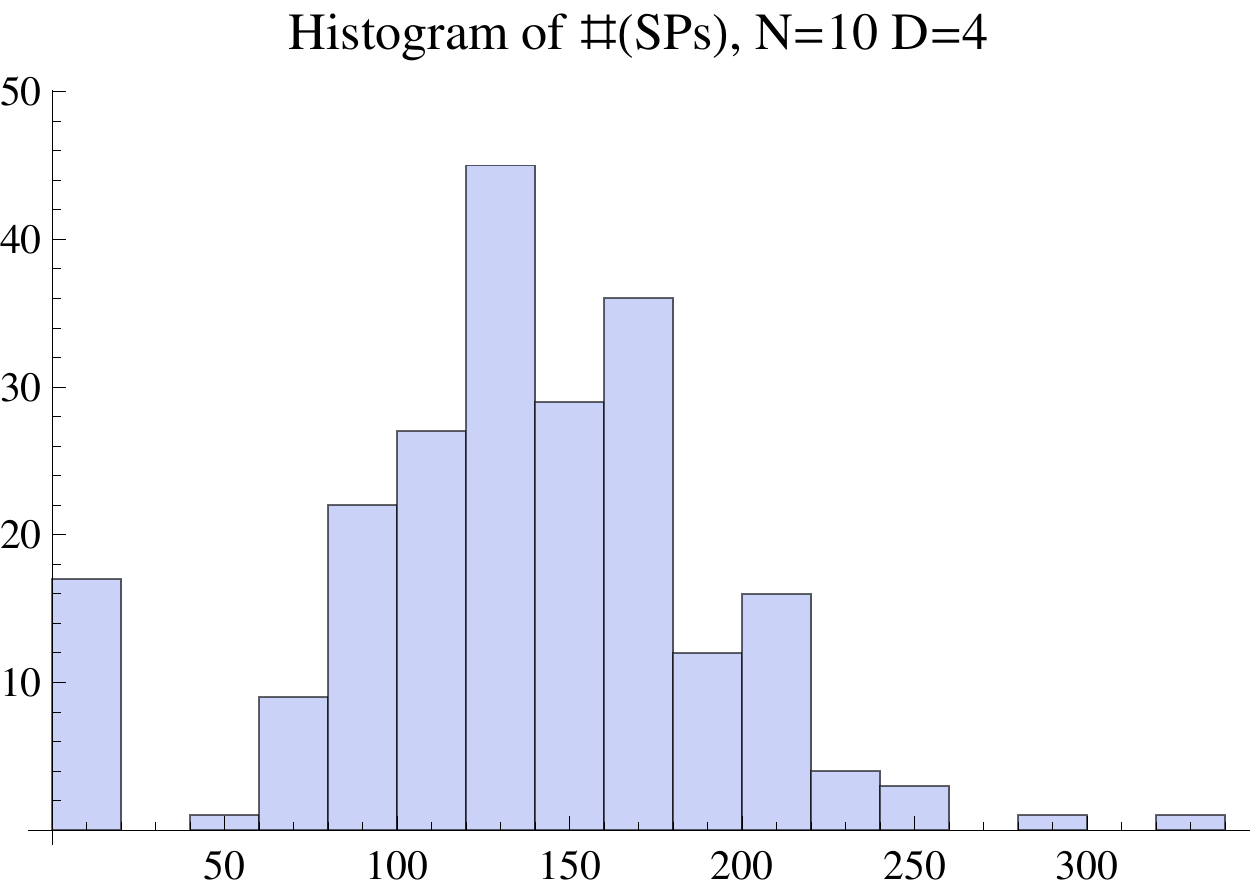}\\
\caption{Histogram of the numbers of SPs for $D=4$. $N=5, 6$ in the top row, $N=7,8$ in the middle row, and 
$N=9,10$ in the bottom row.} \label{HistSPD4}
\end{figure}

\begin{figure}[htp]
\includegraphics[width=4cm]{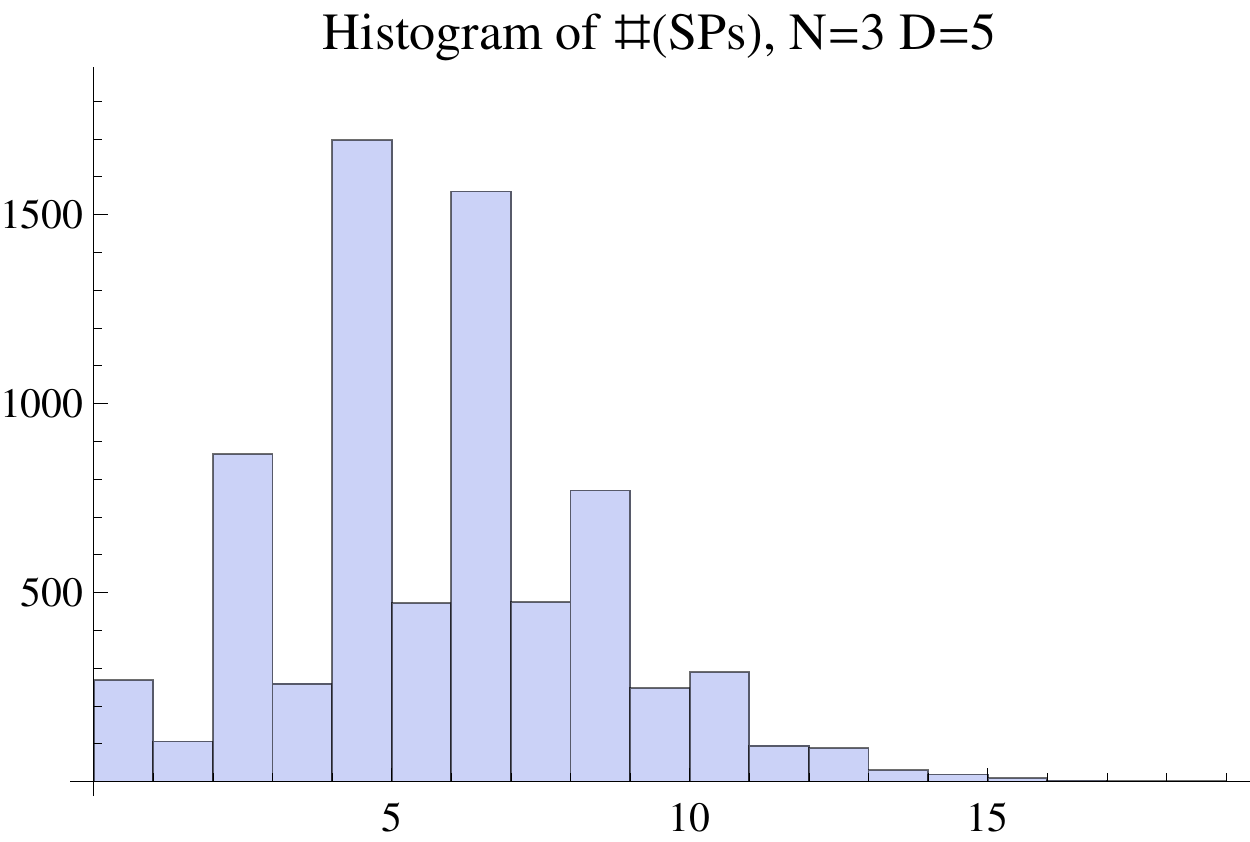}
\includegraphics[width=4cm]{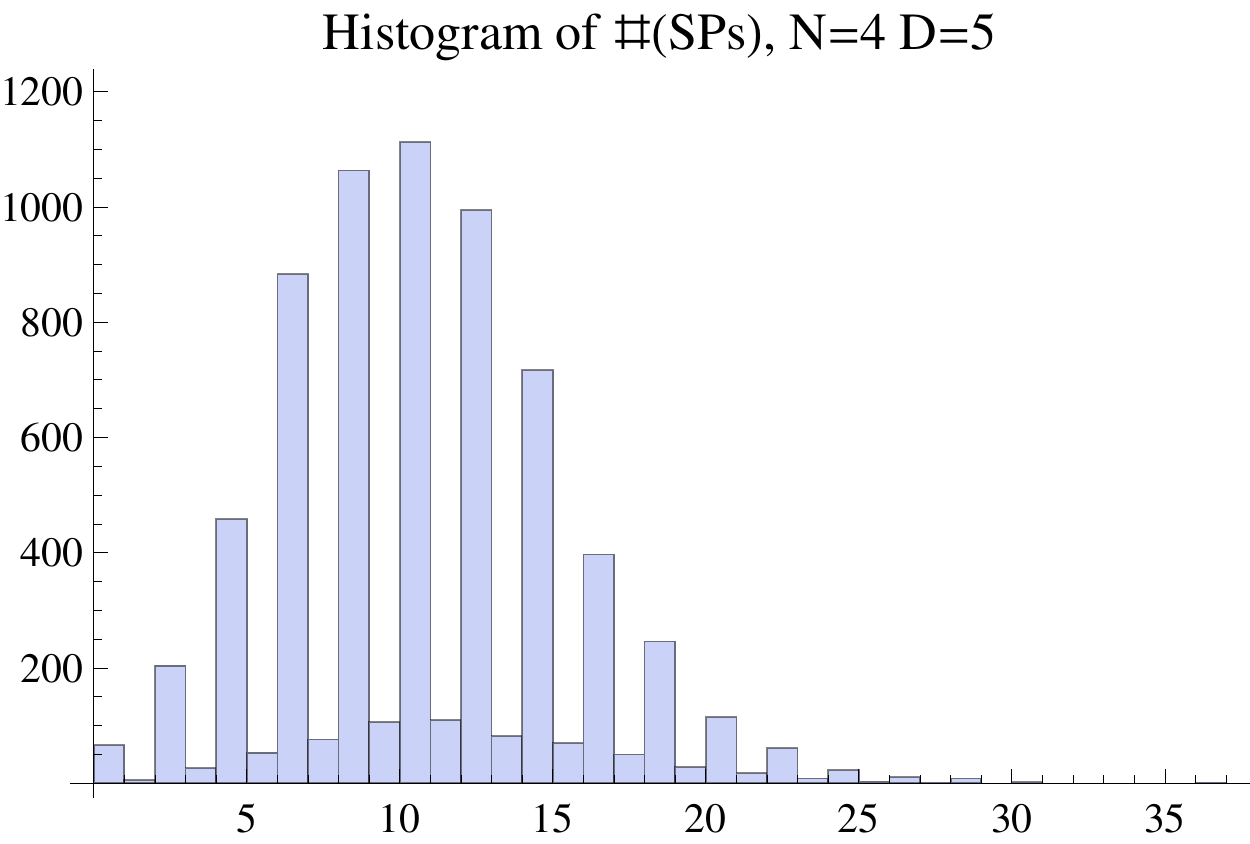}\\
\includegraphics[width=4cm]{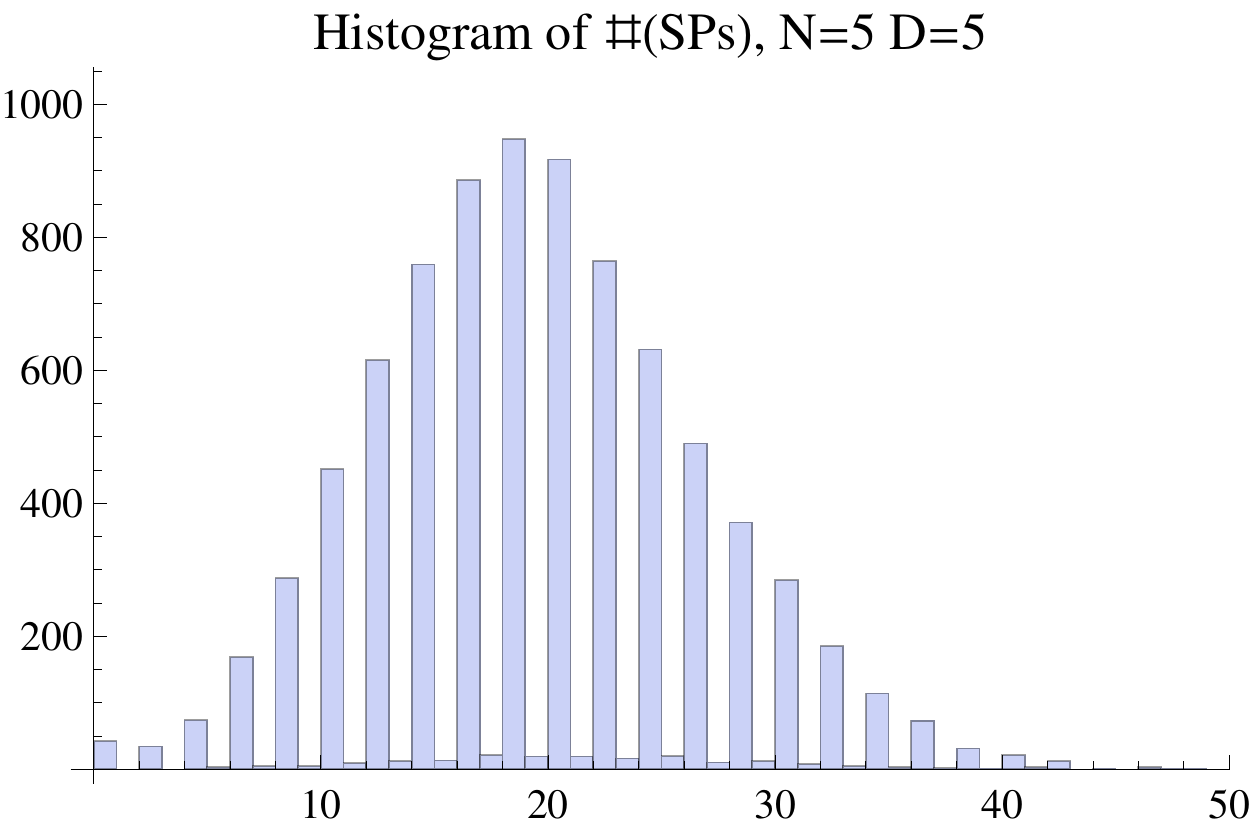}
\includegraphics[width=4cm]{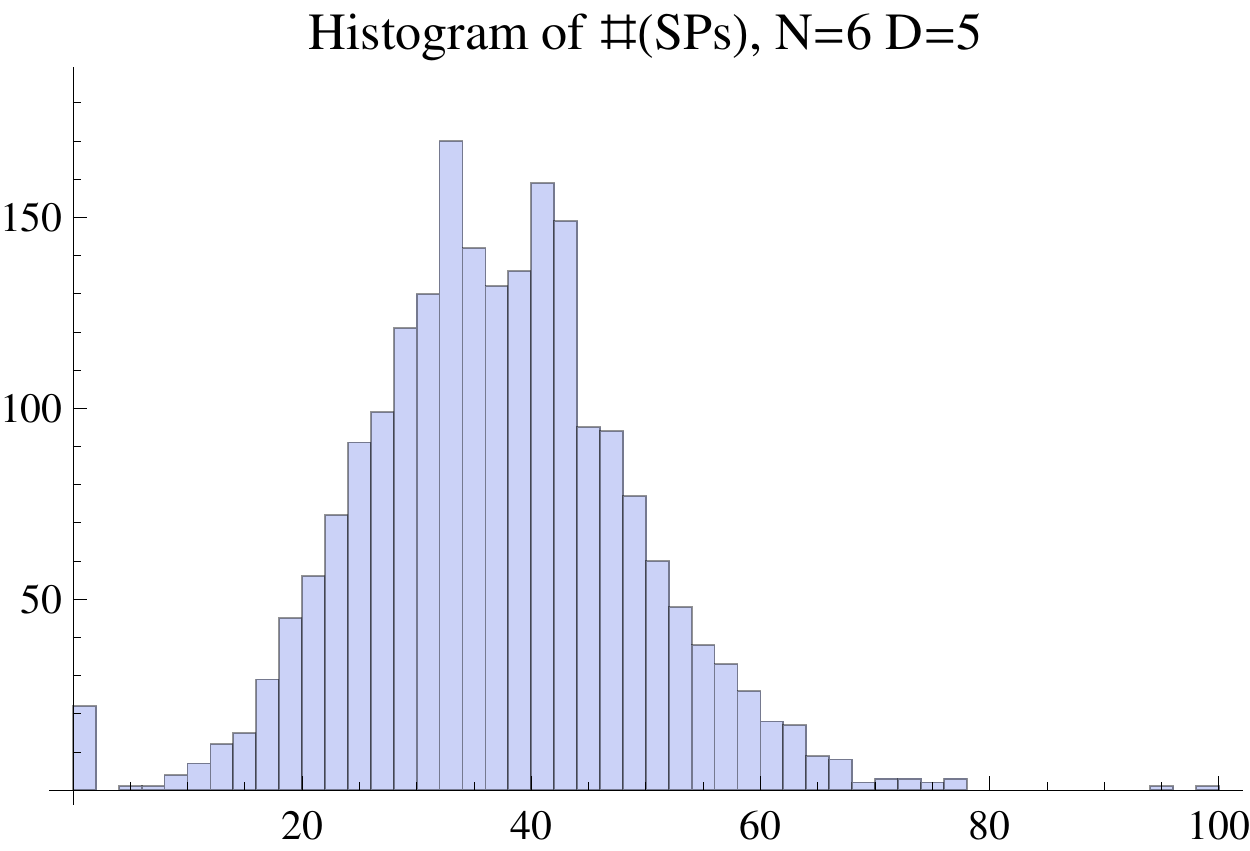}\\
\caption{Histogram of the numbers of SPs for $D=5$. $N=3, 4$ in the top row, and $N=5, 6$ in the bottom row.} \label{HistSPD5}
\end{figure}

As is true with most other analytical computations related to random systems, analytically computing the variance of the number of 
SPs is almost a prohibitively difficult task. Figure \ref{VarRealSP} shows that the variance also behaves in qualitatively
the same fashion as the mean indicating that as $N$ increases the mean of the number of real SPs are 
very spread out from each other and from the mean. To further investigate the spread, we plot histograms of the number of real SPs in 
Figures \ref{HistSPD3}-\ref{HistSPD5}
which show that the peak in the histograms is indeed around the mean value of the number of SPs, i.e., a unimodal distribution.
Away from the peak, the histograms
spread out smoothly. However, our results suggest that these histograms are not always symmetric with respect to the peak but often exhibit
right-skewedness.

\subsection{Average number of minima} 

\begin{figure}[htp]
\includegraphics[width=8.0cm]{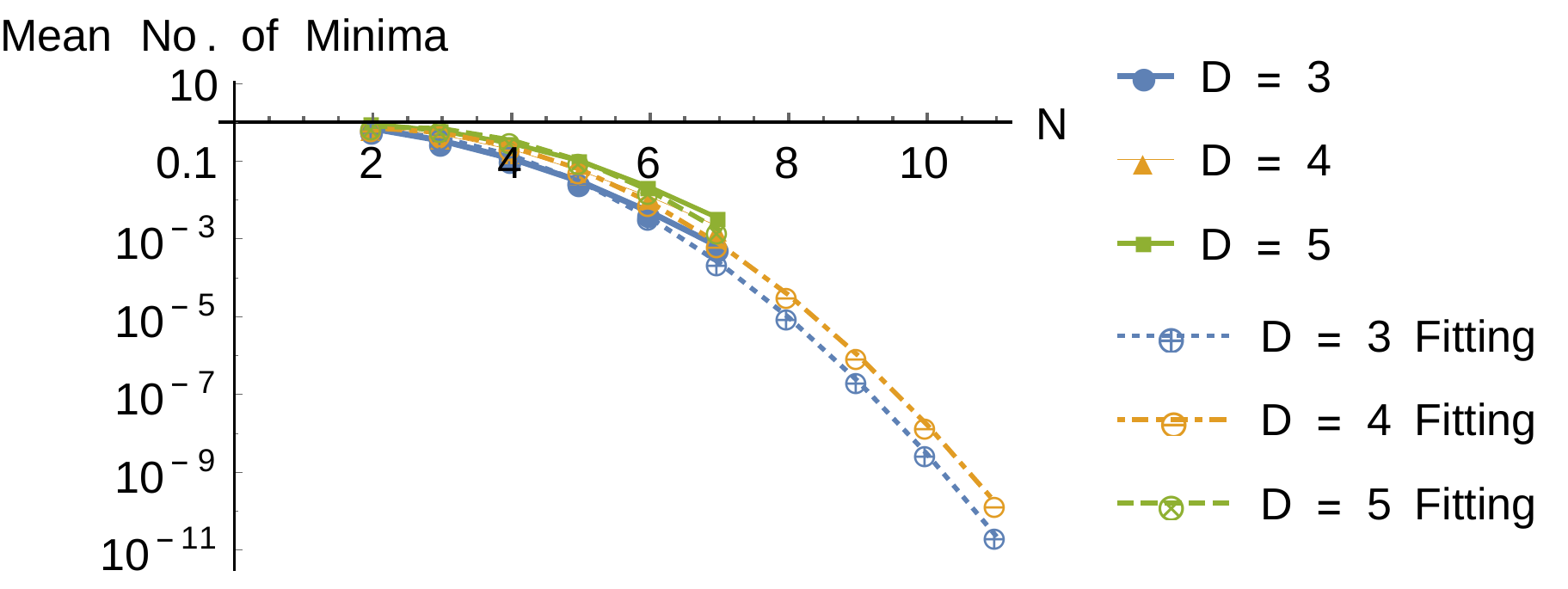}
\caption{Mean number of minima and 
comparing an analytically computed upper bound \cite{dedieu2008number} on the mean number of minima.} \label{MeanMini}
\end{figure}

\begin{figure}[htp]
\includegraphics[width=8.0cm]{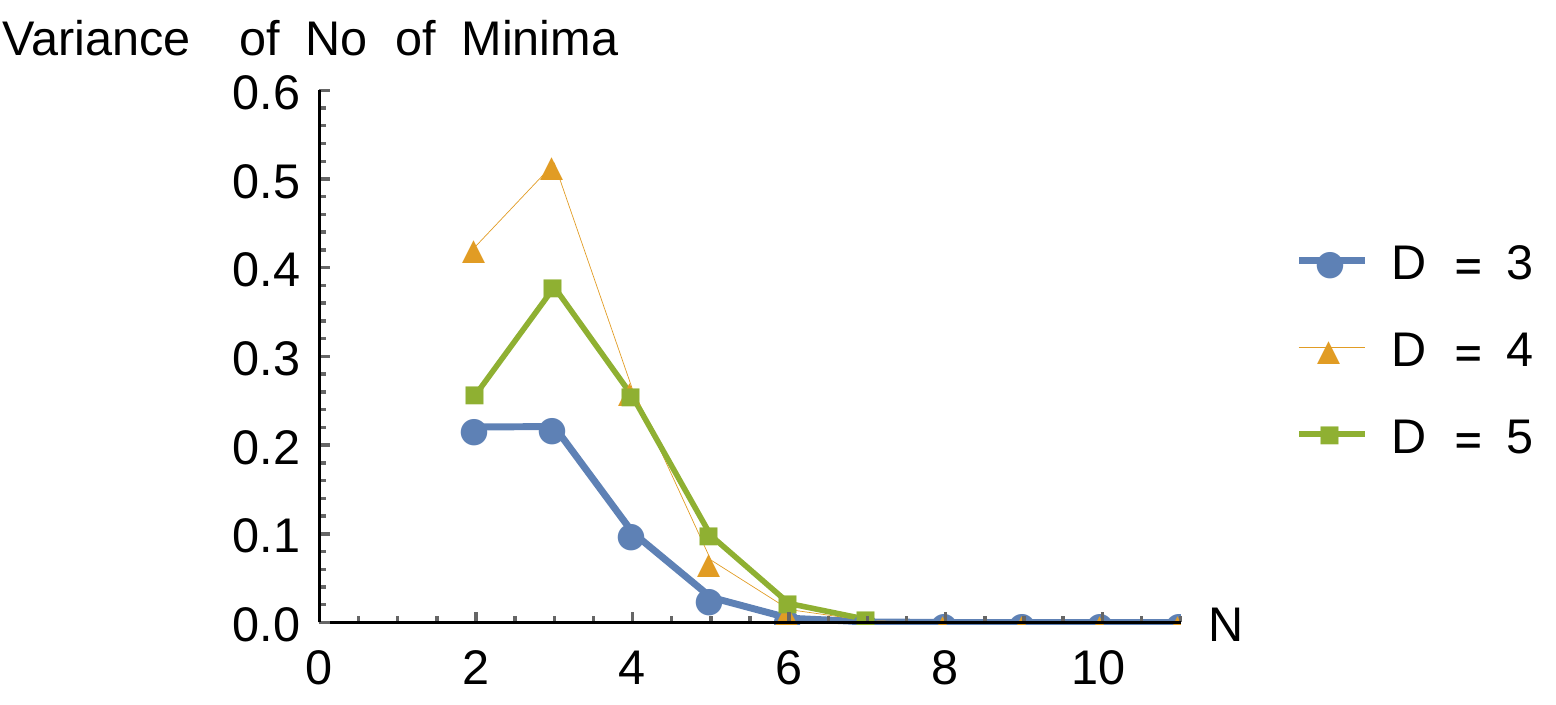}
\caption{Variance of the number of minima. Since the variance is $0$ for $N\geq 7$, we plot the data in the regular scale rather than the log scale.} 
\label{VarMini}
\end{figure}

Since the potential may be unbounded from below (or above),
we must clarify that by minima we simply mean SPs at which all the Hessian eigenvalues are positive definite. 
Though the mean number of real SPs increases exponentially with increasing $N$, the mean number of minima decays exponentially with 
increasing $N$, as shown in Figures \ref{MeanMini} and \ref{VarMini}. We compare our results with an analytically computed upper bound 
\cite{dedieu2008number}, 
namely, $K e^{\big(-N^2 \frac{\ln 3}{4} + \frac{(N + 1)}{2} \ln(D - 1)\big)}$, 
where $K$ is a positive constant. Since a prescription for computing $K$ is not available
in \cite{dedieu2008number}, we obtain it by fitting the upper bound with the numerical data. Our results
obeys the upper bound. In fact, the mean number of minima itself appears to qualitatively behave as
$e^{\big(-N^2 \frac{\ln 3}{4} + \frac{(N + 1)}{2} \ln(D - 1)\big)}$.
We must point out that as a function of $D$, while keeping $N$ fixed,
the mean number of minima appears to be a slowly increasing function. This is not
surprising since in this case the total number of complex solutions themselves $(D-1)^N$ increases drastically.
We do not have enough data in $D$ to extract the precise behaviour of the increase. However, the aforementioned upper bound
with now fixing $N$ and varying $D$ implies that the increase of mean number of minima should be much slower than 
the increase when $D$ is fixed and $N$ is varied because of the natural logarithm term.
The variance of the number of minima also exponentially decays as $N$ increases indicating that 
the mean number of minima gives a good estimate for the actual number of minima for any randomly picked sample.  
We anticipate that our numerical results will instigate an anlytical study for the variance
of the number of real SPs and minima.

\subsection{Average number of minima at which $V>0$ (i.e., the de Sitter minima)} 
\begin{figure}[htp]
\includegraphics[width=8.0cm]{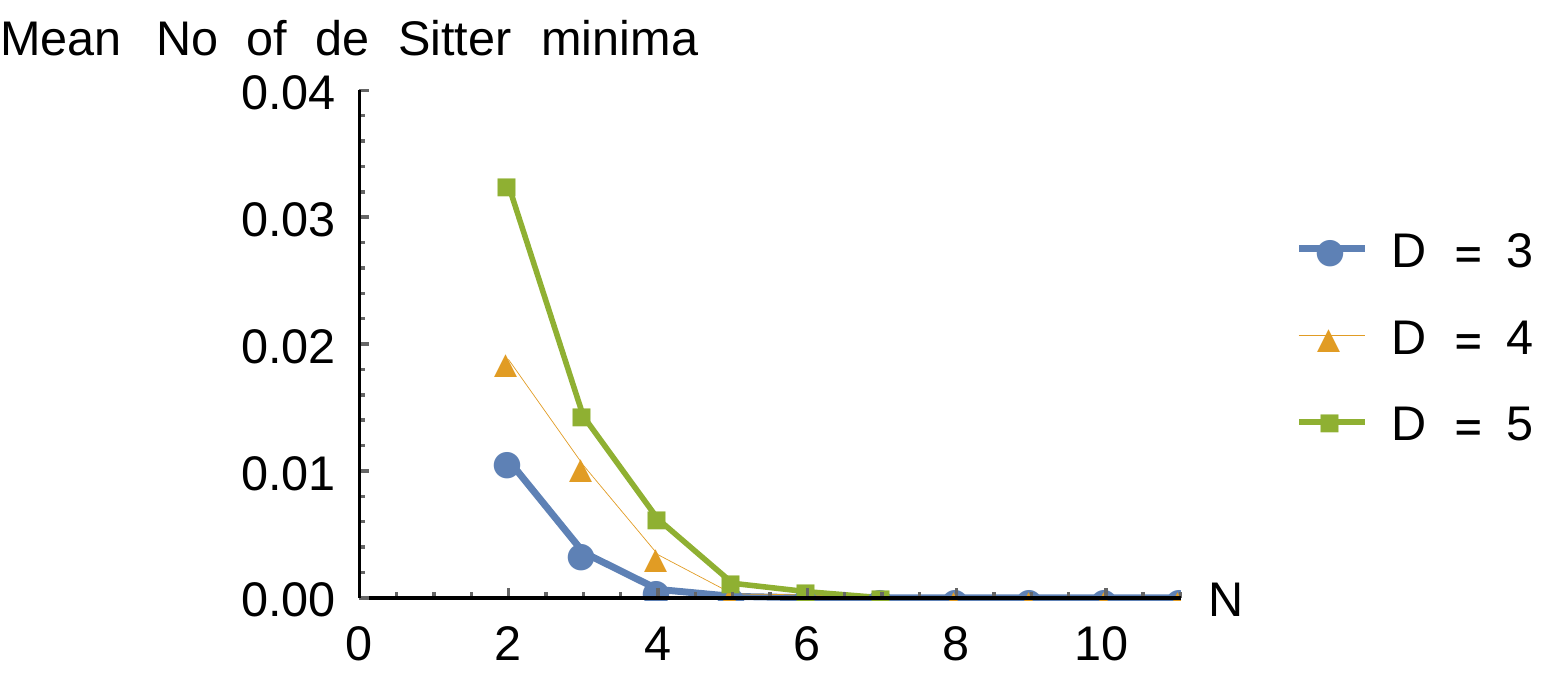}
\caption{Mean number of de Sitter vacua} \label{MeanMinidS}
\end{figure}

\begin{figure}[htp]
\includegraphics[width=8.0cm]{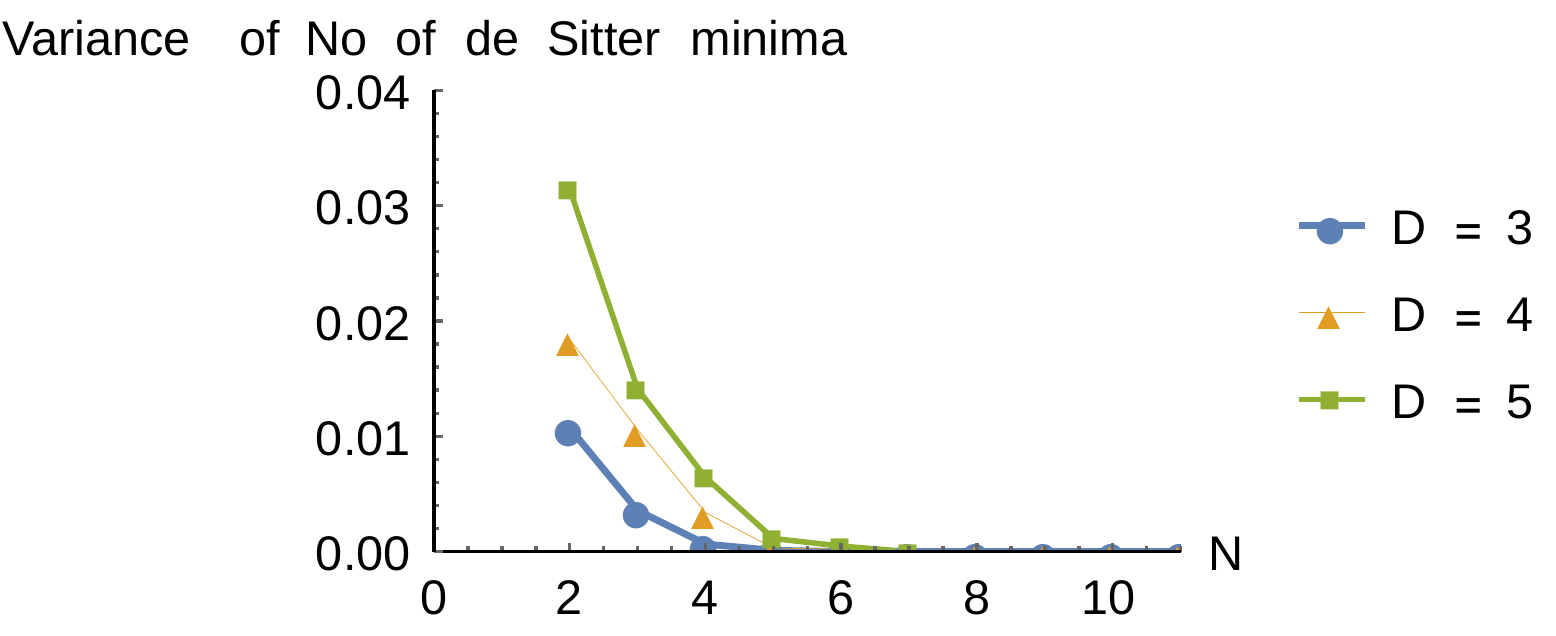}
\caption{Variances of the number of de Sitter vacua} \label{VarMinidS}
\end{figure}

The number of minima at which $V>0$ may be viewed as de Sitter vacua if the potential is considered to mimic the statistical 
aspects of the string theory landscapes. Figures \ref{MeanMinidS} and \ref{VarMinidS} essentially exhibit the same 
qualitative behaviour as the number of minima. The implications of these results on the string theory landscapes may 
be important, i.e.~, the number of de Sitter vacua decreases exponentially with increasing the dimension of the moduli space, 
which is similar to the conclusion
laid out in \cite{Greene:2013ida} (see also \cite{Marsh:2011aa,Aravind:2014aza}): the number of vacua of a RP
with tunneling rates low enough to maintain metastability exponentially decreases
as a function of the moduli space dimension. Though in the latter work, the RP was a bit different than ours in that
in the former the RP was a Taylor expansion of a generic function, with random coefficients i.~i.~d.~picked
from uniform distributions from specific ranges, remarkably, the overall conclusion remains the same even with our investigations with more general 
type of vacua (de Sitter vacua without any further conditions on bubble nucleation rate or else). However, we also point out that 
as we increase the degree of the potential (which would correspond to the truncation degree in the Taylor expanded version of \cite{Greene:2013ida})
the number of de Sitter vacua does rise exponentially though with a linear rate.

\subsection{Average number of Index-sorted SPs}
\begin{figure}[htp]
\includegraphics[width=7.0cm]{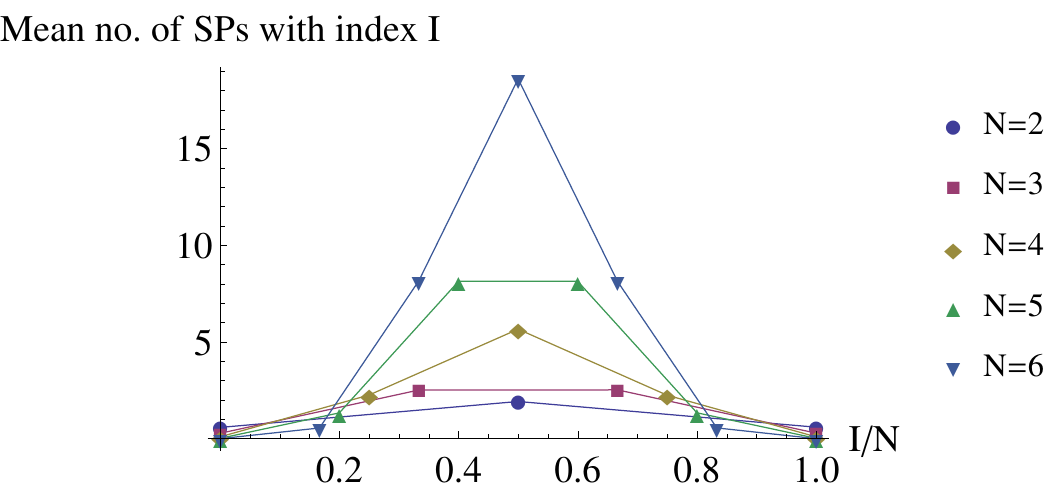}
\caption{Mean number of SPs as a function of $I/N$, 
where $I$ is the Hessian index which runs from $0$ to $N$, for $D=5$ and various $N$. The corresponding 
plots for other values of $D$ exhibit the same qualitative behaviour.} \label{SPwithIndices}
\end{figure}

\begin{figure}[htp]
\includegraphics[width=7.0cm]{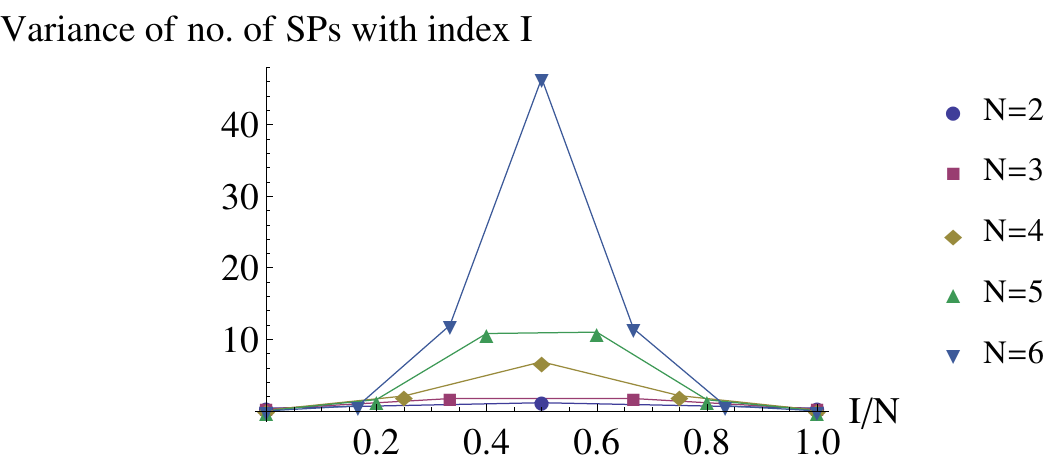}
\caption{Variance of the number of SPs of index $I$ as a function of $I/N$, 
where $I$ is the Hessian index which runs from $0$ to $N$, for $D=5$ and various $N$. The corresponding 
plots for other values of $D$ exhibit the same qualitative behaviour.} \label{VarSPwithIndices}
\end{figure}

In Figures \ref{SPwithIndices} and \ref{VarSPwithIndices}, we further explore
the properties of the real SPs by sorting them according to their Hessian indices. Since for the RP, generic 
random samples do not possess
singular solutions according to Sard's theorem \cite{mumford1995algebraic}, 
the number of negative eigenvalues of the Hessian is indeed the correct Morse 
index of the given
real solution. Because we have moderate size systems in $N$ and $D$ at our disposal, we can not predict the precise behaviour 
for general $N$ and $D$. However, it is clear from our results that the plots for $I$ vs number of real SPs of index $I$ tend to be bell-shaped
curves. Such a plot is observed for many other potentials such as the Lennard-Jones potential \cite{2002JChPh.116.3777D,2003JChPh.11912409W}, 
nearest-neighbour 
XY model \cite{Mehta:2009,Mehta:2009zv,Hughes:2012hg,Hughes:2014moa}, 
spherical $3$-spin model \cite{Mehta:2013fza}, the Kuramoto model \cite{mehta2014algebraic}, 
etc. In \cite{2003JChPh.11912409W}, such a behaviour of the number of real SPs with a given index for general potential
was analytically derived. In short, in the landscape of RPs, there are vast number of SPs with non-zero index compared
to the number of minima, which has also been analytically observed in random set ups \cite{dedieu2008number}.

\subsection{Histograms of the Hessian eigenvalues}
\begin{figure}[htp]
\includegraphics[width=4cm]{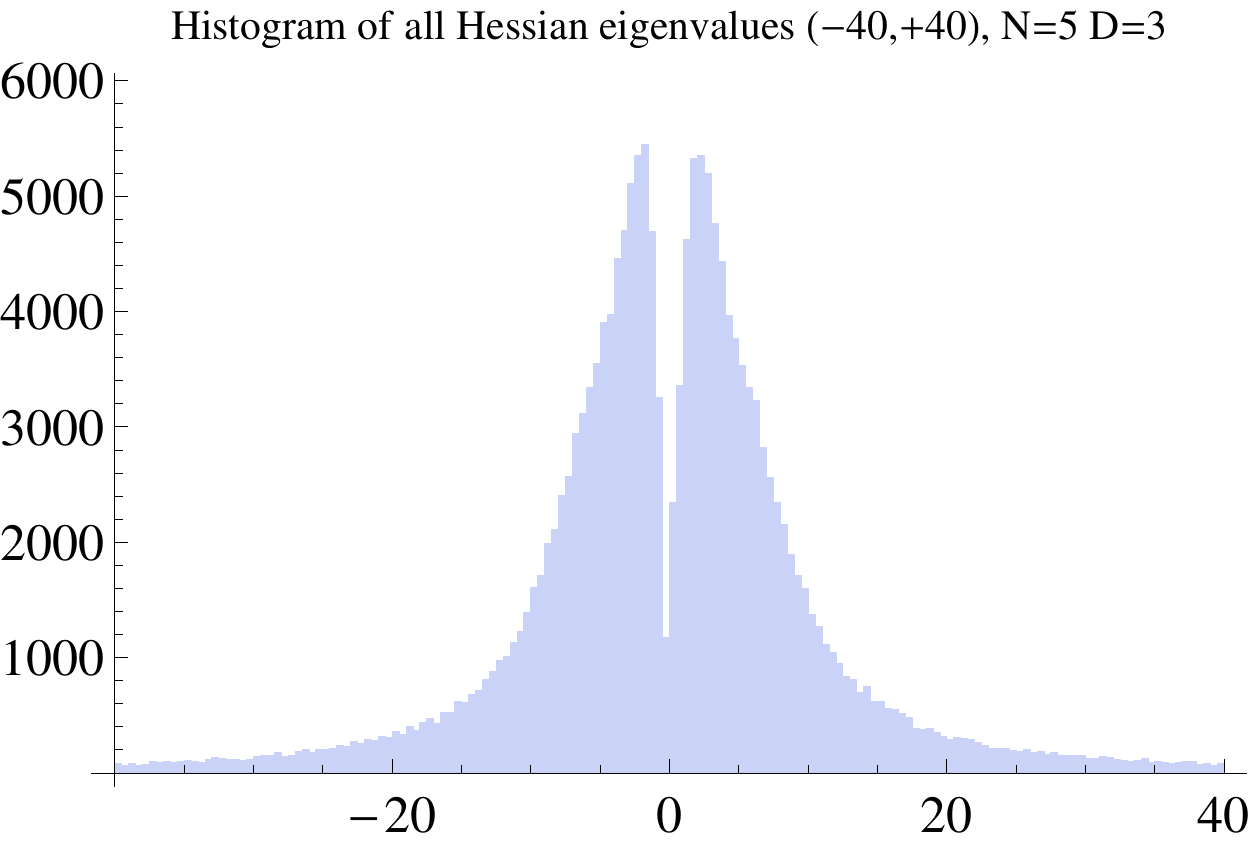}
\includegraphics[width=4cm]{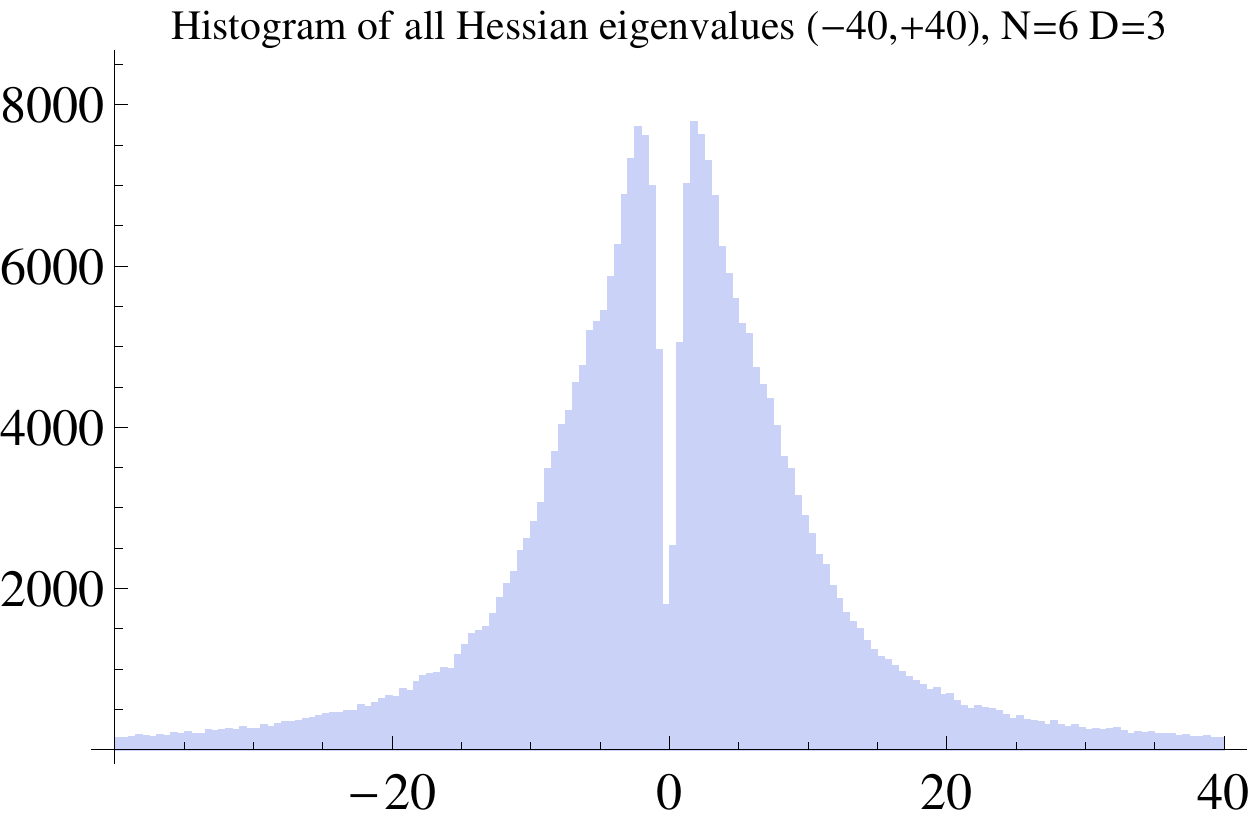}\\
\includegraphics[width=4cm]{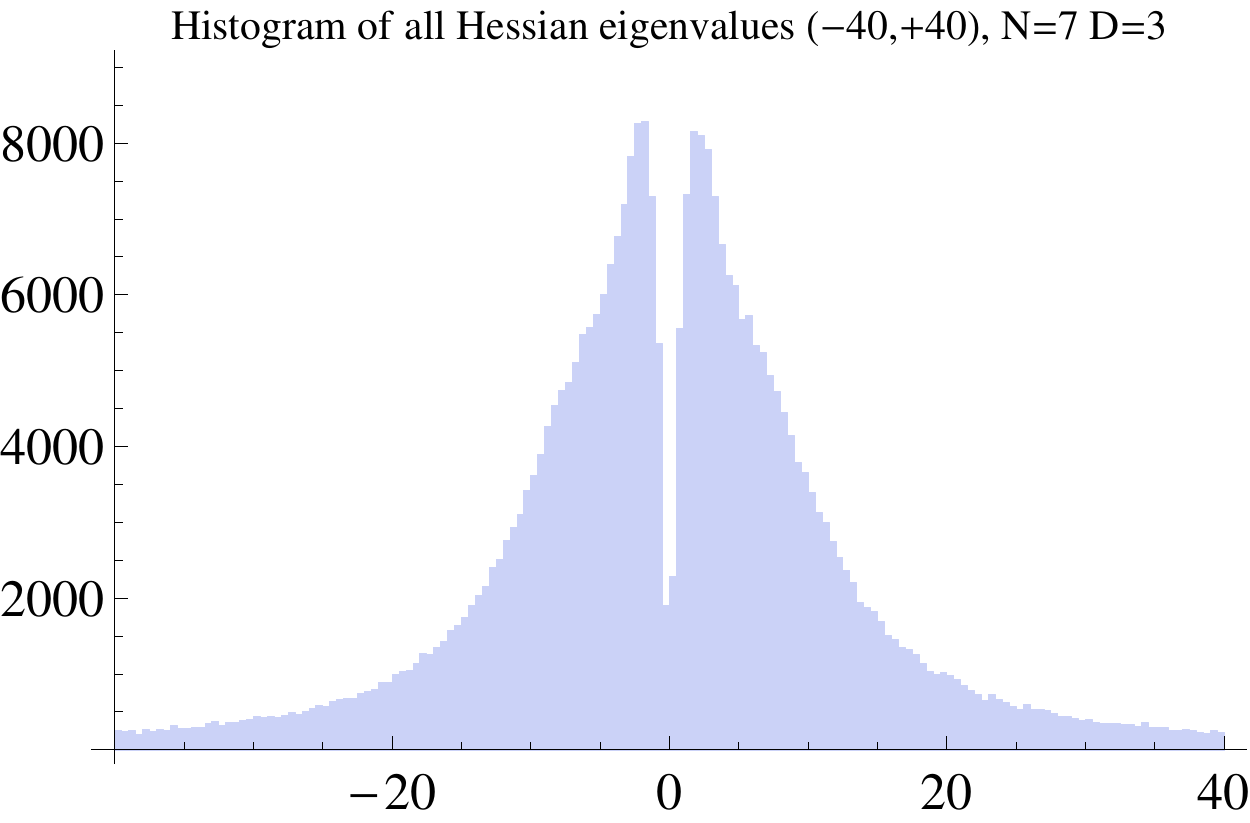}
\includegraphics[width=4cm]{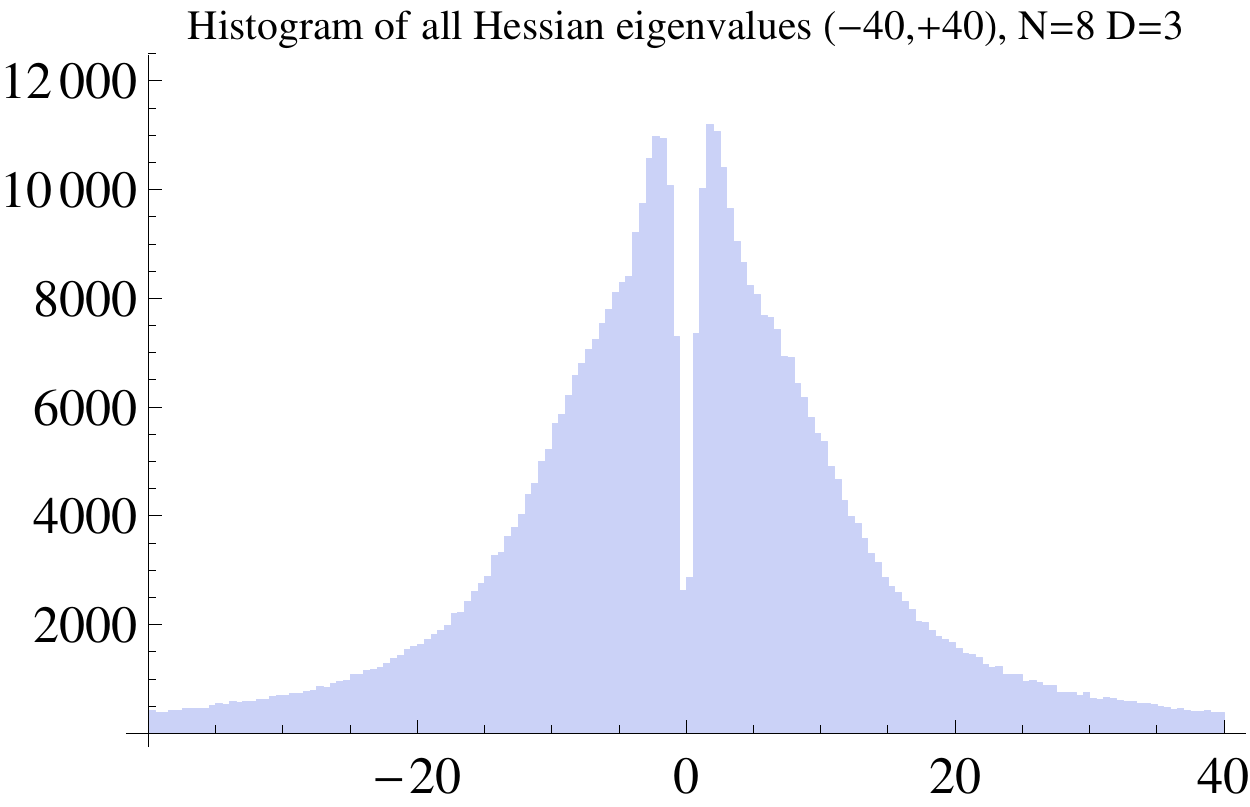}\\
\includegraphics[width=4cm]{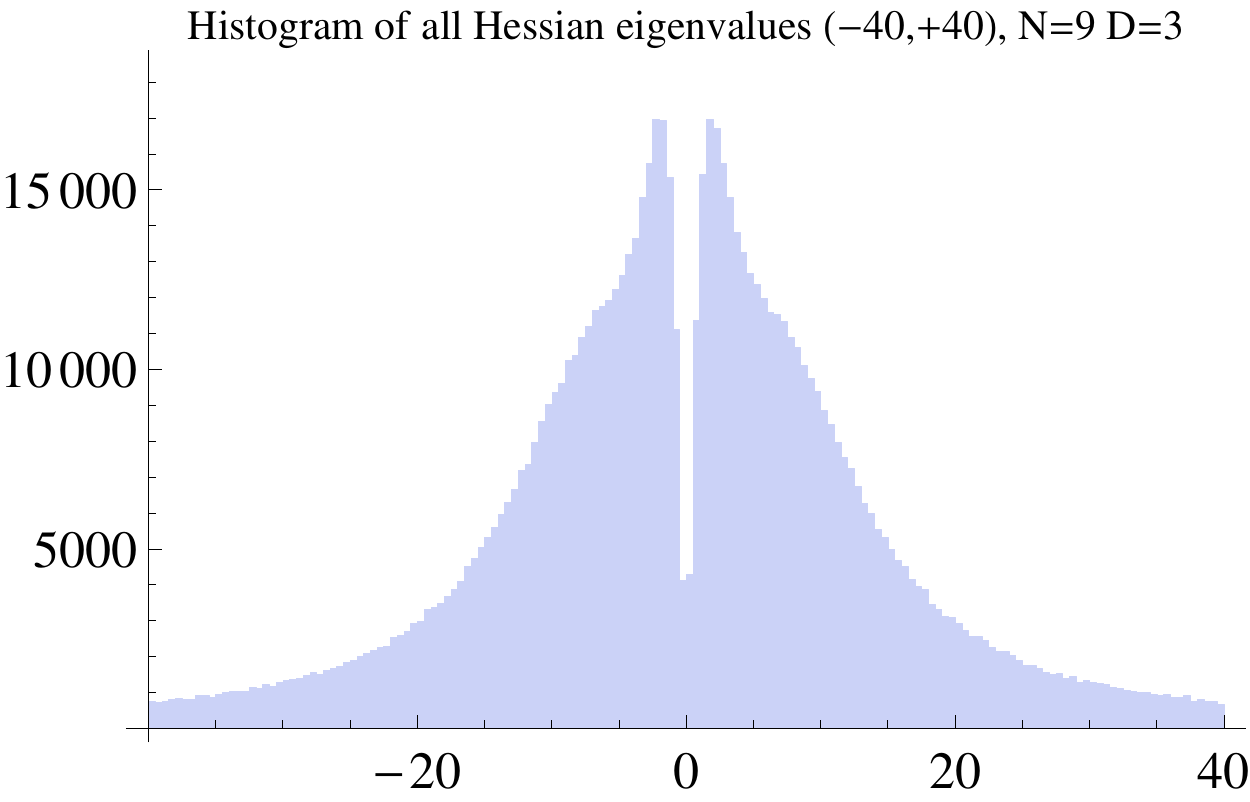}
\includegraphics[width=4cm]{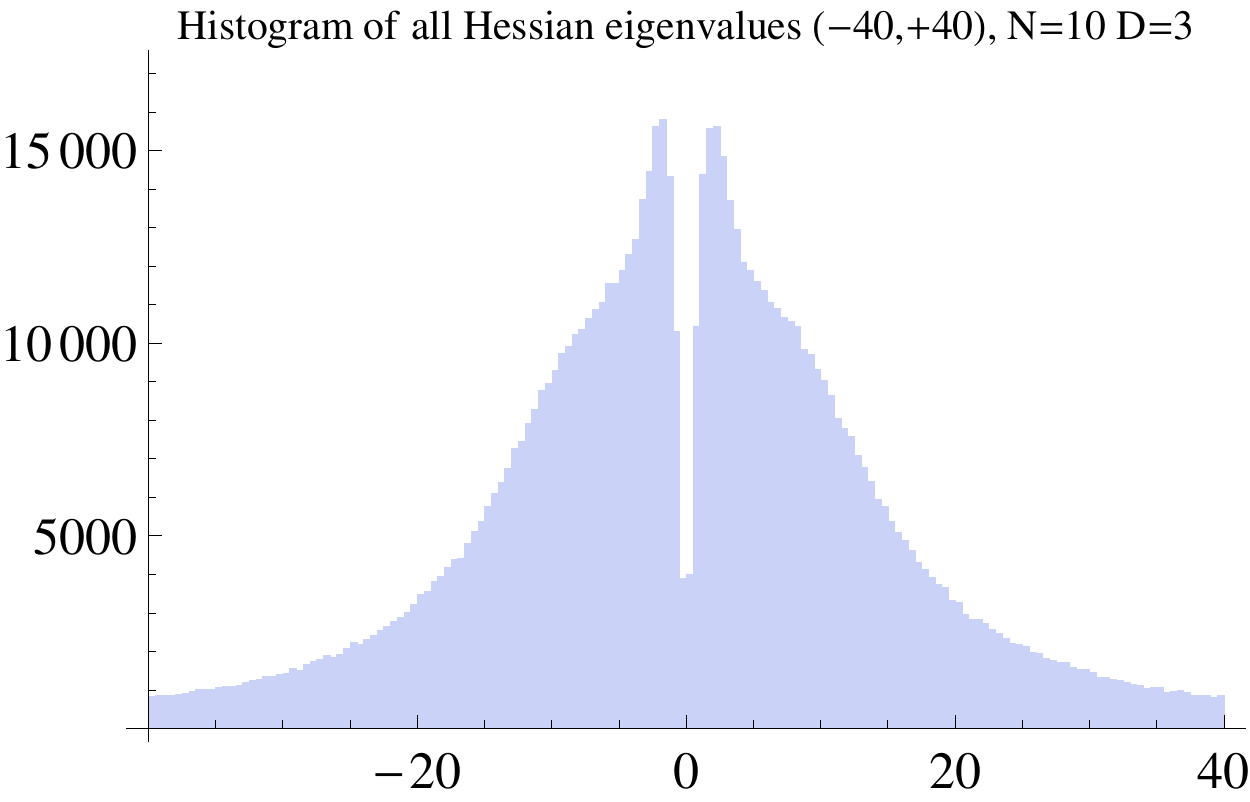}\\
\caption{Histogram of eigenvalues, $D=3$} \label{HistEigensD3}
\end{figure}

\begin{figure}[htp]
\includegraphics[width=4cm]{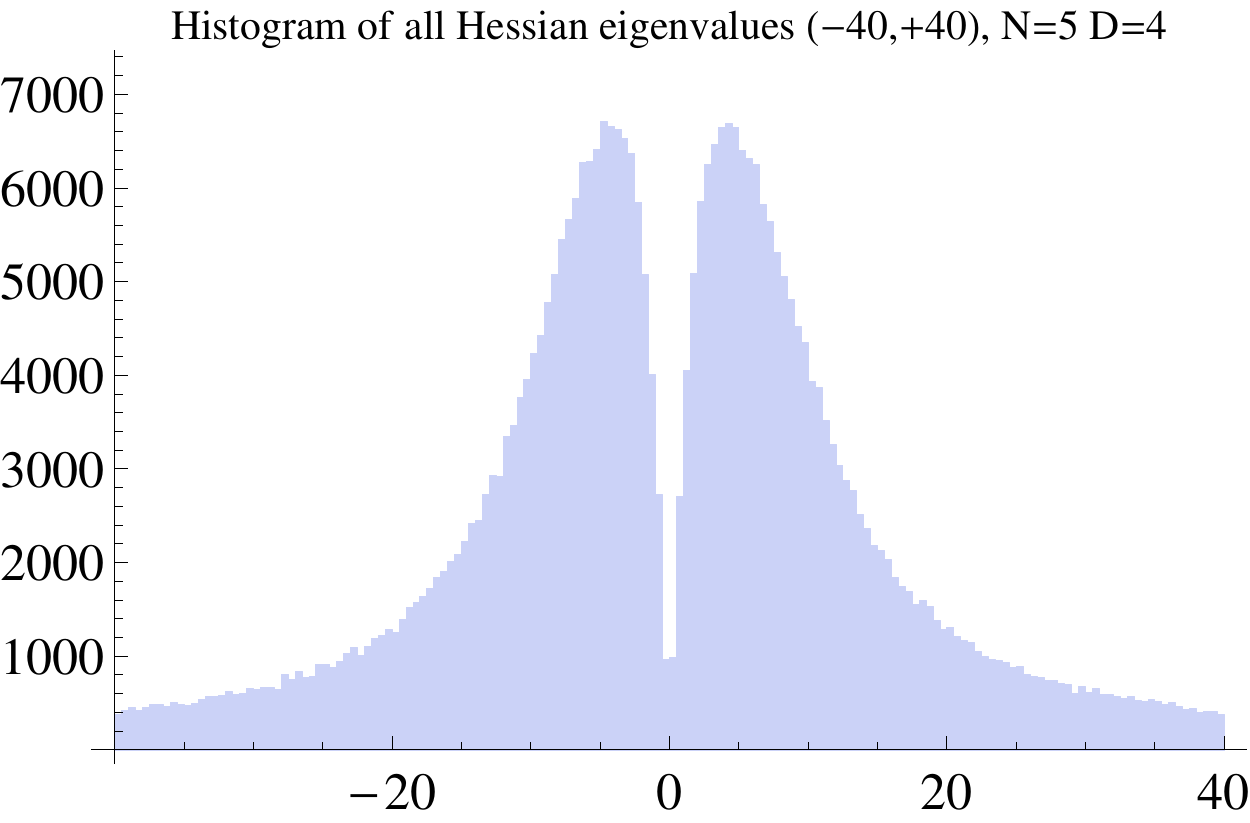}
\includegraphics[width=4cm]{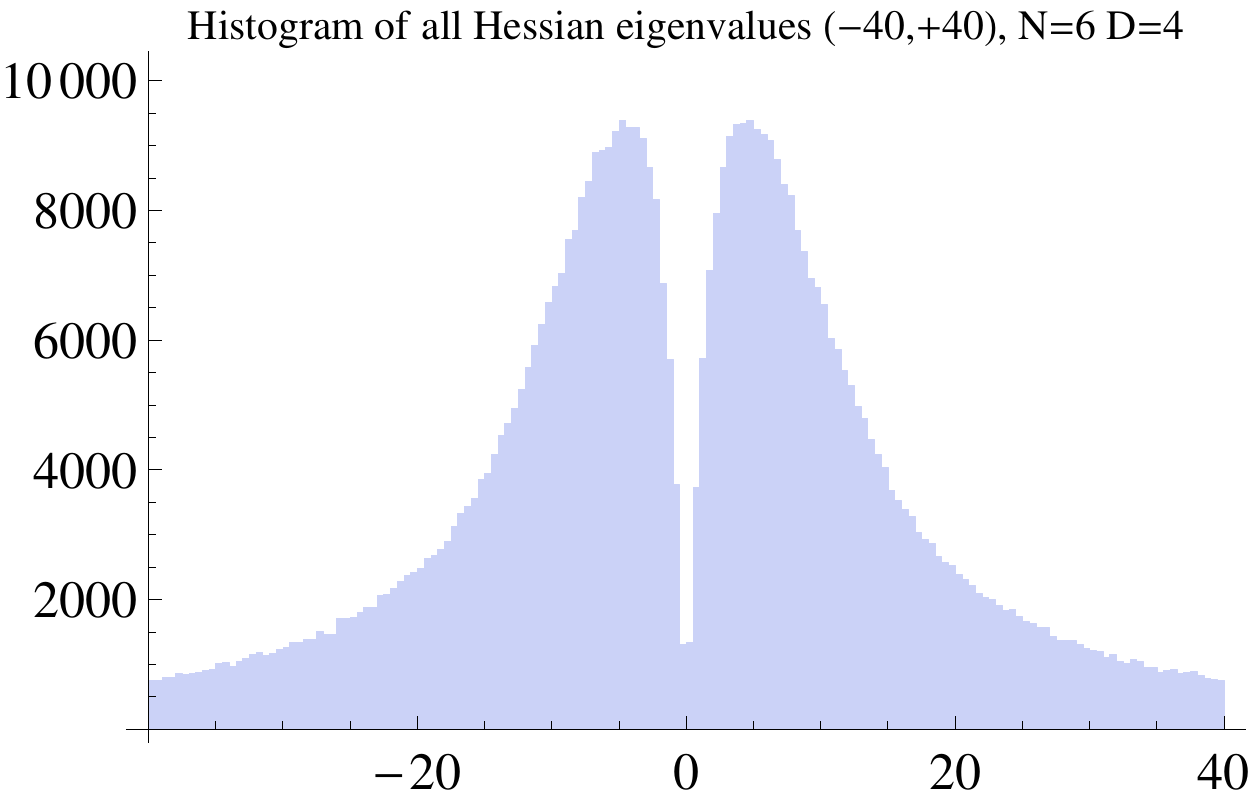}\\
\includegraphics[width=4cm]{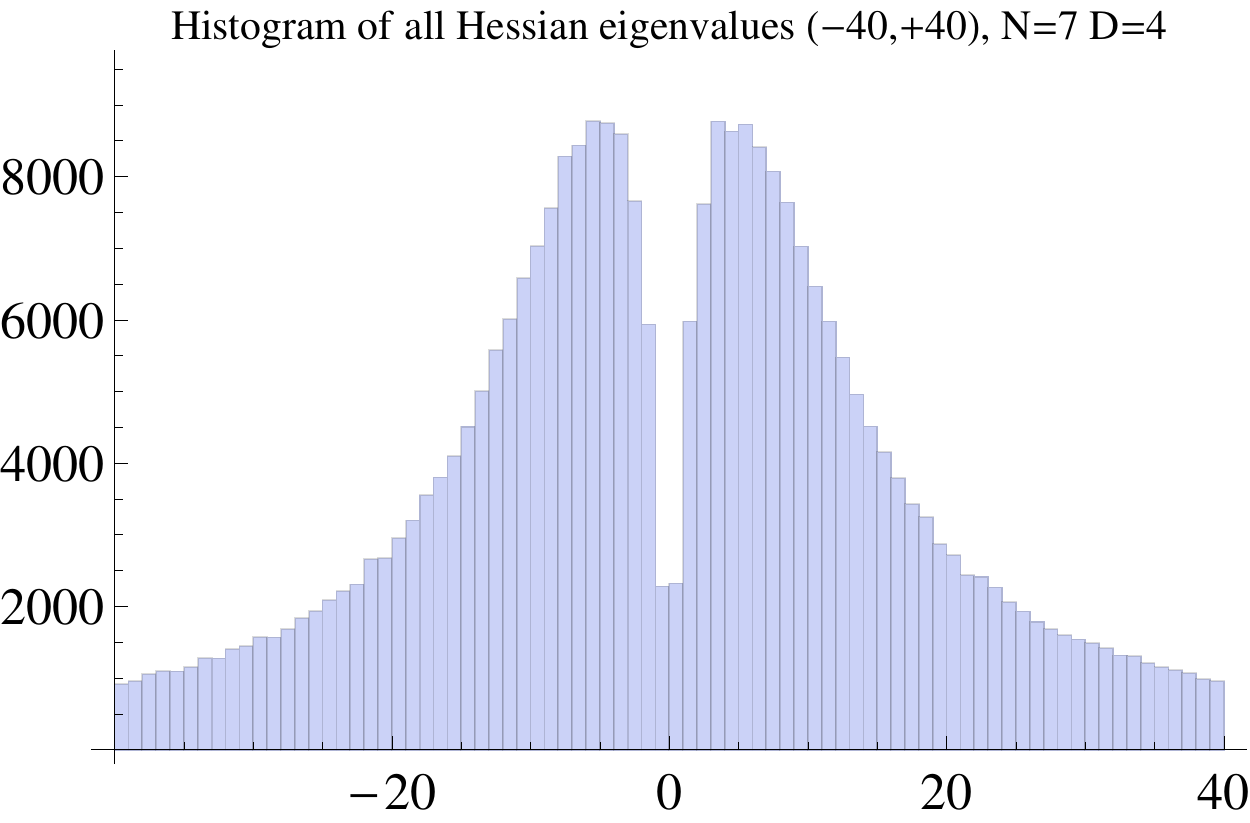}
\includegraphics[width=4cm]{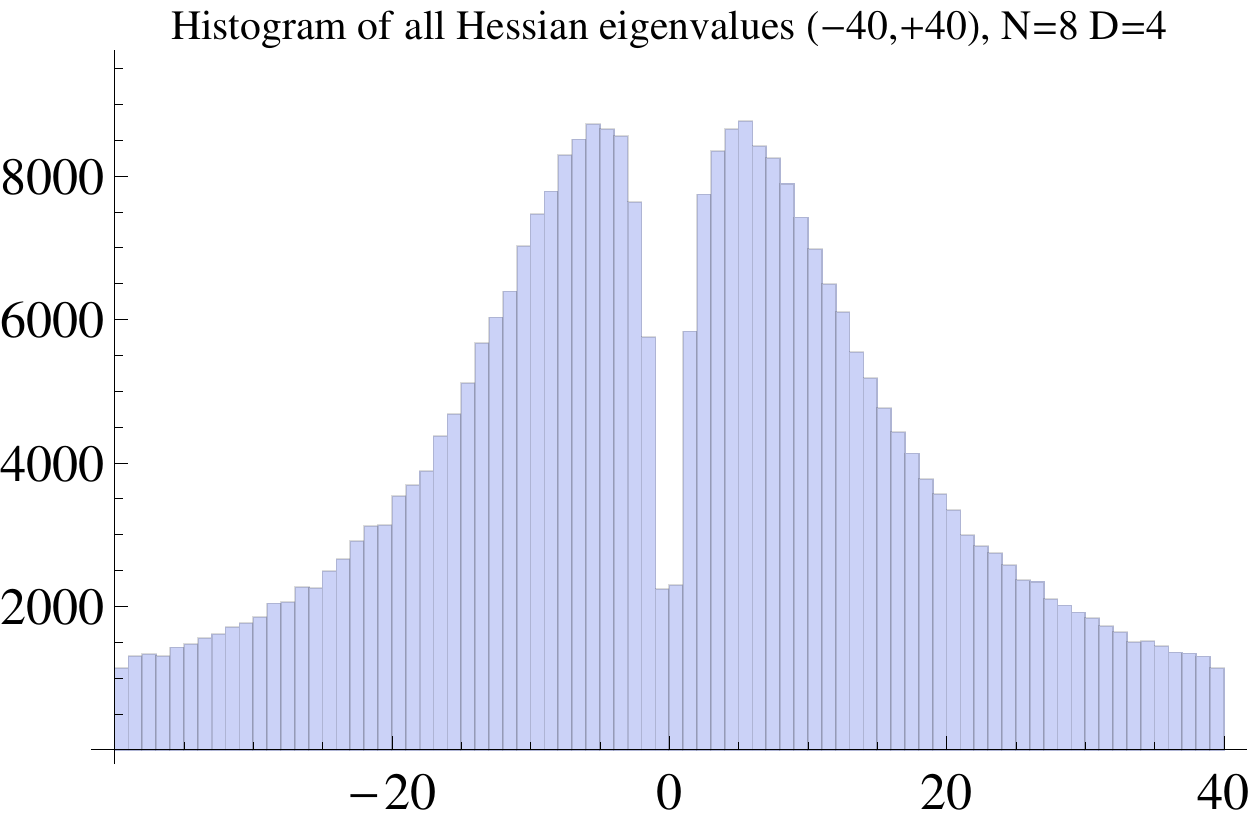}\\
\includegraphics[width=4cm]{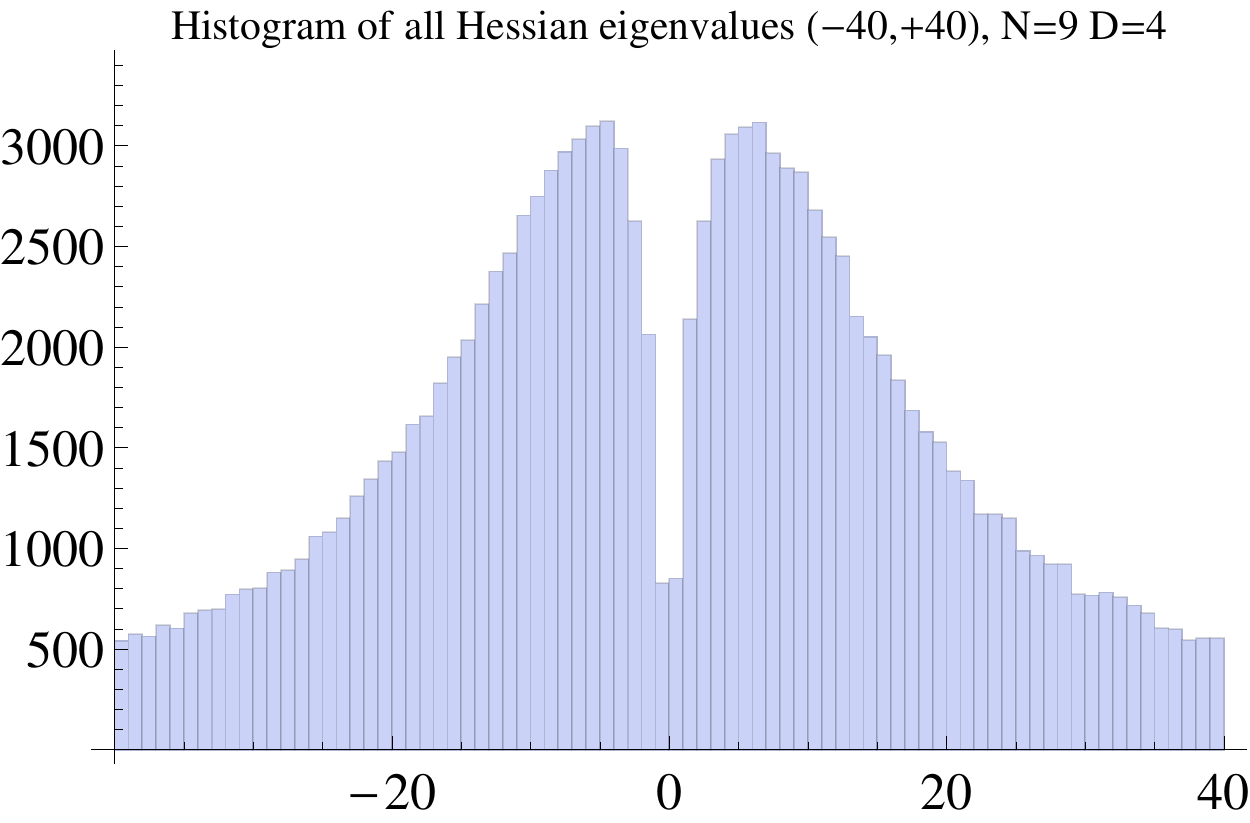}
\includegraphics[width=4cm]{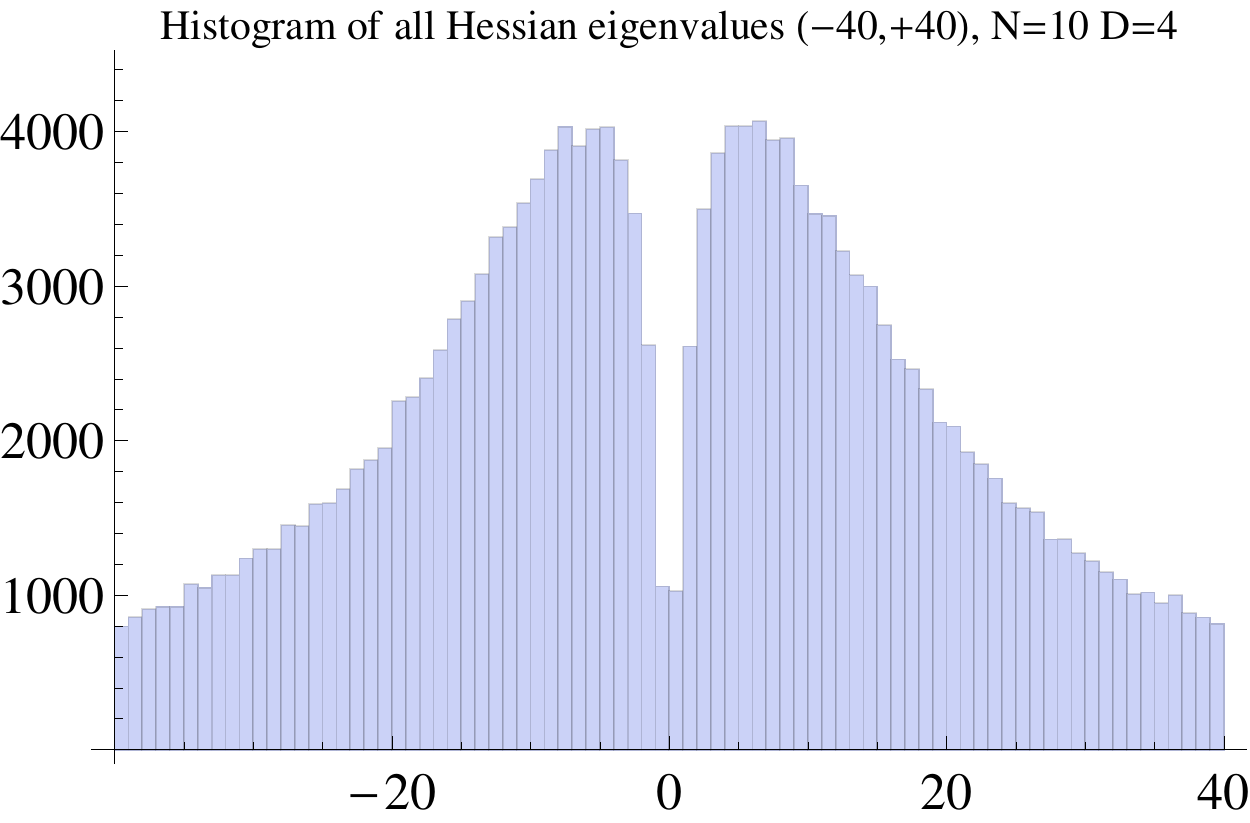}\\
\caption{Histogram of eigenvalues, $D=4$} \label{HistEigensD4}
\end{figure}

\begin{figure}[htp]
\includegraphics[width=4cm]{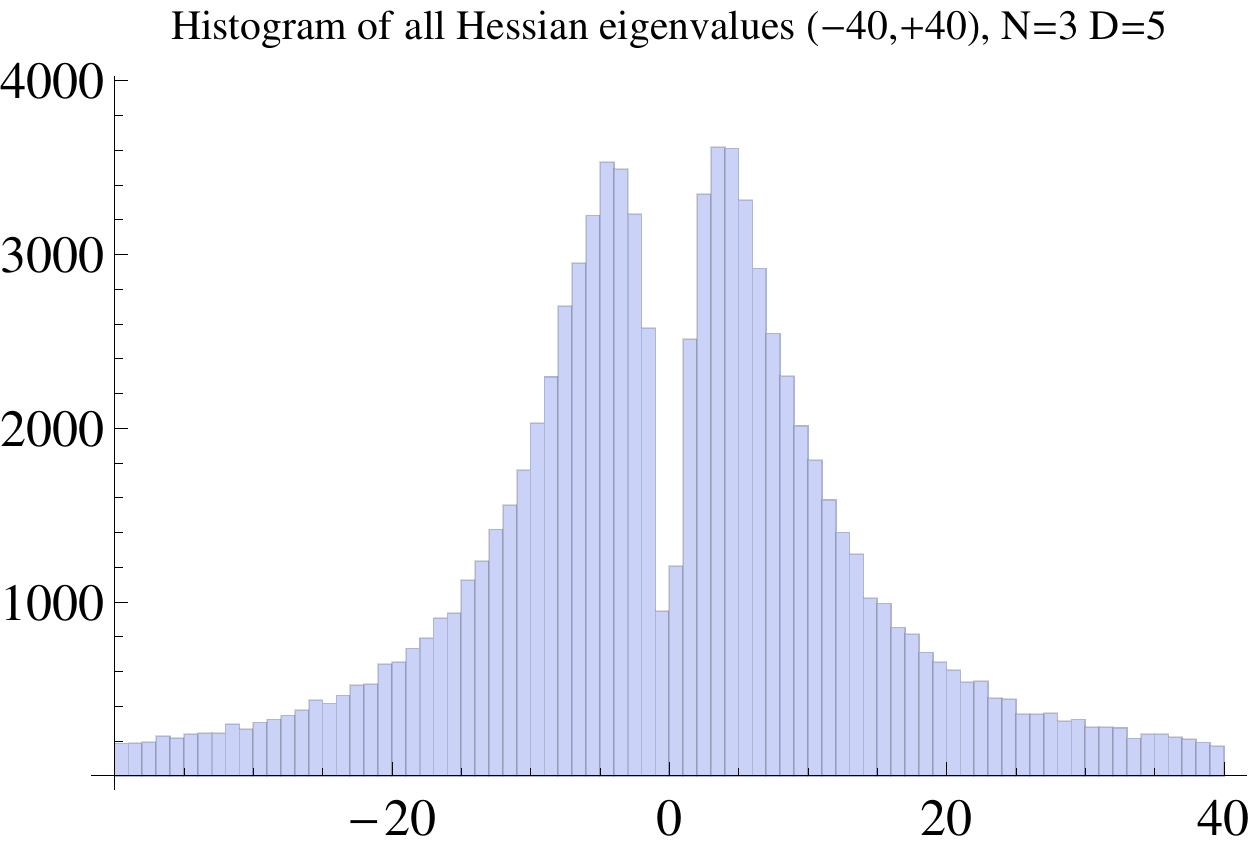}
\includegraphics[width=4cm]{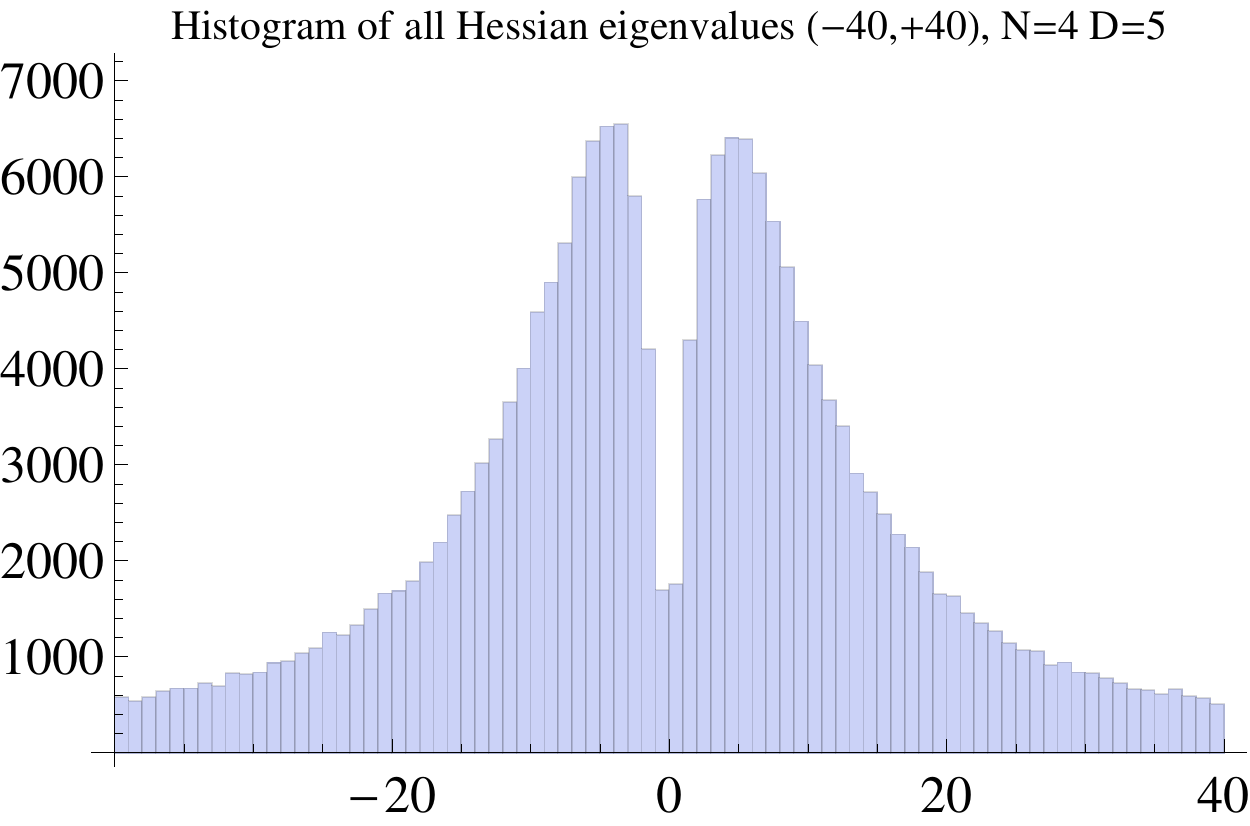}\\
\includegraphics[width=4cm]{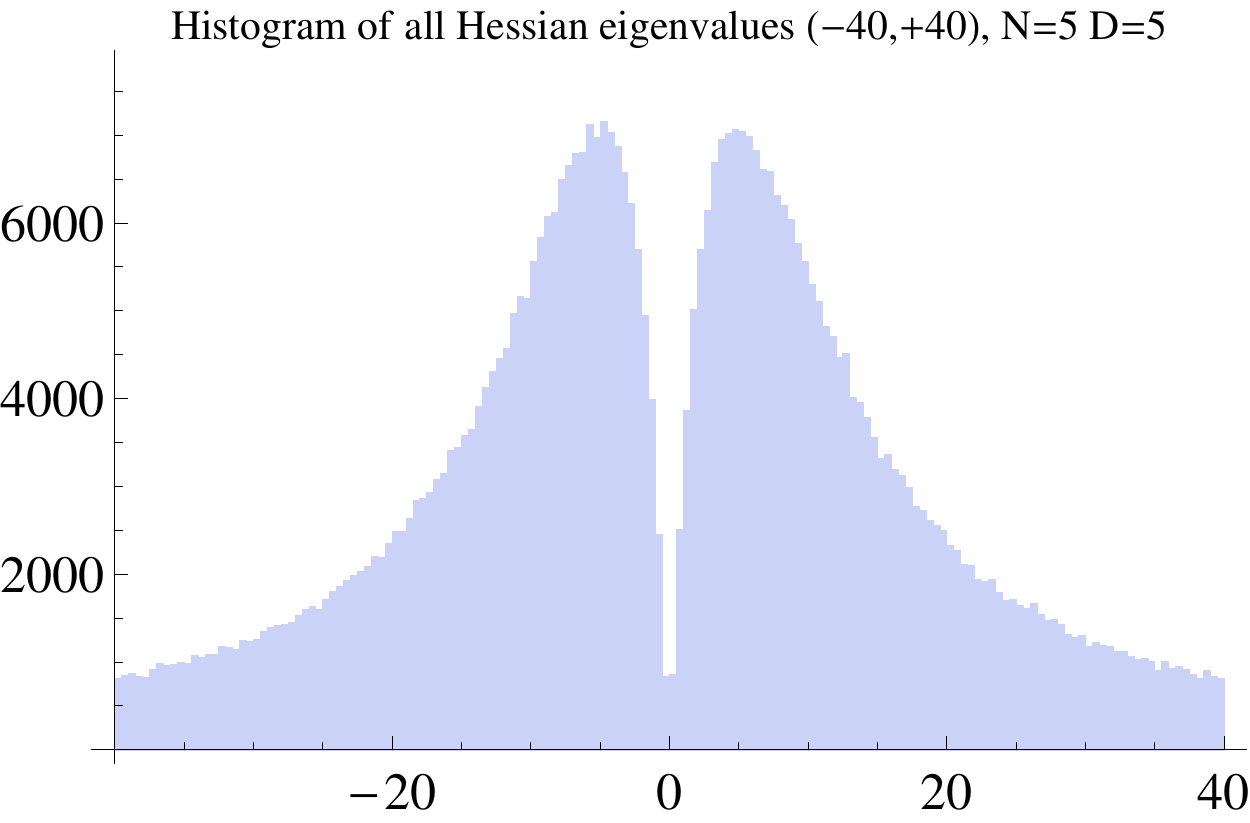}
\includegraphics[width=4cm]{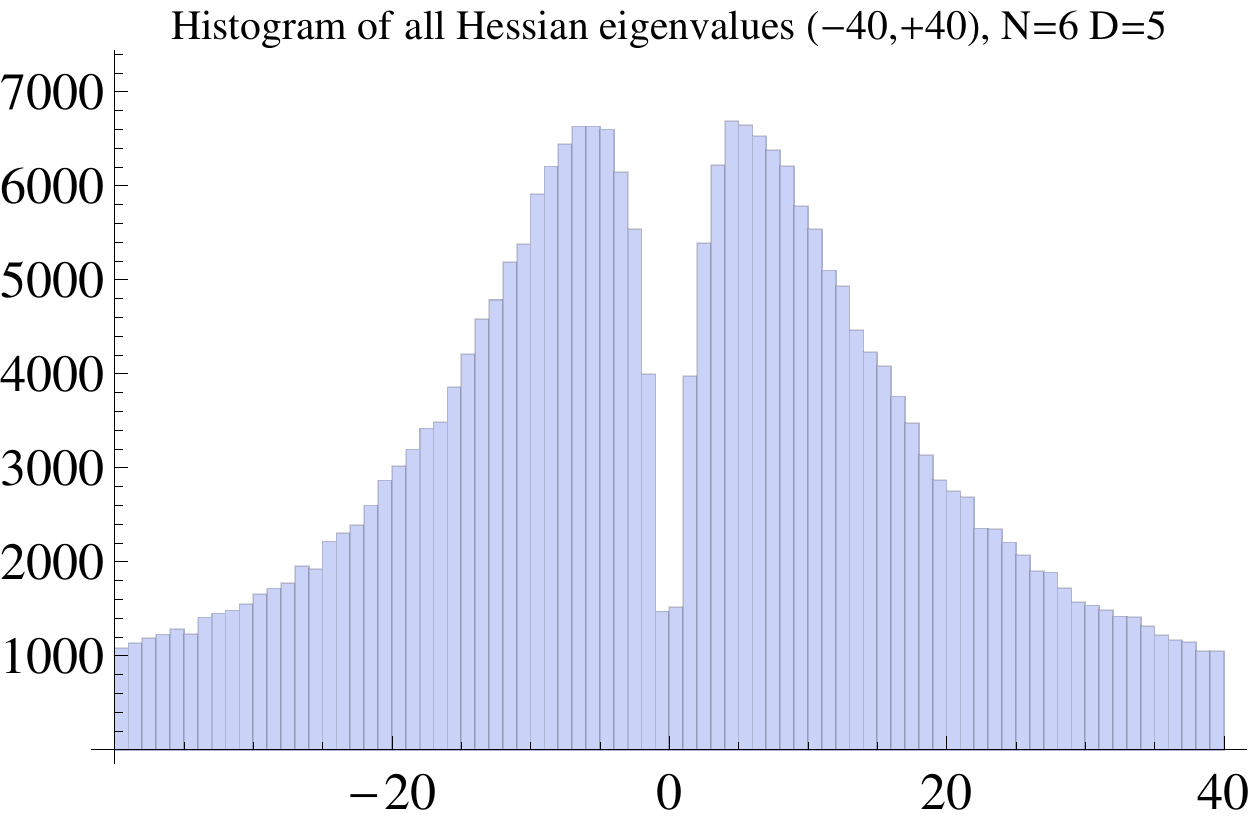}\\
\caption{Histogram of eigenvalues, $D=5$} \label{HistEigensD5}
\end{figure}

In Figures \ref{HistEigensD3}-\ref{HistEigensD5}, we plot histograms of the Hessian eigenvalues
where the x-ranges in the plots are chopped beyond $\pm40$ to show the behaviour of the plots in the middle range. 
Such histograms of eigenvalues
of RPs are widely studied, usually by considering the Hessian matrices of RPs
as random symmetric matrices and then applying Wigner's semicircle law or other
results. Our results, however, are distinct in that they represent the histograms of Hessian eigenvalues
computed \textit{at all the SPs} for each sample. The clefts in the histograms are a departure from Wigner's semicircle law. The 
reason behind the cleft can be explained using Sard's theorem \cite{mumford1995algebraic}, which yields that our dense random system of stationary equations will 
not possess singular solutions for almost all values of coefficients. This means that we are not likely to have zero eigenvalues, though
there are indeed SPs having eigenvalues closer to zero, creating the cleft at the zero eigenvalue in the histograms. From the statistical
physics point of view, as argued in \cite{Marsh:2011aa}, a possible reason of the cleft may be that the
Hessian matrices may have a component that can be described by the Altland-Zirnbauer CI ensemble \cite{altland1997nonstandard}, though
investigating this aspect further is beyond the scope of the present work.  


\subsection{Histograms of the lowest Hessian eigenvalues}
\begin{figure}[htp]
\includegraphics[width=4cm]{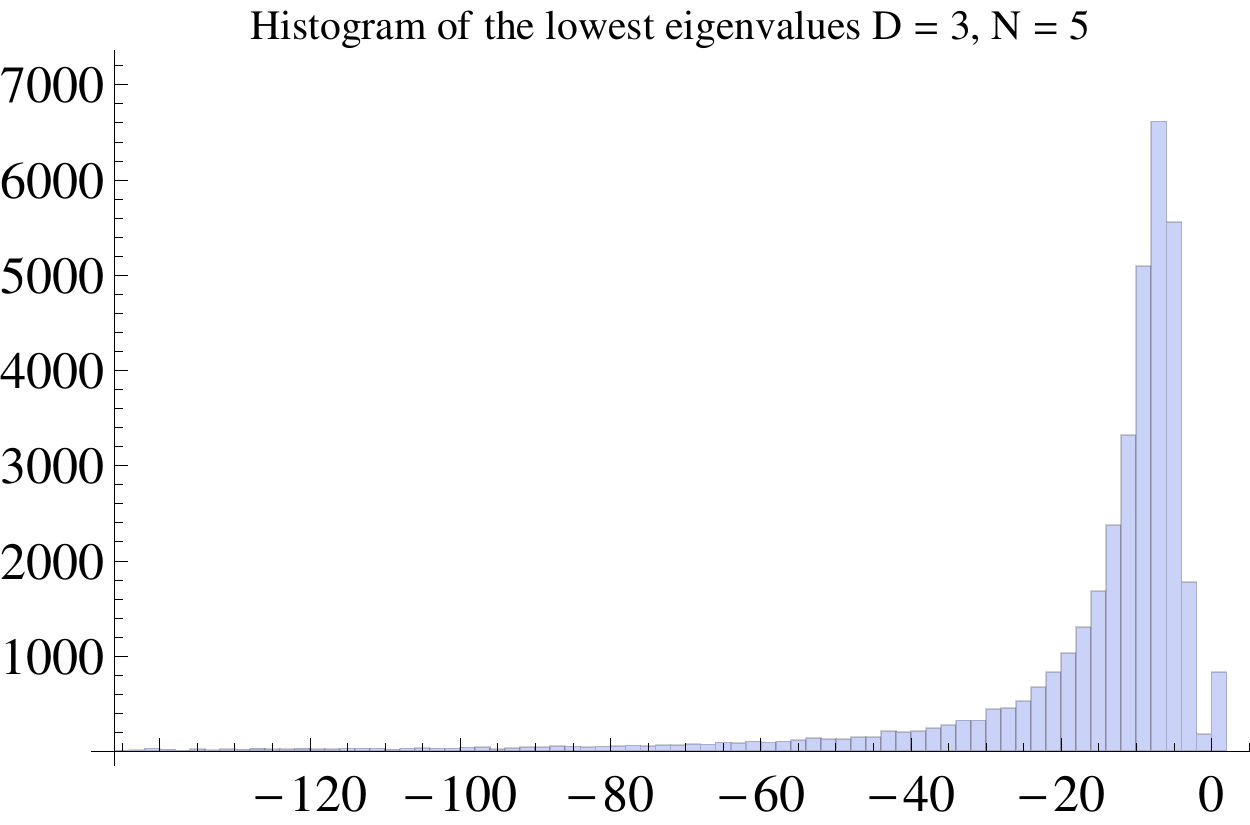}
\includegraphics[width=4cm]{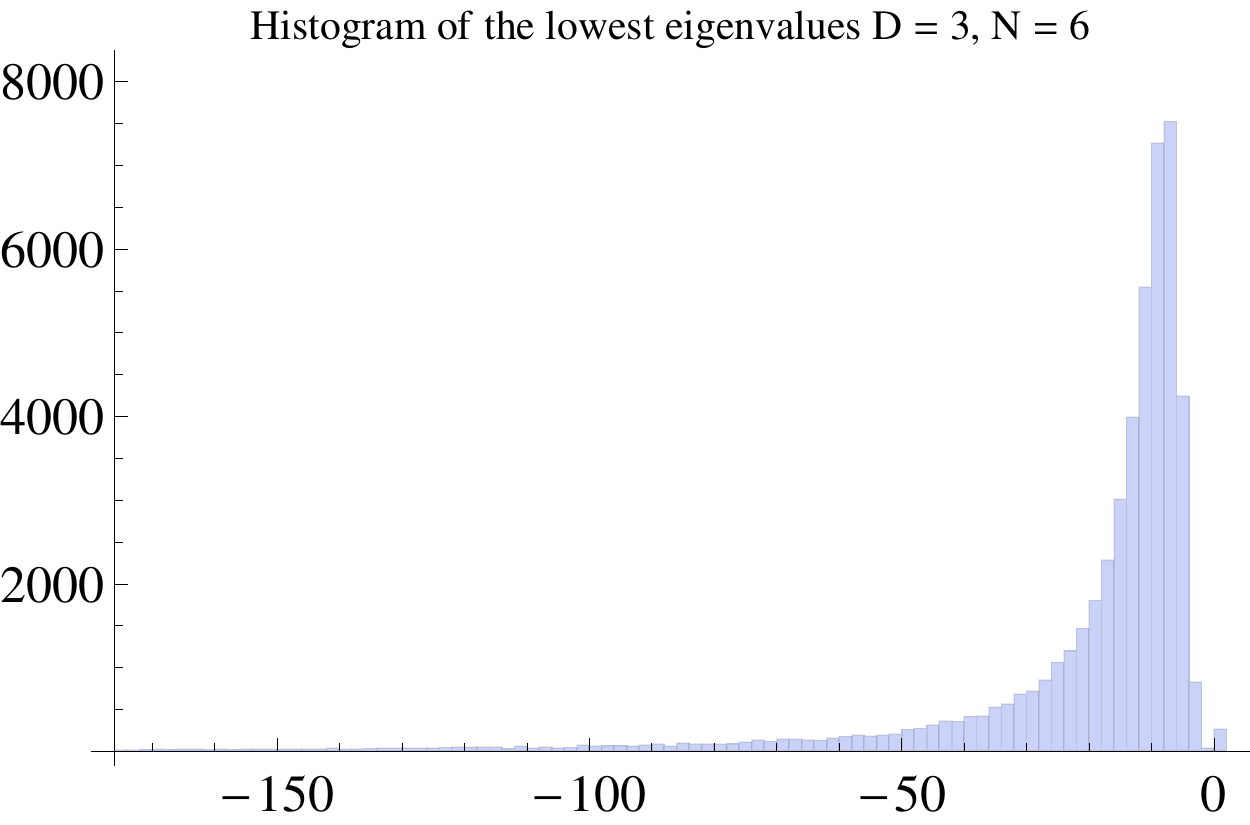}\\
\includegraphics[width=4cm]{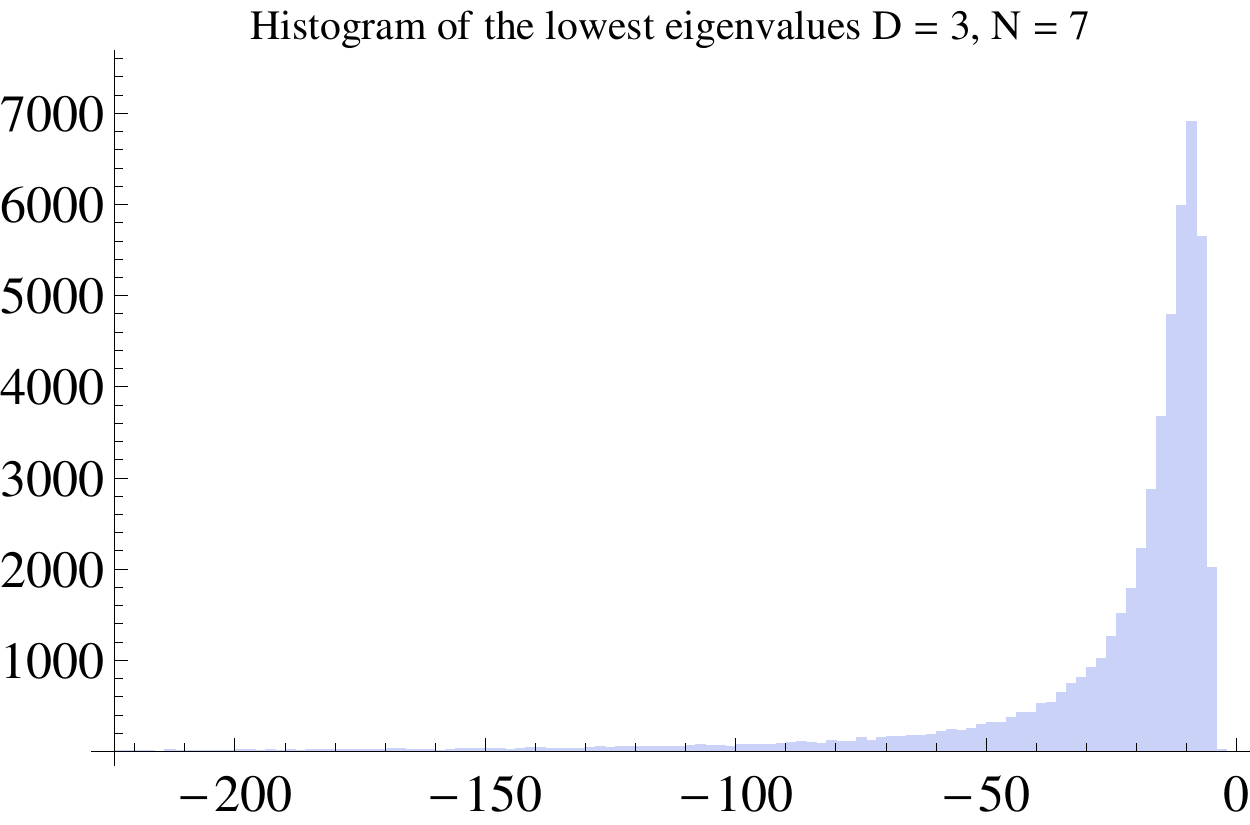}
\includegraphics[width=4cm]{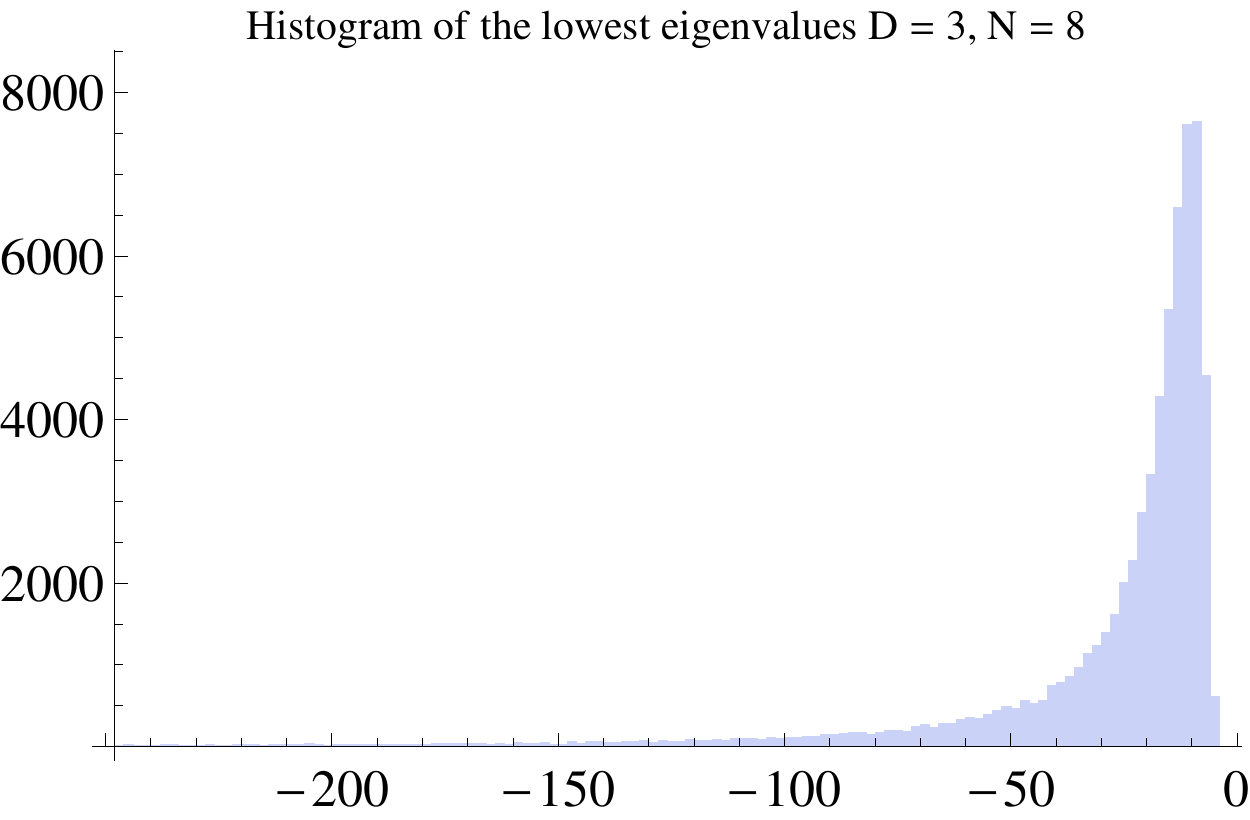}\\
\includegraphics[width=4cm]{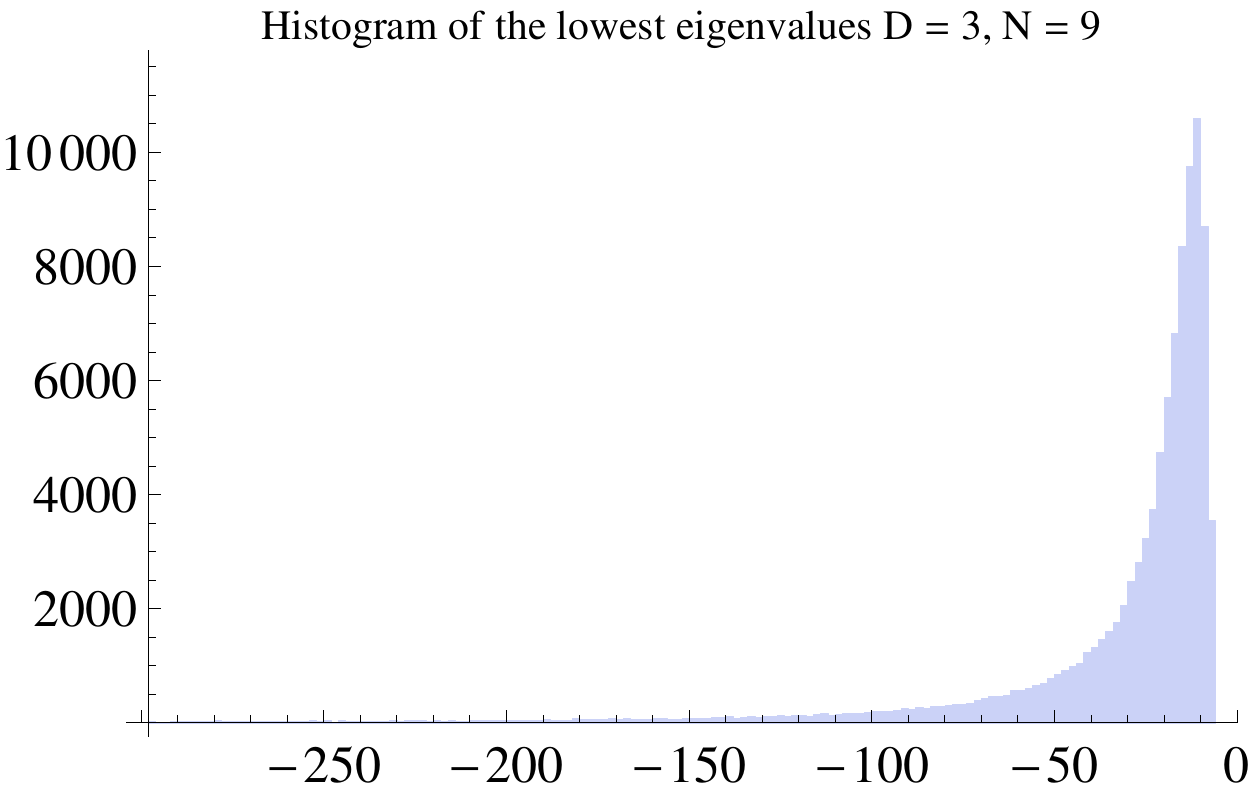}
\includegraphics[width=4cm]{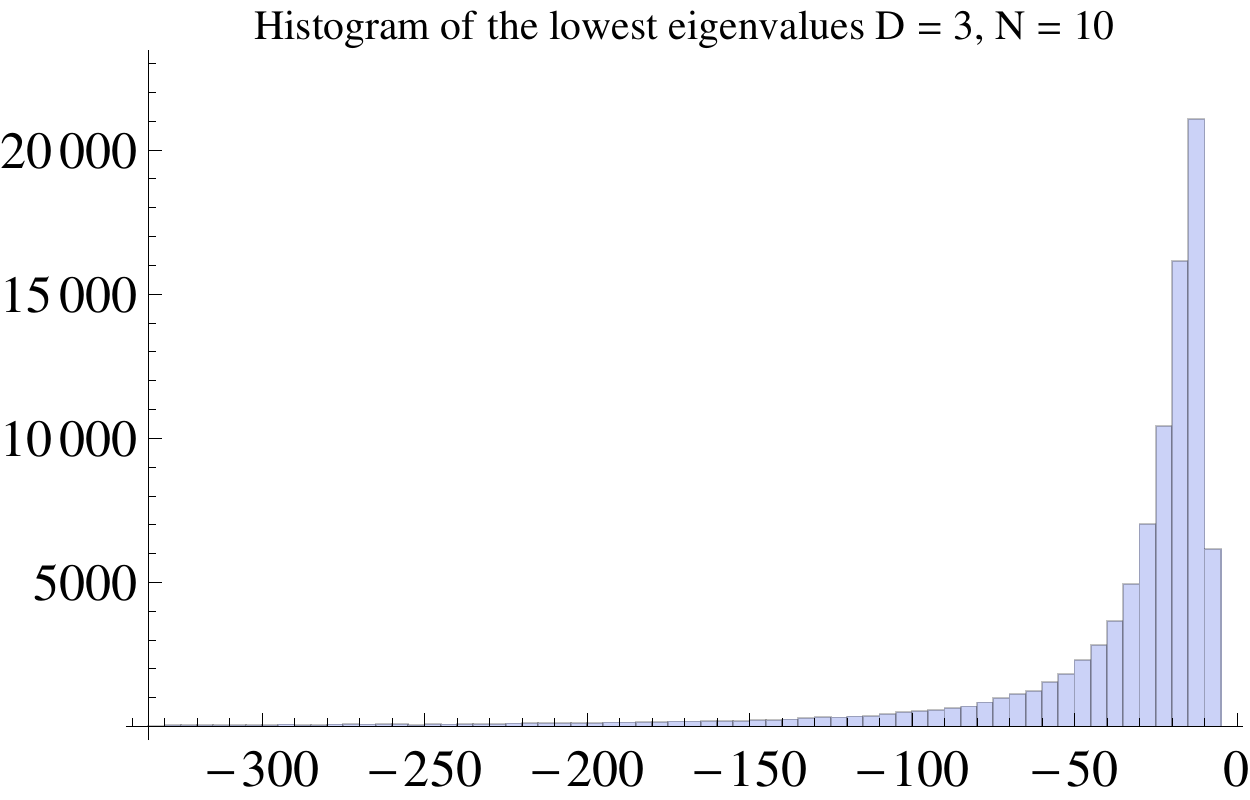}\\
\caption{Histogram of the lowest eigenvalues for $D=3$. Data for smaller eigenvalues than the shown in the histograms
are chopped away for the presentational purposes.} \label{HistLowEigensD3}
\end{figure}

\begin{figure}[htp]
\includegraphics[width=4cm]{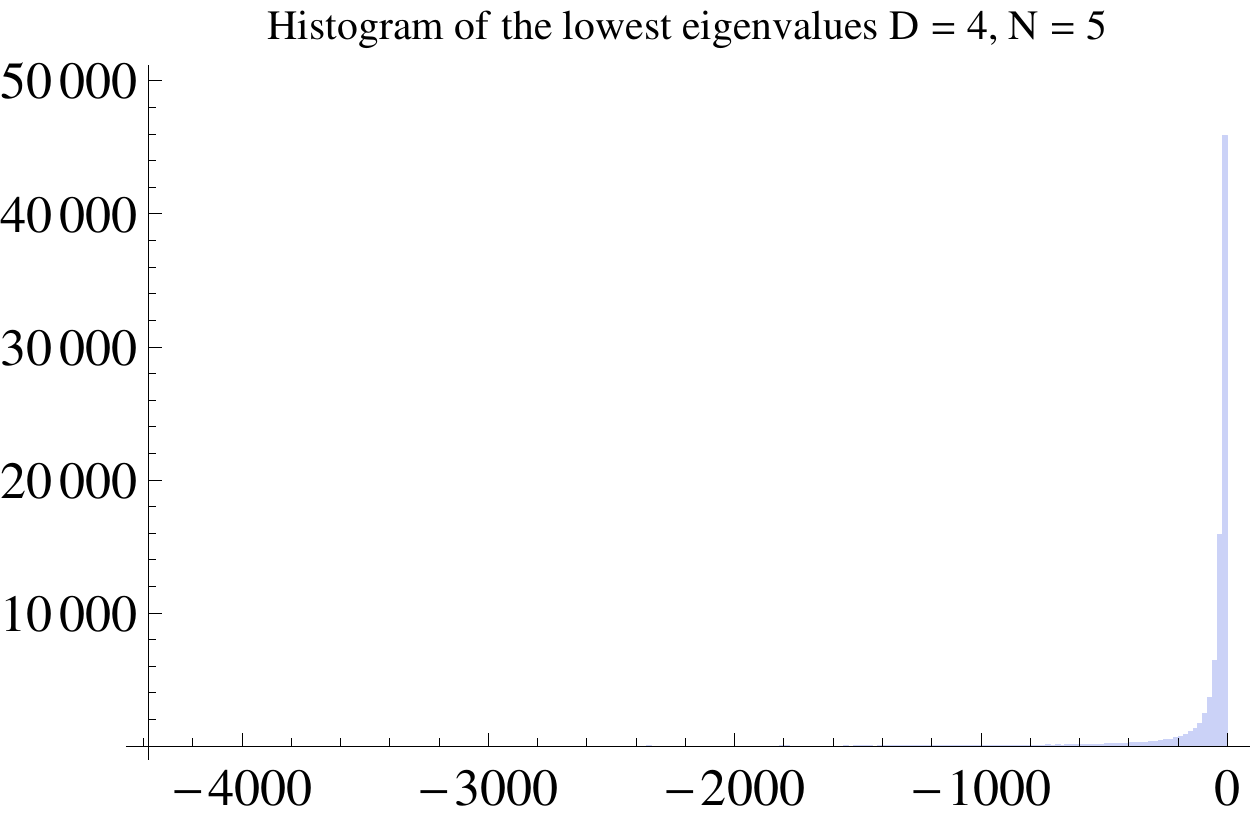}
\includegraphics[width=4cm]{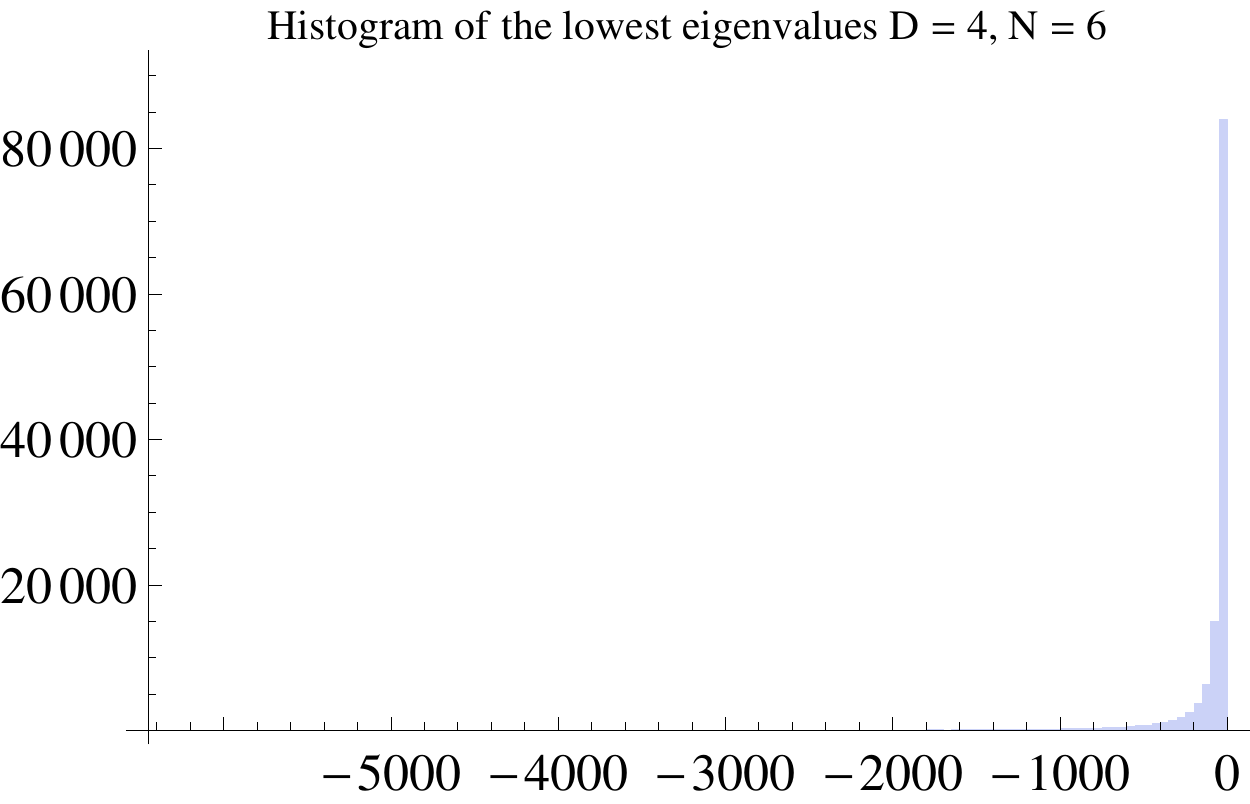}\\
\includegraphics[width=4cm]{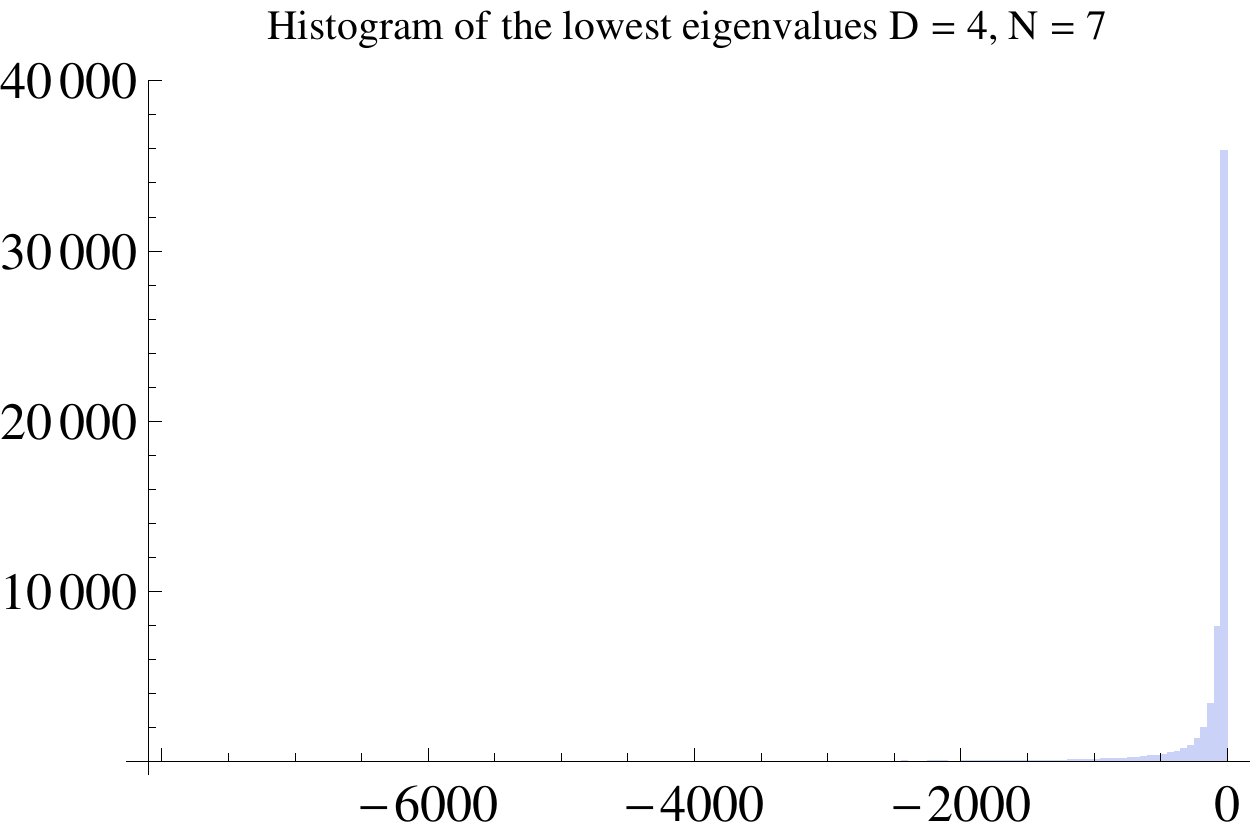}
\includegraphics[width=4cm]{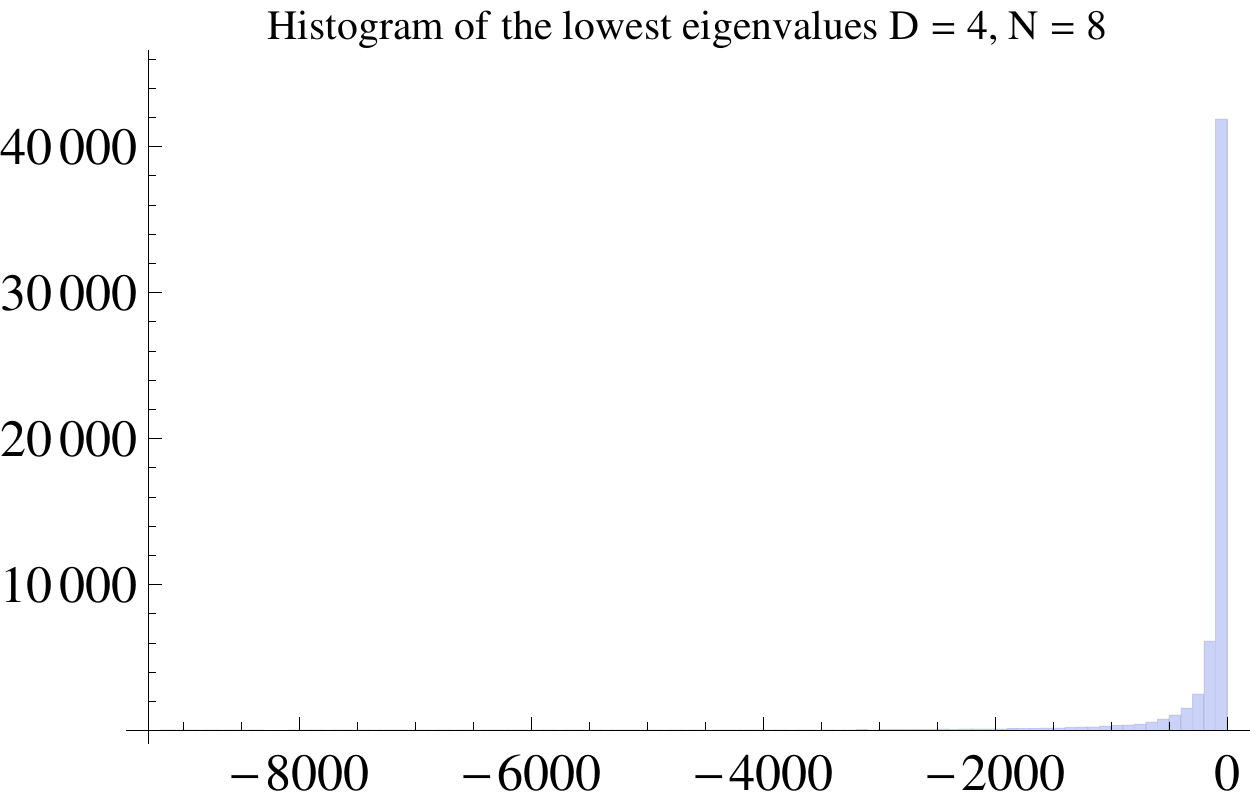}\\
\includegraphics[width=4cm]{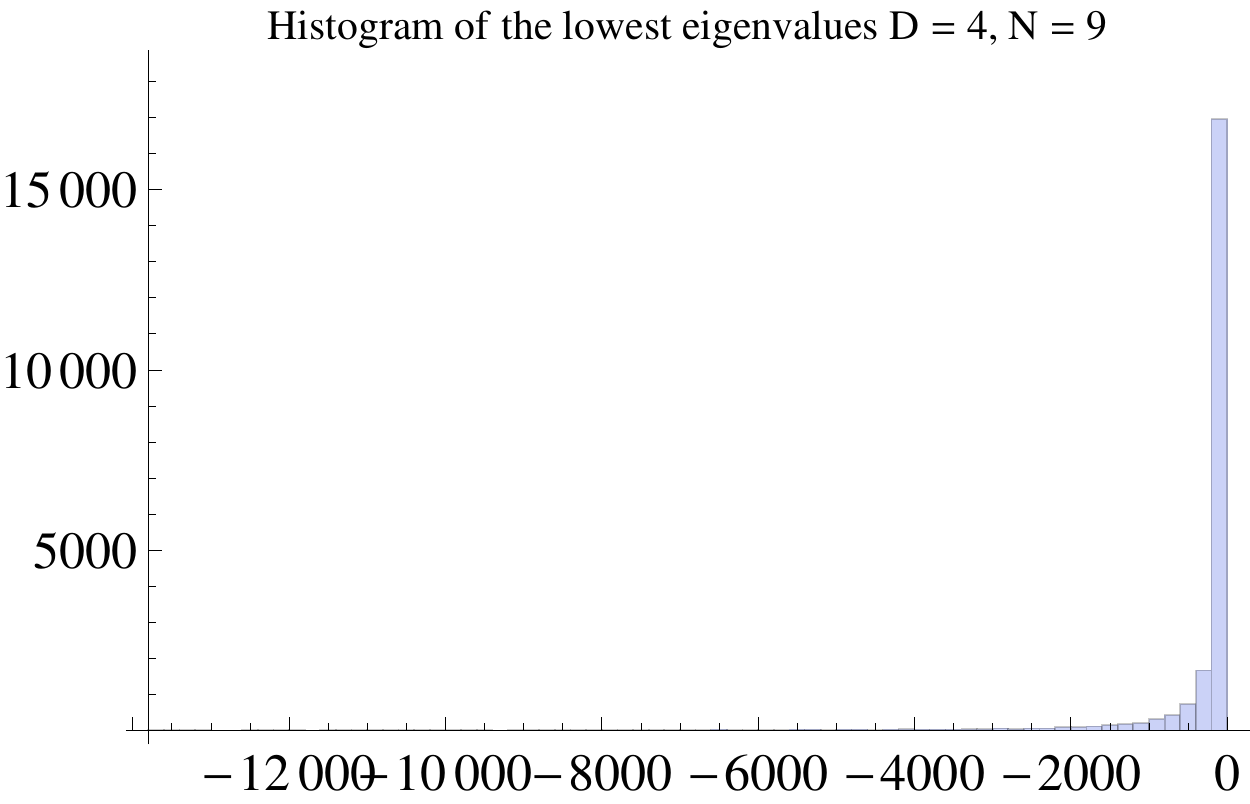}
\includegraphics[width=4cm]{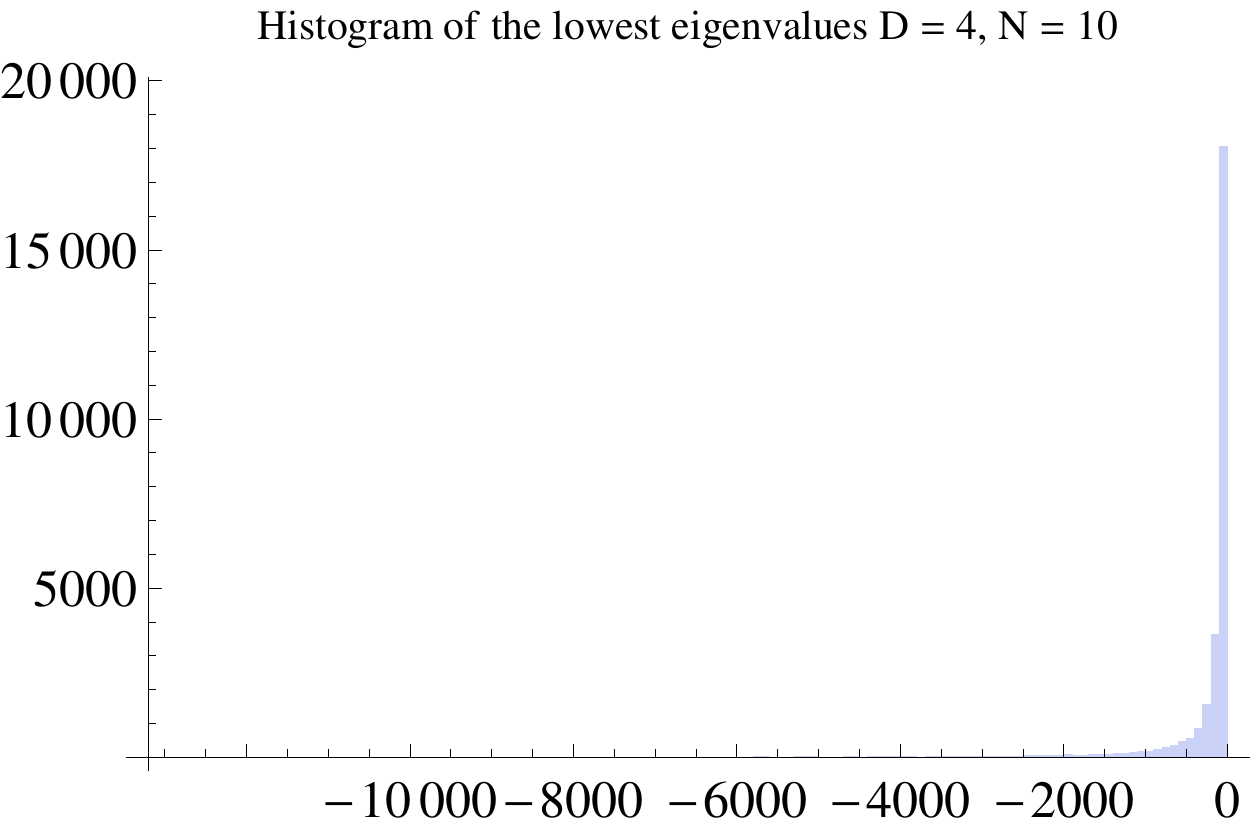}\\
\caption{Histogram of the lowest eigenvalues for $D=4$. Data for smaller eigenvalues than the shown in the histograms
are chopped away for the presentational purposes.} \label{HistLowEigensD4}
\end{figure}

\begin{figure}[htp]
\includegraphics[width=4cm]{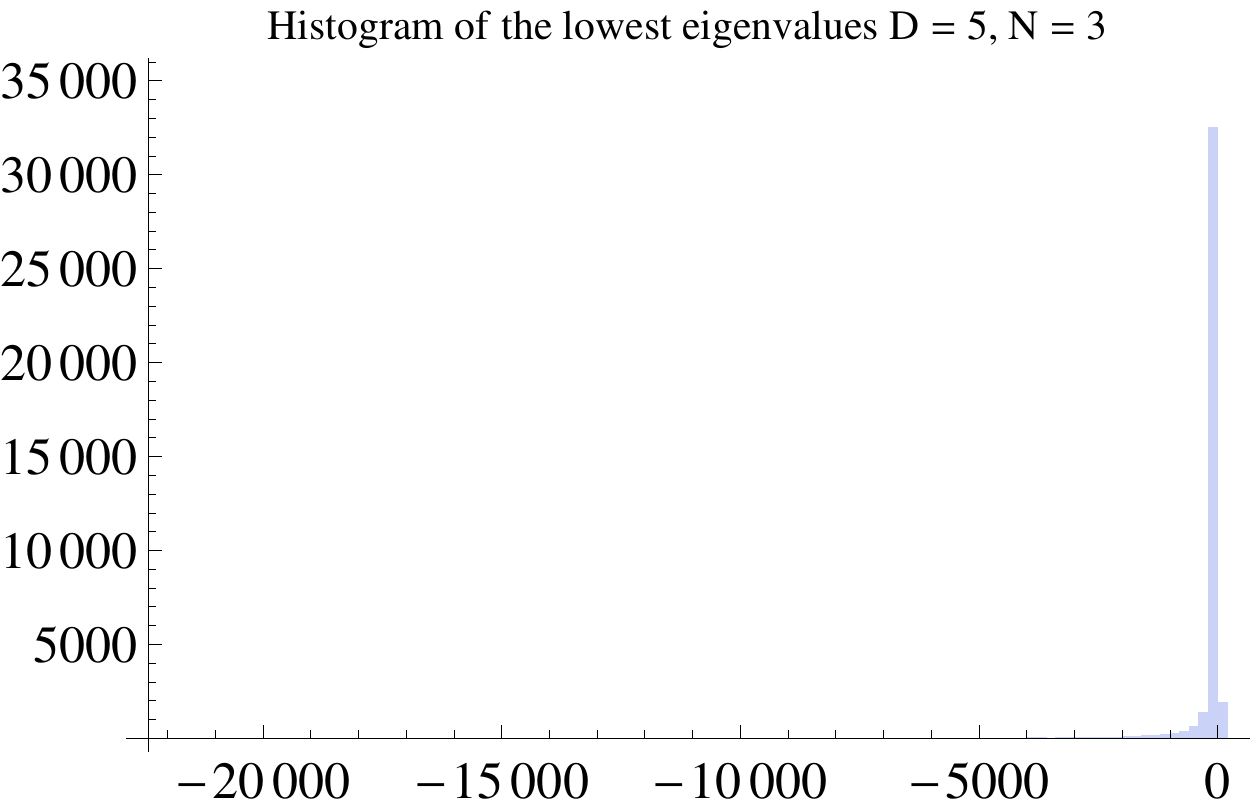}
\includegraphics[width=4cm]{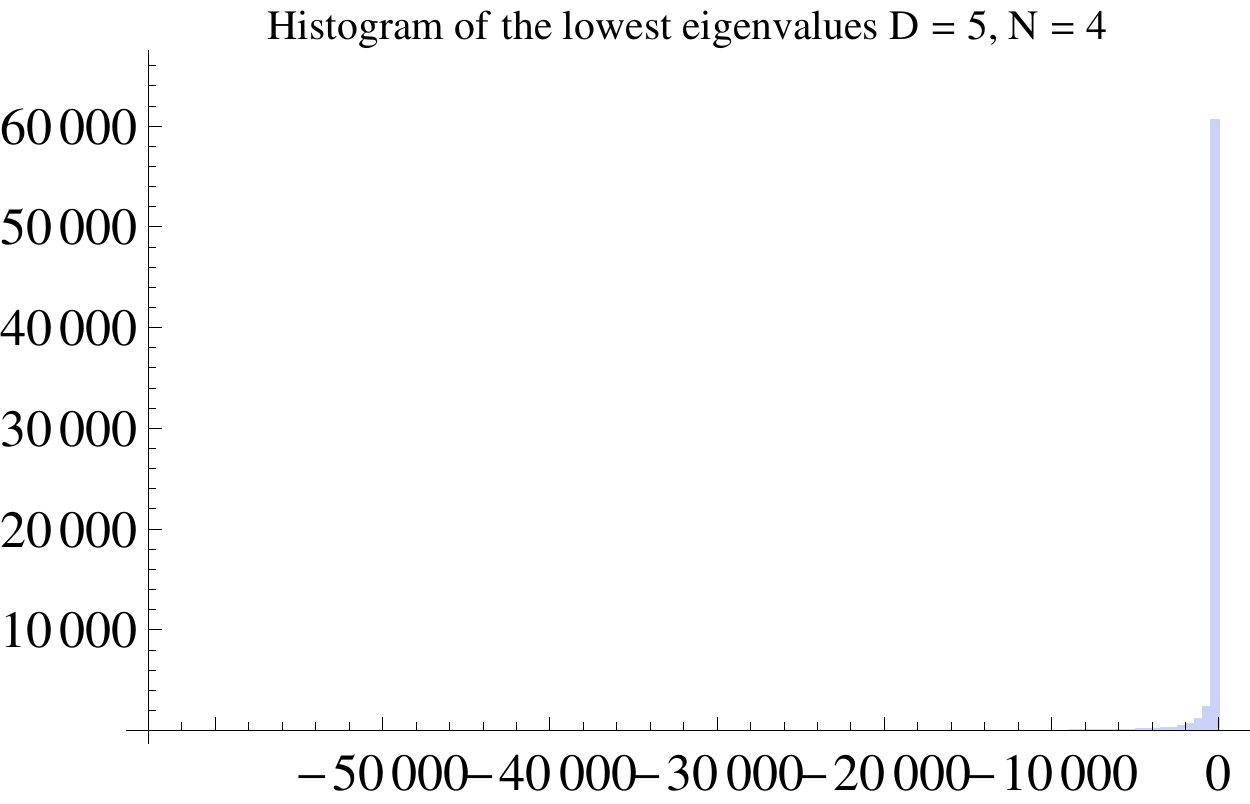}\\
\includegraphics[width=4cm]{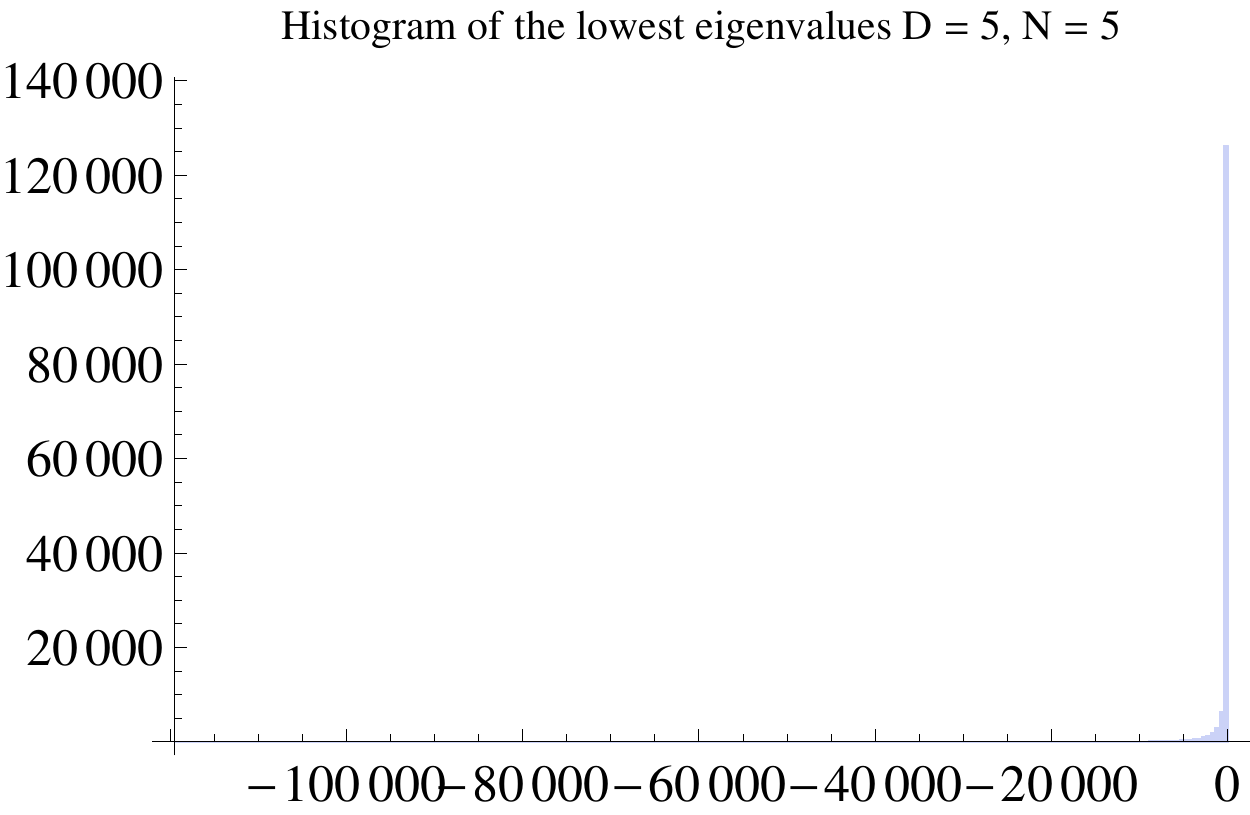}
\includegraphics[width=4cm]{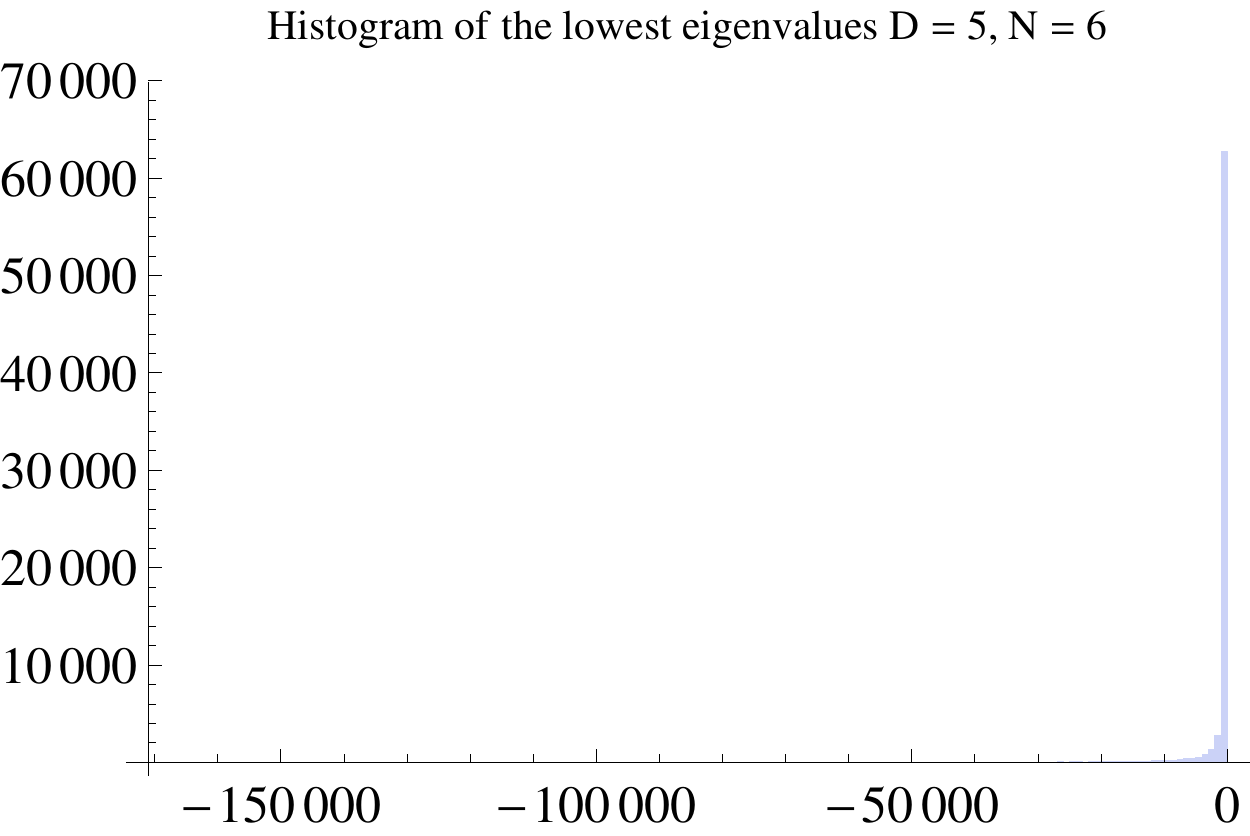}\\
\caption{Histogram of the lowest eigenvalues for $D=5$. Data for smaller eigenvalues than the shown in the histograms
are chopped away for the presentational purposes.} \label{HistLowEigensD5}
\end{figure}

The lowest eigenvalue of the Hessian matrix evaluated at a real SP is one of the most important physical quantities
as it determines physics up to certain extent. Many interesting properties of the lowest eigenvalues
of different potentials also have deep connection to catastrophe theory \cite{wales2001microscopic}. In Figures
\ref{HistLowEigensD3}-\ref{HistLowEigensD5}, we plot histograms of the lowest eigenvalues computed 
at all real SPs for all samples for various values of $N$ and $D$. Our results yield that the tail of the lowest eigenvalues 
is long, though there are only a few events at the far end.

\subsection{Real vs Imaginary Plots of $V$ at Complex Solutions}
\begin{figure}[htp]
\includegraphics[width=4cm]{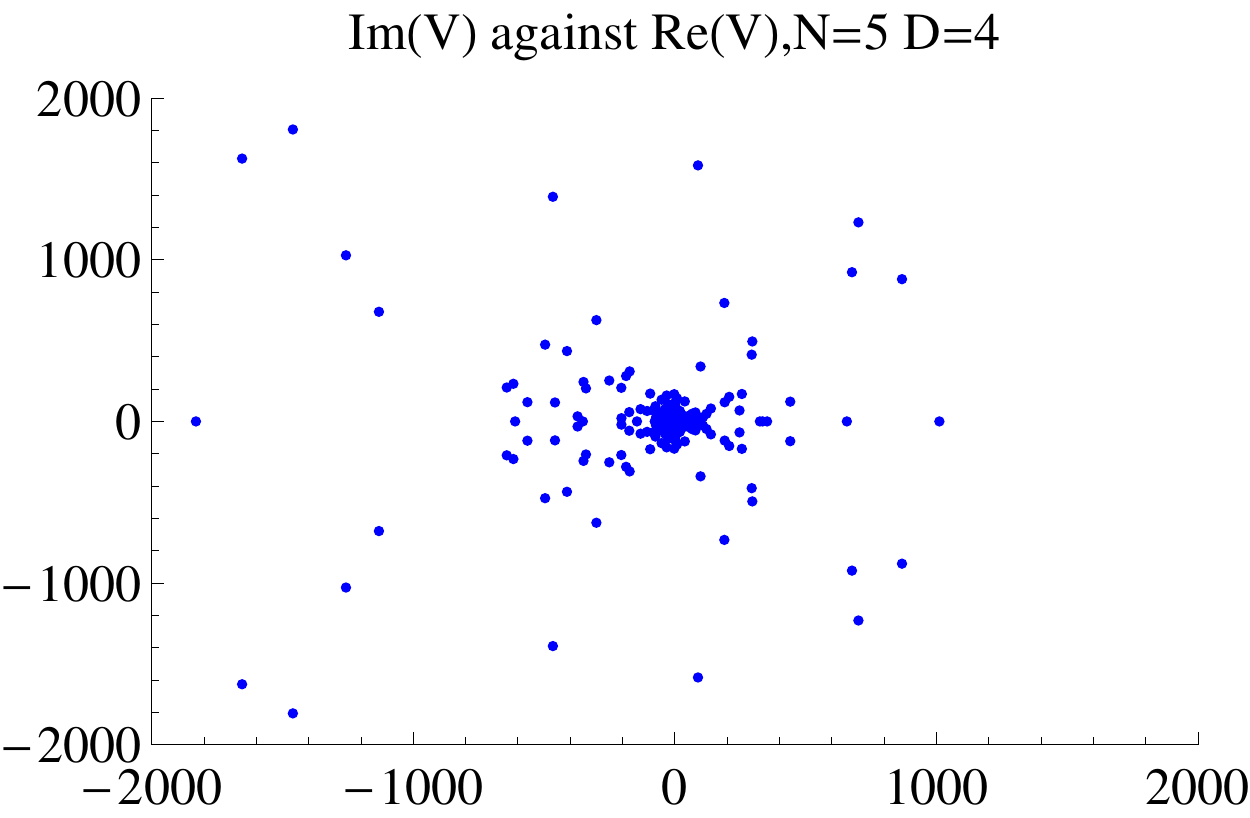}
\includegraphics[width=4cm]{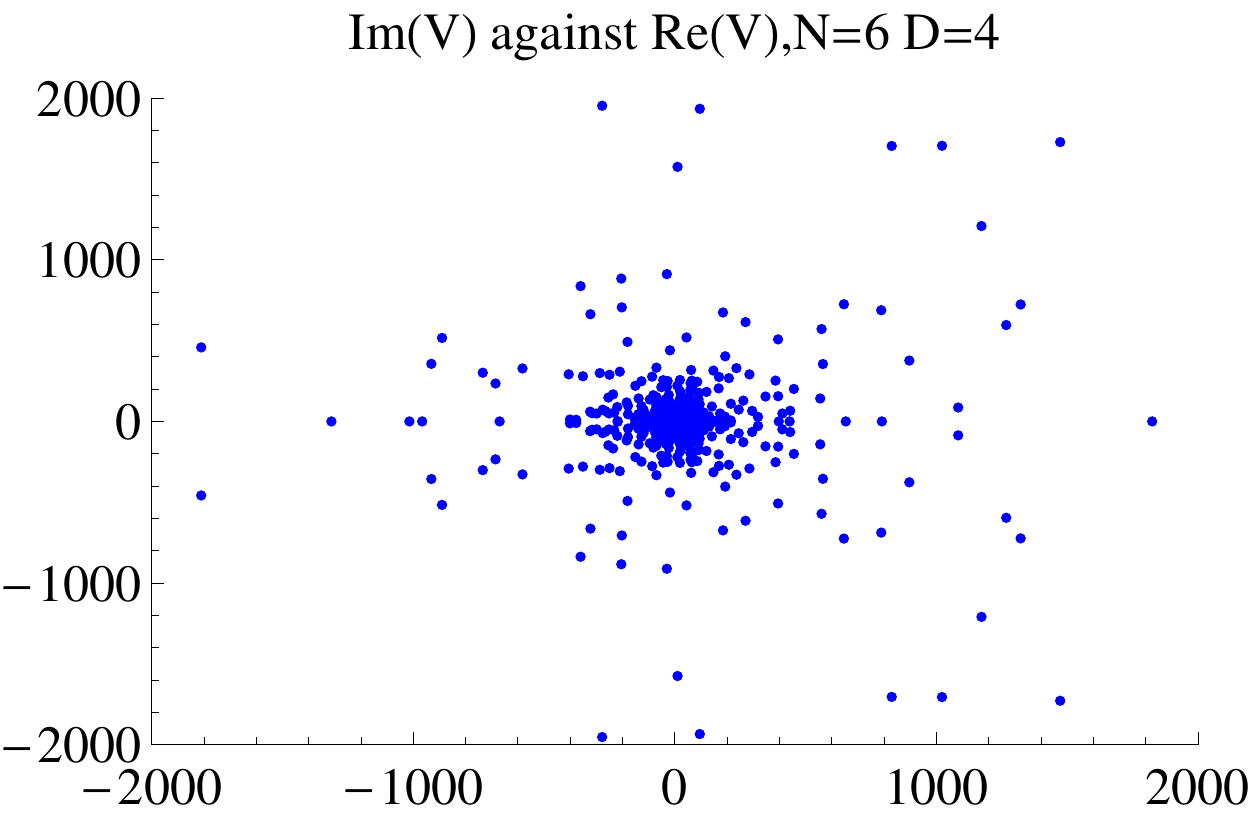}\\
\includegraphics[width=4cm]{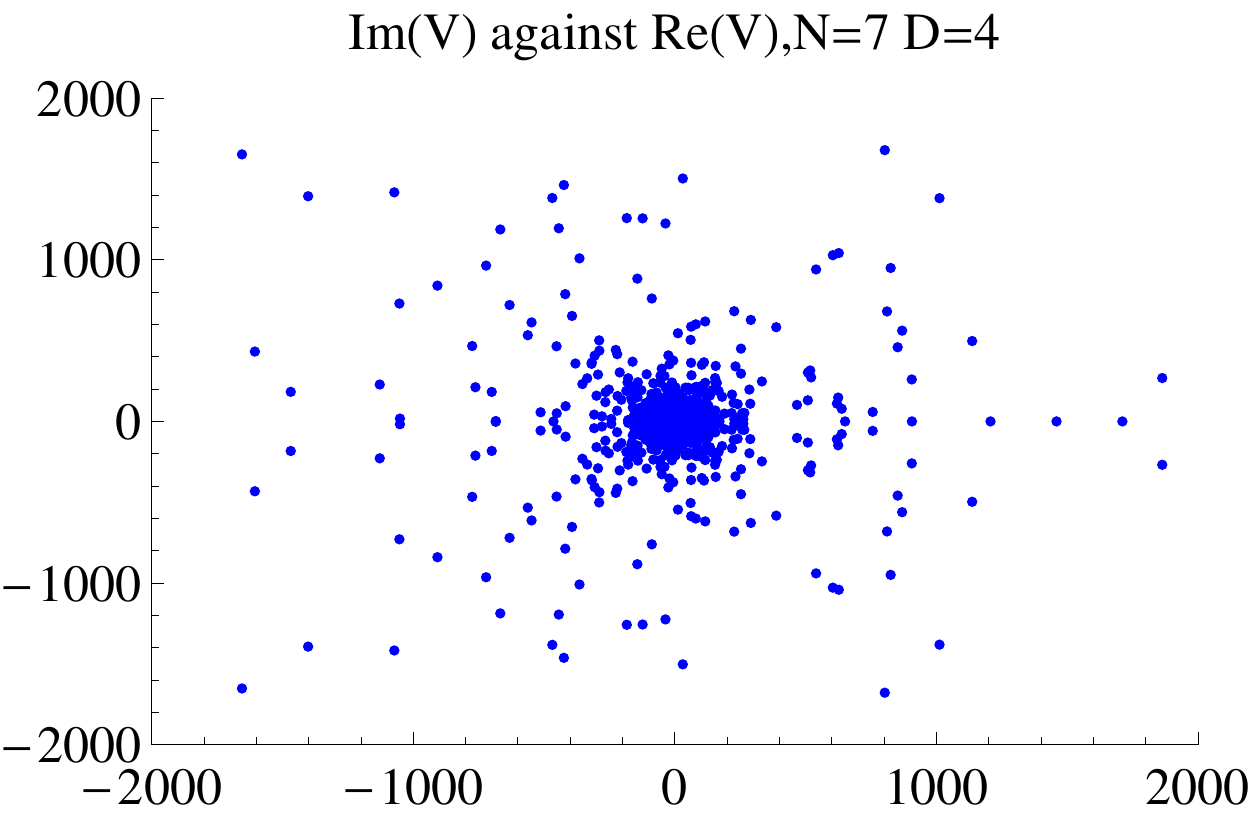}
\includegraphics[width=4cm]{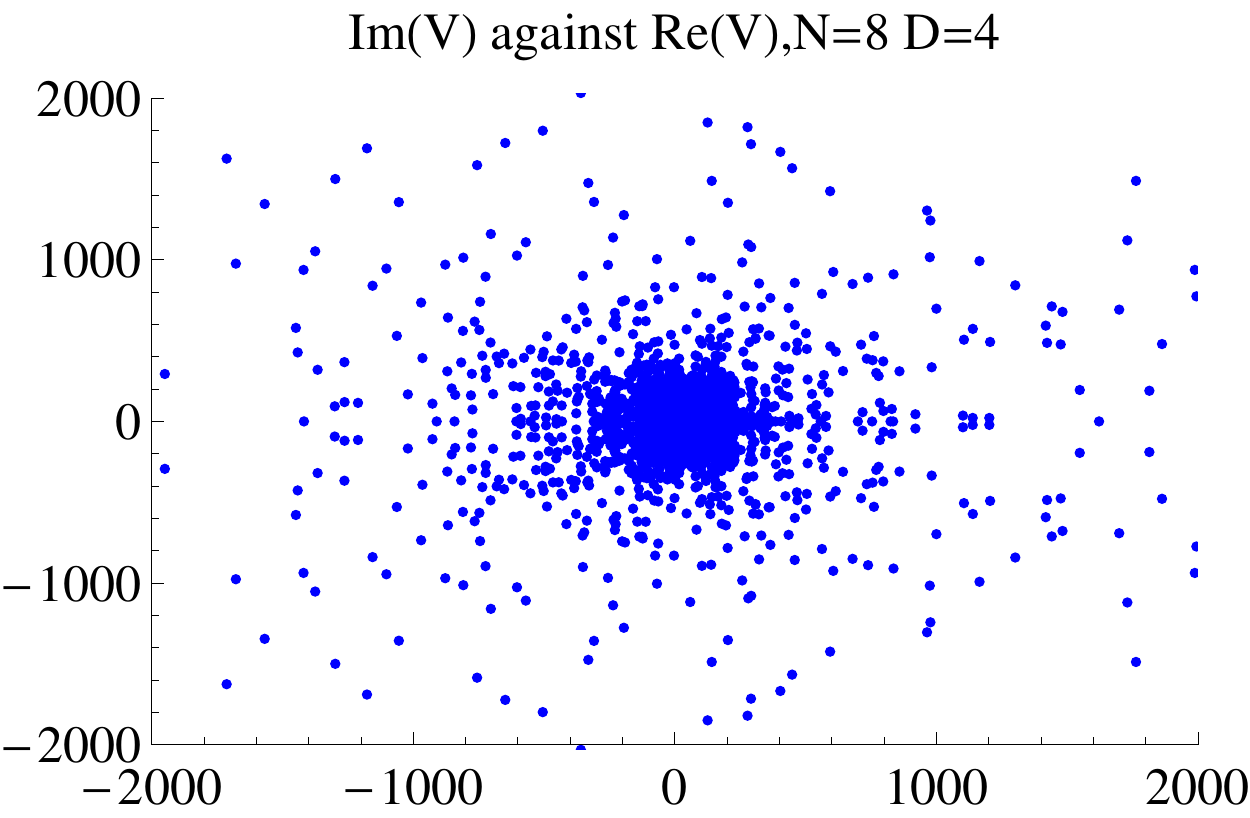}\\
\includegraphics[width=4cm]{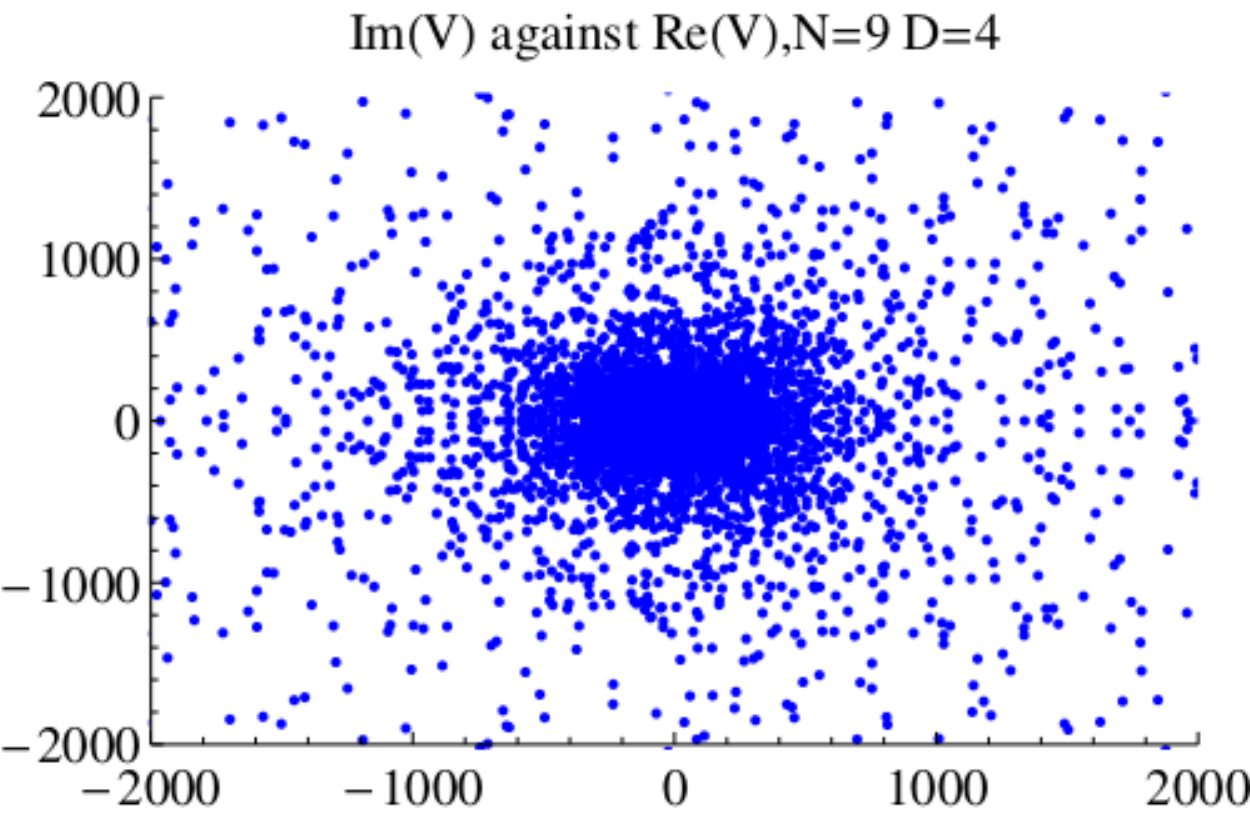}
\includegraphics[width=4cm]{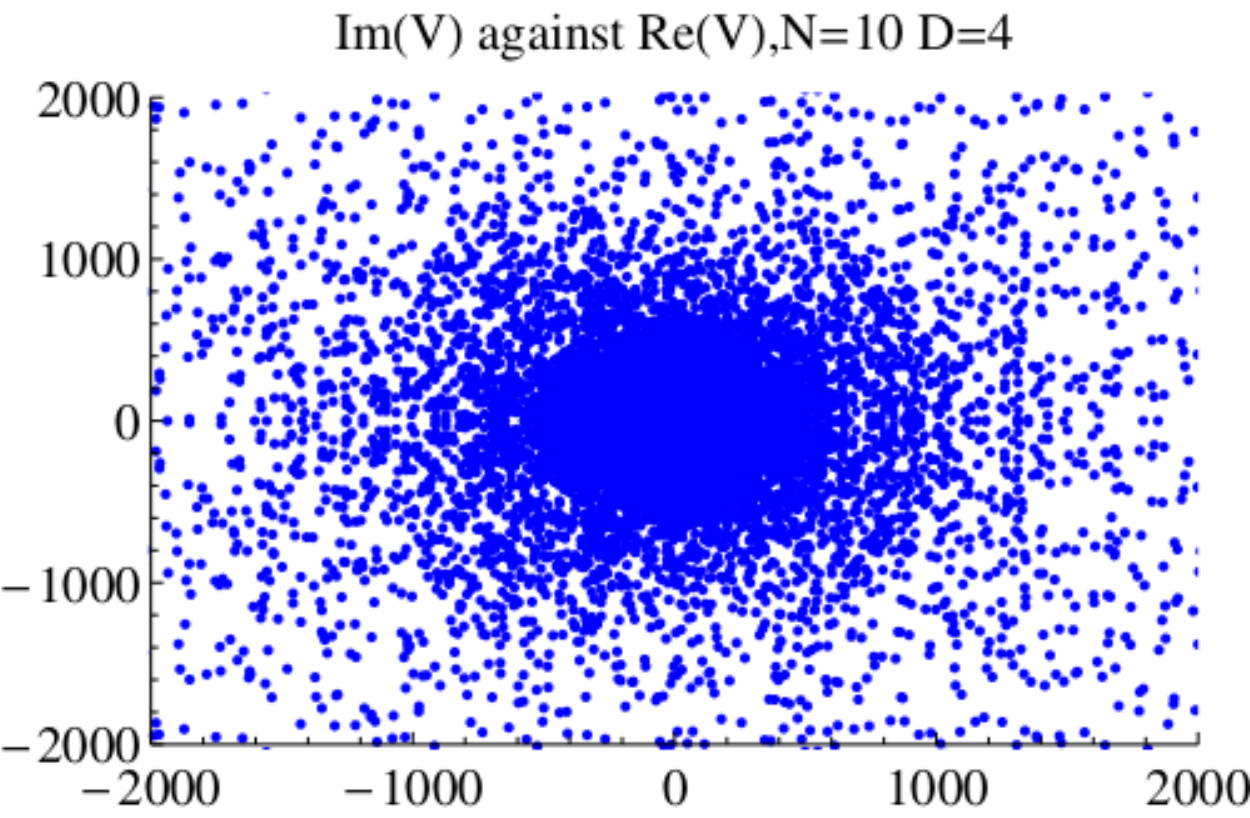}\\
\caption{Real vs Imaginary values of $V$ at all complex solutions, $D=4$} \label{ReVsImD4}
\end{figure}

In most applications we require the knowledge of real SPs, however, motivated by \cite{Mehta:2012qr,PhysRevE.91.022133},
we plot real vs imaginary parts of the potential at complex SPs in Figures \ref{ReVsImD4}, for $D=4$, which yield 
that there are many complex SPs at which the potential $V$ itself is a real quantity. Though in many physical models, both coordinates
and potential have to be strictly real quantities and the complex solutions may appear just due to the complexification of
the system in order to be able to use the complex algebraic geometry methods,
in string phenomenology complex moduli do have physical meaning. We hope that our observations put forward in these figures may provide
interesting avenues for further research to understand and interepret \textit{complexified} fields.

\subsection{Average Timing}
Although not directly attacked, our experiments are related to the 17th problem from the list of the eighteen unsolved 
problems for the 21st century that Smale laid down in Ref.~\cite{smale1998mathematical} which we quote verbatim:
``Can a zero of $N$ complex polynomial equations in $N$ unknowns be found approximately,
on the average, in polynomial time with a uniform algorithm?'' 

In order to gain some insight into this difficult problem, we consider the average time the \emph{Bertini 1.4} software 
takes to find one solution and consider this value for fixed fixed $D$.  

\begin{figure}[htp]
\includegraphics[width=8cm]{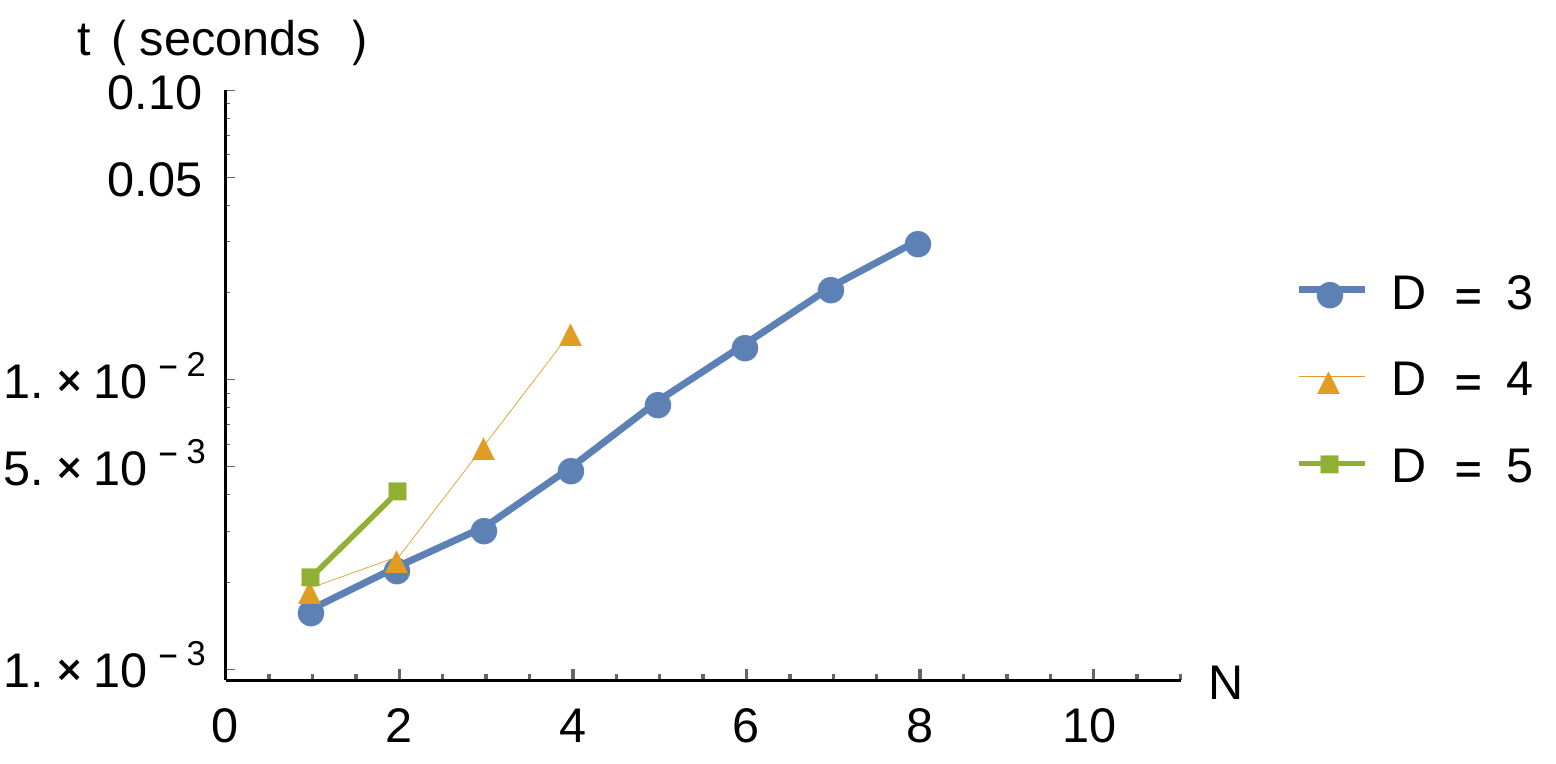}
\caption{$N$ vs average time (in seconds) to track each solution path on a single machine. 
As explained in the text, this plot only provides a very crude estimate on average timing. Here, we start from $N=1$
to get more data points.} \label{fig:Nvst}
\end{figure}

For $D = 3-5$, we note that for the systems that we considered, the average time to track each path
seems to grow exponentially in $N$ as seen in Figure \ref{fig:Nvst}. However, 
it is very important to note that the timings here includes the run-time of the homotopy path-tracking algorithm
as well other nonessential functions. In addition, the timing information in Figure \ref{fig:Nvst} 
was obtained using a single core processor (a 2300 MhZ AMD processor)  
to avoid timing involved during parallel processes, though all the results presented in the previous subsections were carried out
using a computing cluster of 64 processor.
Hence, Figure \ref{fig:Nvst} in no way gives a good estimate of the actual averate time for path tracking, but just gives a very crude
estimate in the hope to provide a rough guide for the practitioners.

\section{Conclusions and Outlook}\label{sec:conclusions}
In this article, we find all the critical points, complex and real, of the most general polynomial potential 
whose coefficients take i.~i.~d.~ values from the Gaussian distribution with mean 0 and variance 1.  With this set up, 
we extract statistical results on random potentials (RPs). We have employed the numerical polynomial 
homotopy continuation (NPHC) method which guarantees that we have found all the critical points. 
Since the NPHC is a numerical method,
choosing specific values for the tolerances to classify when 
the solutions are real becomes a delicate issue when $N$ and $D$ are large enough. Hence, we combine
a \textit{certification} procedure which provides a sufficient condition as to when a given approximate solution converges to a distinct real 
finite solution
using Smale's $\alpha$-theorem \cite{BCSS,hauenstein2012algorithm}. 
Together with this fact and that the stationary equations of our RP always have $(D-1)^N$ complex critical points for generically chosen coefficients,
our numerical results can be treated as good as exact results.

Due to various involved computations (solving equations using the NPHC method, certifying and sorting them, and computing Hessian eigenvalues),
and limited computational resources, we have restricted ourselves to 
the number of variables $N \in \{2,\dots,11\}$ and degree $D \in \{3,\dots,5\}$, and $1000$ samples for each $(N,D)$.

We have shown that the mean number of real SPs exponentially increases, whereas the mean number of minima exponentially decreases,
with increasing $N$ for various values of $D$. Our results also yield that the mean number of minima
increases for fixed $N$ while increasing $D$, though the increase is linear. With an additional constraint on the minima
that the potential evaluated at minima is positive definite, we investigate the statistical aspect of the
string theory landscape where our results may have important consequences in counting the string vacua. These results compare well
with an analytically computed upper bounds for these two average quantities \cite{dedieu2008number}.
One of the main achievements of this paper is a numerical computation of the variance of the number of real SPs and minima since it
is prohibitively difficult to obtain analytical results for the variance.  We observed that the variances exhibit
the same qualitative behaviour as the mean. To further investigate the spread of the number of real SPs,
we also plotted histograms of the number of real SPs which exhibit clear unimodality and right-skewedness.

By sorting the real SPs according to their Hessian indices, we show that the mean number of real SPs having index 
$I+1$ is significantly more than the those having index $I$. Extrapolating our results to higher $N$ and $D$ yields that 
the plot of $I/N$ vs mean number of SPs with index $I$ tend to be a bell-shaped curve, which appears to be a common feature for 
many other potential energy landscapes such as the Lennard-Jones clusters \cite{2002JChPh.116.3777D,2003JChPh.11912409W}, 
XY model \cite{Mehta:2009,Mehta:2009zv,Hughes:2012hg,Hughes:2014moa,mehta2014algebraic}, 
spherical $3$-spin mean-field model \cite{Mehta:2013fza}, etc. A milder version of this result, that the ratio between
the number of SPs with index $I>0$ and the number of minima exponentially blows up when $N$ increases, has gained
attention recently in both mathematics and string theory \cite{dedieu2008number,Marsh:2011aa}.

Yet another novel result of our investigation is 
histogram of the Hessian eigenvalues computed at all real SPs since it is a departure 
from the traditional eigenvalue histogram studies which consider Hessian matrices evaluated at 
arbitrary points and may arrive at Wigner's semicircle law or else. Our histograms show clear bimodal symmetric 
behaviour with the cleft near the zero eigenvalue is pronounced as $N$ increases. We explained this result 
using Sard's theorem. We also observe that the histograms of lowest eigenvalues are long-tailed.

Though usually physically interesting solutions are the real
solutions, we observe in our numerical experiments
that the potential may be real when evaluated at many 
of the complex (i.e., non-real) solutions. Such a phenomenon is observed and discussed previously for 
different potentials \cite{Mehta:2012qr,PhysRevE.91.022133}. Since in string theory, complex moduli 
play a prominent role in determining the physics of the model, our results may prompt a different interpretation
of the \textit{feasible} SPs.

We also provided average timing information for our computation for finding solutions using homotopy continuation method in Figure \ref{fig:Nvst}.
Though we do not claim to have directly worked on Smale's 17th problem ~\cite{smale1998mathematical}, nor to have an accurate timing 
estimate due to the overlaps with several other computations
we anticipate that our 
crude estimates will guide future computations of the other practitioners of the homotopy continuation methods.

\section*{Acknowledgement}
DM was supported by a DARPA Young Faculty Award 
and an Australian Research Council DECRA fellowship. This work was also 
a part of National Science Foundation ACI program grant no. 1460032.
MN was supported in part by the
National Basic Research Program of China Grants 2011CBA00300, 2011CBA00301, 
the Natural Science Foundation of China Grants 61044002, 61361136003.
All the authors would like to thank Carlos Beltran, John Cremona,
Yan Fyodorov, Jonathan Hauenstein, Gregirio Malajovich, Liam McAllister, Liviu Nicolaescu, Sonia Paban and Nicolas Simm for their
feedbacks on various stages of this work.

\bibliographystyle{unsrt}
\bibliography{bibliography_NPHC_NAG,bibliography_NPHC_NAG_1}

\begin{thebibliography}{10}

\bibitem{Wales:04}
David Wales.
\newblock {\em Energy Landscapes : Applications to Clusters, Biomolecules and
  Glasses (Cambridge Molecular Science)}.
\newblock {Cambridge University Press}, January 2004.

\bibitem{RevModPhys.80.167}
Michael Kastner.
\newblock Phase transitions and configuration space topology.
\newblock {\em Rev. Mod. Phys.}, 80(1):167--187, 2008.

\bibitem{Denef:2004ze}
Frederik Denef and Michael~R. Douglas.
\newblock {Distributions of flux vacua}.
\newblock {\em JHEP}, 0405:072, 2004.

\bibitem{Douglas:2006es}
Michael~R. Douglas and Shamit Kachru.
\newblock {Flux compactification}.
\newblock {\em Rev.Mod.Phys.}, 79:733--796, 2007.

\bibitem{Aazami:2005jf}
Amir Aazami and Richard Easther.
\newblock {Cosmology from random multifield potentials}.
\newblock {\em JCAP}, 0603:013, 2006.

\bibitem{Tye:2008ef}
S.-H.~Henry Tye, Jiajun Xu, and Yang Zhang.
\newblock {Multi-field Inflation with a Random Potential}.
\newblock {\em JCAP}, 0904:018, 2009.

\bibitem{Marsh:2011aa}
David Marsh, Liam McAllister, and Timm Wrase.
\newblock {The Wasteland of Random Supergravities}.
\newblock {\em JHEP}, 1203:102, 2012.

\bibitem{Aravind:2014aza}
Aditya Aravind, Dustin Lorshbough, and Sonia Paban.
\newblock {Lower bound for the multifield bounce action}.
\newblock {\em Phys.Rev.}, D89(10):103535, 2014.

\bibitem{Battefeld:2012qx}
Diana Battefeld, Thorsten Battefeld, and Sebastian Schulz.
\newblock {On the Unlikeliness of Multi-Field Inflation: Bounded Random
  Potentials and our Vacuum}.
\newblock {\em JCAP}, 1206:034, 2012.

\bibitem{Bach:2014}
Bachlechner~Th. C.
\newblock {On Gaussian Random Supergravity.}
\newblock {\em JHEP}, 2014:54:32pp, 2014.

\bibitem{PhysRevLett.92.240601}
Yan~V. Fyodorov.
\newblock Complexity of random energy landscapes, glass transition, and
  absolute value of the spectral determinant of random matrices.
\newblock {\em Phys. Rev. Lett.}, 92:240601, Jun 2004.

\bibitem{PhysRevLett.93.149901}
Yan~V. Fyodorov.
\newblock Erratum: Complexity of random energy landscapes, glass transition,
  and absolute value of spectral determinant of random matrices [phys. rev.
  lett. \textbf{92} , 240601 (2004)].
\newblock {\em Phys. Rev. Lett.}, 93:149901, Sep 2004.

\bibitem{PhysRevLett.98.150201}
A.~J. Bray and Dean D.S.
\newblock Statistics of critical points of gaussian fields on large-dimensional
  spaces.
\newblock {\em Phys. Rev. Lett.}, 98:150201, Apr 2007.

\bibitem{PhysRevLett.109.167203}
Y.~V. Fyodorov and C.~Nadal.
\newblock Critical behavior of the number of minima of a random landscape at
  the glass transition point and the tracy-widom distribution.
\newblock {\em Phys. Rev. Lett.}, 109:167203, Jul 2012.

\bibitem{auf:2013a}
A.~Auffinger, G.~Ben~Arous, and J.~Cerny.
\newblock {Random matrices and complexity of spin glasses.}
\newblock {\em Comm. Pure. Appl. Math.}, 66:165--201, 2013.

\bibitem{auf:2013b}
A.~Auffinger and G.~Ben~Arous.
\newblock {Complexity of random smooth functions on the high-dimensional
  sphere.}
\newblock {\em Ann. Prob.}, 41:4214--4247, 2013.

\bibitem{2013JSP...tmp..205F}
Y.~V. {Fyodorov} and P.~{Le Doussal}.
\newblock {Topology Trivialization and Large Deviations for the Minimum in the
  Simplest Random Optimization}.
\newblock {\em Journal of Statistical Physics}, September 2013.

\bibitem{wainrib2013topological}
Gilles Wainrib and Jonathan Touboul.
\newblock Topological and dynamical complexity of random neural networks.
\newblock {\em Physical review letters}, 110(11):118101, 2013.

\bibitem{Cheng:2015}
Dan Cheng and Armin Schwartzman.
\newblock {On the Explicit Height Distribution and Expected Number of Local
  Maxima of Isotropic Gaussian Random Fields}.
\newblock {\em ArXiv e-prints}, July 2015.

\bibitem{Nicolaescu:2014}
L.~I. Nicolaescu.
\newblock {Complexity of random smooth functions on compact manifolds.}
\newblock {\em Indiana Univ. Math. J.}, 63:1037--1065, 2014.

\bibitem{FLL:2014}
Yan~V. Fyodorov, A~Lerario, and E.~Lundberg.
\newblock {On the number of connected components of random algebraic
  hypersurfaces}.
\newblock {\em ArXiv e-prints}, 2014.

\bibitem{PhysRevLett.108.170601}
Yan~V. Fyodorov, Ghaith~A. Hiary, and Jonathan~P. Keating.
\newblock Freezing transition, characteristic polynomials of random matrices,
  and the riemann zeta function.
\newblock {\em Phys. Rev. Lett.}, 108:170601, Apr 2012.

\bibitem{kac1948average}
Mark Kac.
\newblock On the average number of real roots of a random algebraic equation
  (ii).
\newblock {\em Proceedings of the London Mathematical Society}, 2(1):390--408,
  1948.

\bibitem{farahmand1986average}
Kambiz Farahmand.
\newblock On the average number of real roots of a random algebraic equation.
\newblock {\em The Annals of Probability}, pages 702--709, 1986.

\bibitem{bogomolny1992distribution}
E~Bogomolny, O~Bohigas, and P~Leboeuf.
\newblock Distribution of roots of random polynomials.
\newblock {\em Physical Review Letters}, 68(18):2726, 1992.

\bibitem{edelman1995many}
Alan Edelman and Eric Kostlan.
\newblock How many zeros of a random polynomial are real?
\newblock {\em Bulletin of the American Mathematical Society}, 32(1):1--37,
  1995.

\bibitem{kostlan2002expected}
Eric Kostlan.
\newblock On the expected number of real roots of a system of random polynomial
  equations.
\newblock {\em Foundations of computational mathematics (Hong Kong, 2000),
  World Sci. Publishing, River Edge, NJ}, pages 149--188, 2002.

\bibitem{azais2005roots}
Jean-Marc Aza{\"\i}s and Mario Wschebor.
\newblock On the roots of a random system of equations. the theorem of shub and
  smale and some extensions.
\newblock {\em Foundations of Computational Mathematics}, 5(2):125--144, 2005.

\bibitem{armentano2009random}
Diego Armentano, Mario Wschebor, et~al.
\newblock Random systems of polynomial equations. the expected number of roots
  under smooth analysis.
\newblock {\em Bernoulli}, 15(1):249--266, 2009.

\bibitem{malajovich2004high}
Gregorio Malajovich and J~Maurice Rojas.
\newblock High probability analysis of the condition number of sparse
  polynomial systems.
\newblock {\em Theoretical computer science}, 315(2):525--555, 2004.

\bibitem{rojas1996average}
J~Maurice Rojas.
\newblock On the average number of real roots of certain random sparse
  polynomial systems.
\newblock {\em LECTURES IN APPLIED MATHEMATICS-AMERICAN MATHEMATICAL SOCIETY},
  32:689--700, 1996.

\bibitem{dean2006large}
David~S Dean and Satya~N Majumdar.
\newblock Large deviations of extreme eigenvalues of random matrices.
\newblock {\em Physical review letters}, 97(16):160201, 2006.

\bibitem{tao2012random}
Terence Tao and Van Vu.
\newblock Random matrices: Universality of local spectral statistics of
  non-hermitian matrices.
\newblock {\em arXiv preprint arXiv:1206.1893}, 2012.

\bibitem{bhargava2015probability}
M~Bhargava, JE~Cremona, TA~Fisher, NG~Jones, and JP~Keating.
\newblock What is the probability that a random integral quadratic form in $ n
  $ variables has an integral zero?
\newblock {\em arXiv preprint arXiv:1502.05992}, 2015.

\bibitem{2013arXiv1307.2379F}
Y.~V {Fyodorov}.
\newblock {High-Dimensional Random Fields and Random Matrix Theory}.
\newblock {\em ArXiv e-prints}, July 2013.

\bibitem{Note1}
See \cite {bleher2015two} for recent progress on this problem.

\bibitem{Li:2003}
T~Y Li.
\newblock Solving polynomial systems by the homotopy continuation method.
\newblock {\em Handbook of numerical analysis}, XI:209--304, 2003.

\bibitem{SW:05}
Andrew~J Sommese and Charles~W Wampler.
\newblock {\em The numerical solution of systems of polynomials arising in
  Engineering and Science}.
\newblock World Scientific Publishing Company, 2005.

\bibitem{BHSW13}
D.J. Bates, J.D. Hauenstein, A.J. Sommese, and C.W. Wampler.
\newblock {\em Numerically solving polynomial systems with Bertini}, volume~25.
\newblock SIAM, 2013.

\bibitem{BCSS}
Lenore Blum, Felipe Cucker, Michael Shub, and Steve Smale.
\newblock Complexity and real computation. 1998.

\bibitem{smale1986newton}
Steve Smale.
\newblock {\em Newton’s method estimates from data at one point}.
\newblock Springer, 1986.

\bibitem{hauenstein2012algorithm}
Jonathan~D Hauenstein and Frank Sottile.
\newblock Algorithm 921: alphacertified: certifying solutions to polynomial
  systems.
\newblock {\em ACM Transactions on Mathematical Software (TOMS)}, 38(4):28,
  2012.

\bibitem{Mehta:2013zia}
Dhagash Mehta, Jonathan~D Hauenstein, and David~J Wales.
\newblock Communication: Certifying the potential energy landscape.
\newblock {\em The Journal of chemical physics}, 138(17):171101, 2013.

\bibitem{Mehta:2014gia}
Dhagash Mehta, Jonathan~D. Hauenstein, and David~J. Wales.
\newblock {Certification and the Potential Energy Landscape}.
\newblock {\em J.Chem.Phys.}, 140:224114, 2014.

\bibitem{Mehta:2009}
Dhagash Mehta.
\newblock {Lattice vs. Continuum: Landau Gauge Fixing and 't Hooft-Polyakov
  Monopoles}.
\newblock {\em Ph.D. Thesis, The Uni. of Adelaide, Australasian Digital Theses
  Program}, 2009.

\bibitem{Mehta:2009zv}
Dhagash Mehta, Andre Sternbeck, Lorenz von Smekal, and Anthony~G Williams.
\newblock {Lattice Landau Gauge and Algebraic Geometry}.
\newblock {\em PoS}, QCD-TNT09:025, 2009.

\bibitem{Mehta:2011xs}
Dhagash Mehta.
\newblock {Finding All the Stationary Points of a Potential Energy Landscape
  via Numerical Polynomial Homotopy Continuation Method}.
\newblock {\em Phys.Rev.}, E84:025702, 2011.

\bibitem{Mehta:2011wj}
Dhagash Mehta.
\newblock {Numerical Polynomial Homotopy Continuation Method and String Vacua}.
\newblock {\em Adv.High Energy Phys.}, 2011:263937, 2011.

\bibitem{Kastner:2011zz}
Michael Kastner and Dhagash Mehta.
\newblock {Phase Transitions Detached from Stationary Points of the Energy
  Landscape}.
\newblock {\em Phys.Rev.Lett.}, 107:160602, 2011.

\bibitem{Maniatis:2012ex}
Markos Maniatis and Dhagash Mehta.
\newblock {Minimizing Higgs Potentials via Numerical Polynomial Homotopy
  Continuation}.
\newblock {\em Eur.Phys.J.Plus}, 127:91, 2012.

\bibitem{Mehta:2012wk}
Dhagash Mehta, Yang-Hui He, and Jonathan~D. Hauenstein.
\newblock {Numerical Algebraic Geometry: A New Perspective on String and Gauge
  Theories}.
\newblock {\em JHEP}, 1207:018, 2012.

\bibitem{Hughes:2012hg}
Ciaran Hughes, Dhagash Mehta, and Jon-Ivar Skullerud.
\newblock {Enumerating Gribov copies on the lattice}.
\newblock {\em Annals Phys.}, 331:188--215, 2013.

\bibitem{Mehta:2012qr}
Dhagash Mehta, Jonathan~D. Hauenstein, and Michael Kastner.
\newblock Energy-landscape analysis of the two-dimensional nearest-neighbor
  $\phi^{4}$ model.
\newblock {\em Phys. Rev. E}, 85:061103, Jun 2012.

\bibitem{He:2013yk}
Yang-Hui He, Dhagash Mehta, Matthew Niemerg, Markus Rummel, and Alexandru
  Valeanu.
\newblock {Exploring the Potential Energy Landscape Over a Large
  Parameter-Space}.
\newblock {\em JHEP}, 1307:050, 2013.

\bibitem{morgan1987homotopy}
Alexander Morgan and Andrew Sommese.
\newblock A homotopy for solving general polynomial systems that respects
  m-homogeneous structures.
\newblock {\em Applied Mathematics and Computation}, 24(2):101--113, 1987.

\bibitem{morgan1987computing}
Alexander Morgan and Andrew Sommese.
\newblock Computing all solutions to polynomial systems using homotopy
  continuation.
\newblock {\em Applied Mathematics and Computation}, 24(2):115--138, 1987.

\bibitem{BHSW06}
Daniel~J. Bates, Jonathan~D. Hauenstein, Andrew~J. Sommese, and Charles~W.
  Wampler.
\newblock Available at www.nd.edu/$\sim$sommese/bertini.

\bibitem{Greene:2013ida}
Brian Greene, David Kagan, Ali Masoumi, Dhagash Mehta, Erick~J. Weinberg,
  et~al.
\newblock {Tumbling through a landscape: Evidence of instabilities in
  high-dimensional moduli spaces}.
\newblock {\em Phys.Rev.}, D88(2):026005, 2013.

\bibitem{dedieu2008number}
Jean-Pierre Dedieu and Gregorio Malajovich.
\newblock On the number of minima of a random polynomial.
\newblock {\em Journal of Complexity}, 24(2):89--108, 2008.

\bibitem{mumford1995algebraic}
David Mumford.
\newblock {\em ALgebraic Geometry: Complex projective varieties. vol. 1},
  volume~1.
\newblock Springer Science \& Business Media, 1995.

\bibitem{2002JChPh.116.3777D}
J.~P.~K. {Doye} and D.~J. {Wales}.
\newblock {Saddle points and dynamics of Lennard-Jones clusters, solids, and
  supercooled liquids}.
\newblock {\em Journal of Chem. Phys.}, 116:3777--3788, 2002.

\bibitem{2003JChPh.11912409W}
D.~J. {Wales} and J.~P.~K. {Doye}.
\newblock {Stationary points and dynamics in high-dimensional systems}.
\newblock {\em Journal of Chem. Phys.}, 119:12409--12416, December 2003.

\bibitem{Hughes:2014moa}
Ciaran Hughes, Dhagash Mehta, and David~J Wales.
\newblock {An Inversion-Relaxation Approach for Sampling Stationary Points of
  Spin Model Hamiltonians}.
\newblock {\em J.Chem.Phys.}, 140:194104, 2014.

\bibitem{Mehta:2013fza}
Dhagash Mehta, Daniel~A. Stariolo, and Michael Kastner.
\newblock {Energy landscape of the finite-size spherical three-spin glass
  model}.
\newblock {\em Phys.Rev.}, E87(5):052143, 2013.

\bibitem{mehta2014algebraic}
Dhagash Mehta, Noah Daleo, Florian D{\"o}rfler, and Jonathan~D Hauenstein.
\newblock Algebraic geometrization of the kuramoto model: Equilibria and
  stability analysis.
\newblock {\em arXiv preprint arXiv:1412.0666}, 2014.

\bibitem{altland1997nonstandard}
Alexander Altland and Martin~R Zirnbauer.
\newblock Nonstandard symmetry classes in mesoscopic normal-superconducting
  hybrid structures.
\newblock {\em Physical Review B}, 55(2):1142, 1997.

\bibitem{wales2001microscopic}
David~J Wales.
\newblock A microscopic basis for the global appearance of energy landscapes.
\newblock {\em Science}, 293(5537):2067--2070, 2001.

\bibitem{PhysRevE.91.022133}
Dhagash Mehta, Jonathan~D. Hauenstein, Matthew Niemerg, Nicholas~J. Simm, and
  Daniel~A. Stariolo.
\newblock Energy landscape of the finite-size mean-field 2-spin spherical model
  and topology trivialization.
\newblock {\em Phys. Rev. E}, 91:022133, Feb 2015.

\bibitem{smale1998mathematical}
Steve Smale.
\newblock Mathematical problems for the next century.
\newblock {\em The Mathematical Intelligencer}, 20(2):7--15, 1998.

\bibitem{bleher2015two}
Pavel~M Bleher, Yushi Homma, and Roland~KW Roeder.
\newblock Two-point correlation functions and universality for the zeros of
  systems of so (n+ 1)-invariant gaussian random polynomials.
\newblock {\em arXiv preprint arXiv:1502.01427}, 2015.

\end{thebibliography}

\end{document}